\documentclass[aps,pra,twocolumn,superscriptaddress,nofootinbib,longbibliography,floatfix]{revtex4-2}

\usepackage{tikz}
\usetikzlibrary{arrows.meta,positioning,calc}
\usepackage{xcolor}

\usepackage{iftex}
\ifPDFTeX\pdfoutput=1\fi
\usepackage[T1]{fontenc}
\usepackage{amsmath}
\usepackage{amssymb}
\usepackage{amsthm}
\usepackage{bm}
\usepackage{mathtools}
\usepackage{graphicx}
\usepackage{quantikz}
\usepackage{hyperref}
\hypersetup{colorlinks=true, allcolors={blue!70!black}}
\usepackage[capitalise]{cleveref}
\usepackage{placeins}
\usepackage{xcolor}

\newsavebox{\TCLCircuitBox}
\newlength{\CircuitPageWidth}
\newcommand{\CircuitFont}{\fontsize{8}{9.6}\selectfont}
\tikzset{tcl circuit/.style={row sep=0.18cm,column sep=0.28cm}}
\tikzset{tcl dense circuit/.style={tcl circuit,column sep=0.07cm}}
\newcommand{\CircuitDiagram}[2][\linewidth]{%
  \begingroup
  \sbox{\TCLCircuitBox}{\CircuitFont#2}%
  \ifdim\wd\TCLCircuitBox>#1\relax
    \resizebox{#1}{!}{\usebox{\TCLCircuitBox}}%
  \else
    \usebox{\TCLCircuitBox}%
  \fi
  \endgroup
}
\newcommand{\CircuitPanelTitle}[1]{%
  {\footnotesize\bfseries#1\par}\smallskip
}

\newtheorem{theorem}{Theorem}
\newtheorem{lemma}{Lemma}
\newtheorem{proposition}{Proposition}
\theoremstyle{definition}
\newtheorem{definition}{Definition}

\providecommand{\TCL}{}\renewcommand{\TCL}{\mathrm{TCL}}
\providecommand{\ph}{}\renewcommand{\ph}{\mathrm{ph}}
\providecommand{\at}{}\renewcommand{\at}{\mathrm{at}}
\providecommand{\intt}{}\renewcommand{\intt}{\mathrm{int}}
\providecommand{\hop}{}\renewcommand{\hop}{\mathrm{hop}}
\providecommand{\dg}{}\renewcommand{\dg}{\mathrm{diag}}
\providecommand{\loc}{}\renewcommand{\loc}{\mathrm{loc}}
\providecommand{\be}{}\renewcommand{\be}{\mathrm{be}}
\providecommand{\PREP}{}\renewcommand{\PREP}{\operatorname{PREP}}
\providecommand{\SELECT}{}\renewcommand{\SELECT}{\operatorname{SELECT}}
\providecommand{\LOAD}{}\renewcommand{\LOAD}{\operatorname{LOAD}}
\providecommand{\DEST}{}\renewcommand{\DEST}{\operatorname{DEST}}
\providecommand{\sgn}{}\renewcommand{\sgn}{\operatorname{sgn}}
\providecommand{\qph}{}\renewcommand{\qph}{q_{\ph}}
\providecommand{\nmax}{}\renewcommand{\nmax}{n_{\max}}
\providecommand{\ket}{}\renewcommand{\ket}[1]{\left|#1\right\rangle}
\providecommand{\bra}{}\renewcommand{\bra}[1]{\left\langle#1\right|}
\providecommand{\proj}{}\renewcommand{\proj}[1]{\left|#1\right\rangle\left\langle#1\right|}
\providecommand{\norm}{}\renewcommand{\norm}[1]{\left\lVert#1\right\rVert}
\providecommand{\abs}{}\renewcommand{\abs}[1]{\left\lvert#1\right\rvert}

\begin{document}

\title{Simulating closed- and open-system dynamics of Tavis--Cummings lattices on fault-tolerant quantum computers}

\author{Soumya Sarkar}
\affiliation{Centre for Quantum Software and Information, School of Computer Science, Faculty of Engineering \& Information Technology, University of Technology Sydney, NSW 2007, Australia}

\affiliation{BTQ Technologies, 16-104 555 Burrard Street, Vancouver BC, V7X 1M8 Canada}

\author{Rakshit Gharat}
\affiliation{School of Mathematics \& Physical Sciences, Macquarie University, NSW 2109, Australia}

\author{Nam Nguyen}
\affiliation{Department of Physics and Astronomy, George Mason University, Fairfax, VA, USA}
\affiliation{Quantum Science and Engineering Center, George Mason University, Fairfax, VA, USA}
\affiliation{Visa Research, Foster City, CA, USA}

\author{Gavin Brennen}

\affiliation{BTQ Technologies, 16-104 555 Burrard Street, Vancouver BC, V7X 1M8 Canada}

\affiliation{School of Mathematics \& Physical Sciences, Macquarie University, NSW 2109, Australia}

\begin{abstract}
Tavis--Cummings lattices, arrays of coupled cavities each containing several emitters, host polaritonic many-body dynamics whose real-time correlations are costly to compute classically in two dimensions, especially when unequal emitter relaxation or dephasing rates break permutation symmetry. Here we analyze the logical resources that a fault-tolerant quantum computer needs to simulate closed- and open-system dynamics of an atom-resolved lattice with $L$ cavities, $N$ emitters per cavity, and photon cutoff $n_{\max}$, and to estimate target observables. Building on ladder-operator block encoding, we construct circuits for photon hopping and local light--matter exchange and combine them into a block encoding of the complete lattice Hamiltonian. By updating the source and destination photon occupations in place, the hopping circuit eliminates the auxiliary $L\lceil\log_2(n_{\max}+1)\rceil$-qubit photon register used by two-register implementations. Across three uniform-coefficient examples with $16$ to $256$ cavities, the controlled walk uses $4.75$--$8.11\%$ fewer Toffoli gates than the paired two-register implementation under matched shared preparation, precision, control, and workspace conventions. For closed systems, we implement real-time evolution with generalized quantum signal processing and bound the cost of estimating retarded cavity spectral functions. For open systems with photon loss, emitter relaxation, and dephasing, we use a sampled first-order dilation and bound the runs and timesteps needed to estimate transient photon statistics such as $g^{(2)}(0)$. We verify coherent gate and qubit estimates through explicit circuit decompositions, and test operator actions, observable definitions, and open-system timestep convergence using small-instance gate simulations and independent matrix calculations. With state-preparation and measurement costs kept explicit, these logical resource estimates provide a concrete reference for assessing the fault-tolerant cost of simulating light--matter lattices.
\end{abstract}

\maketitle

\section{Introduction}
\label{sec:intro}
An array of coupled cavities combines local light-matter exchange with the transport of photons between sites \cite{hartmann2008review,noh2017review,hartmann2006strongly}. The Jaynes-Cummings model describes the coupling of one two-level emitter to a cavity mode \cite{Jaynes63}, while the Tavis-Cummings model extends this interaction to an ensemble of emitters \cite{TavisCummings1968}. Allowing photons to hop between cavities extends these models to the Jaynes--Cummings lattice~\cite{greentree2006quantum,angelakis2007photon} and the Tavis--Cummings lattice~\cite{knap2010tavis}, respectively, with the resulting polaritonic behavior governed by hopping, detuning, and emitter saturation~\cite{aichhorn2008detuning,koch2009superfluid}. Related proposals use collective cavity--emitter interactions to prepare entangled atomic states and implement nonlocal multiqubit gates~\cite{jandura2024nonlocal,srivastava2025polariton}. Calculations of spectral functions \cite{pippan2009spectra,Knap2011Spectral} and non-equilibrium dynamics~\cite{noh2017review} can help distinguish collective modes, resolve interaction effects, and assess how imperfections \cite{rossini2007mott} modify the response of these arrays. In cavity QED, transmission spectra and equal-time photon correlations diagnose ensemble-induced strong coupling and photon blockade~\cite{schutz2020ensemble}.

Exact calculations of these quantities quickly become expensive. For identical emitters, permutation symmetry reduces each cavity's emitter space, and conservation of the total excitation number splits closed-system evolution into independent excitation sectors. For a $4\times4$ lattice with four emitters per cavity and photon cutoff $\nmax=15$ at unit filling, these reductions take $2^{128}$ basis states down to $2^{40.4}$, but a single state vector still occupies about 23~TB (\cref{sec:classical}). Sign-problem-free quantum Monte Carlo helps with equilibrium properties for suitable hopping conventions~\cite{hohenadler2011jch}, but none of these tools guarantees efficient calculation of real-time many-body correlations in two dimensions.  Open dynamics introduces an additional consideration: when emitters in the same cavity have different relaxation or dephasing rates, permutation symmetry is generally broken, and reductions based on equivalent emitters no longer apply.

Digital quantum simulations of related light--matter systems have used product formulas for photon hopping in Jaynes--Cummings--Hubbard cavity arrays~\cite{tudorovskaya2024cavity}, extended Jaynes--Cummings models~\cite{nguyen2025ejcm}, and coherent or dissipative single-cavity models~\cite{sims2025nonresonant,mezzacapo2014digital,langford2017experimentally,haukisalmi2025noisy}. The open Tavis--Cummings simulation of Ref.~\cite{sims2025nonresonant} uses established Lindblad simulation methods~\cite{cleve2017effective} to study both undriven relaxation and pumped steady states. Ladder-operator block encoding (LOBE) provides a general method for implementing bosonic and fermionic operators directly in their occupation registers~\cite{simon2025ladder}. Kane et al.~\cite{Kane2025blockencodingbosons} built alternative block encodings of lattice-boson Hamiltonians from quantum singular value transformation (QSVT)~\cite{gilyen2019quantum}, quantum eigenvalue transformation of unitary matrices (QETU)~\cite{dong2022ground}, and linear operators via exponentiation and linear combination of unitaries (LOVE-LCU), and compared them at the circuit level. For the lattice model, the remaining resource questions concern the cost of addressing a coherently selected pair of cavities, the amount of occupation-dependent coefficient data that must be loaded, and the cost of combining photon hopping with local exchange. A further question is how these circuit costs contribute to the resources required to estimate a specified observable. Resolving these questions is necessary to turn an asymptotic Hamiltonian-simulation bound into a useful resource estimate, as previous studies have done for condensed-matter and chemistry models~\cite{childs2018toward,Babbush2018,nguyen2026corrosion}.

In this work, we analyze the logical resources required to simulate a two-dimensional atom-resolved Tavis--Cummings lattice. The Hamiltonian allows site-dependent hopping amplitudes, polariton potentials, and light--matter couplings. The emitters within each cavity share the same coherent parameters, but may have different relaxation and dephasing rates. Such differences occur in cavity-QED experiments with individually characterized emitters~\cite{mlynek2014observation}. We therefore represent each emitter by its own qubit, so that its dissipative channels can be applied individually. Building on the ladder-operator block-encoding framework of \cite{simon2025ladder}, we construct circuits for photon hopping and local light--matter exchange and combine them into a block encoding of the lattice Hamiltonian. A photon hop changes only the source and destination occupations, so our circuit decrements the source register and increments the destination register in place. A sparse-access implementation~\cite{berry2015nearly} instead writes the updated photon configuration into a second register. We build both with the same treatment of the bosonic coefficients and local exchange, so the comparison isolates the in-place update. On the largest of our three instances, with 256 cavities, the in-place update removes 1,024 logical qubits and reduces the peak qubit count for one walk from 3,423 to 2,399 (\cref{tab:v3-coherent-compare}). Across the three instances, it also uses $4.75$--$8.11\%$ fewer Toffoli gates per controlled walk step than the cheapest two-register implementation we construct. Two further savings apply to both constructions: the two branches of each walk step share one coefficient preparation, and removing the uniform part of the polariton potential lowers the normalization by 34--46\% in the three instances, which reduces the number of walk steps needed. We use the resulting block encoding with generalized quantum signal processing (GQSP) for real-time evolution~\cite{PRXQuantum.5.020368} and connect its cost to the estimation of photon spectra and transient photon statistics.We report Toffoli-family operations, arbitrary rotations, and logical qubits separately, and account for both the cost of each block-encoding call and the number of calls needed for the simulation. We further implement finite-precision rotations, including the outer selection and GQSP signal rotations, through reversible arithmetic on reusable phase-gradient catalysts~\cite{sunderhauf2024block,Hindlycke2024}. Combining these circuits with a seven-$T$ \cite{selinger2013tdepth,amy2013meet} decomposition of each coherent Toffoli gives explicit logical $T$-count bounds for the specified closed-and sampled open-system constructions, with catalyst preparation accounted for separately. Product formulas exploit the local structure of the Hamiltonian~\cite{childs2019nearly,childs2021trotter} and may offer lower practical costs~\cite{childs2018toward}, even though signal-processing methods scale better with accuracy~\cite{Low2019hamiltonian,low2017optimal,gilyen2019quantum}; we return to this trade-off in \cref{sec:discussion}.

For closed dynamics, we consider the retarded photon spectrum, obtained from correlation functions that describe adding a photon to, or removing one from, a specified stationary state~\cite{somma2002simulating,wecker2015solving}. We account for the evolution time, simulation accuracy, and repeated measurements needed to estimate the spectrum over a finite observation window, and leave the cost of preparing the initial state as a separate term. For open dynamics, we use a sampled first-order unitary dilation~\cite{PRXQuantum.5.020332} to estimate photon populations and equal-time second-order correlations during relaxation. The model includes photon loss, emitter relaxation, and dephasing, but no external drive or pump. We therefore study the transient behavior of an initially excited lattice.

Numerical matrix calculations check the operator identities, while small-instance gate simulations test projected block encodings, controls, adjoints, and catalyst return. Recursive decomposition of executable Qualtran circuits \cite{harrigan2024qualtran} reproduces the specified coherent gate counts and logical-qubit allocations. Exact diagonalization checks the spectral calculation on a small lattice, and a two-cavity simulation checks the open-system method's timestep convergence. These are logical resource
estimates under the stated counting conventions; physical error-correction overhead is not included.

The flowchart in \cref{fig:tcl-workflow} summarizes the organization of the paper. We introduce the atom-resolved Tavis--Cummings lattice in \cref{sec:model} and discuss the cost of its classical simulation in \cref{sec:classical}. In \cref{sec:hop,sec:loc}, we construct block encodings for photon hopping and for the on-site diagonal and light--matter exchange terms, respectively. We combine these constructions in \cref{sec:full} to encode the full Hamiltonian, derive the cost of closed-system simulation, and compare this in-place (direct) photon update with two-register implementations. In \cref{sec:open}, we extend the simulation to open dynamics using a sampled first-order {dilation}. We then analyze the cost of estimating retarded photon spectra and transient photon statistics in \cref{sec:endtoend}. Finally, we discuss our findings and their limitations in \cref{sec:discussion}.

\definecolor{tclWorkflowVisaBlue}{HTML}{1434CB}
\definecolor{tclWorkflowVisaGold}{HTML}{F7B600}
\colorlet{tclWorkflowTeal}{teal}
\colorlet{tclWorkflowBerry}{purple}
\begin{figure*}
\centering
\resizebox{\textwidth}{!}{\begin{tikzpicture}[
  >=Latex,
  font=\small,
  flow/.style={draw=black!65, line width=0.6pt,
    rounded corners=2pt, align=center, inner xsep=5pt,
    inner ysep=4pt, minimum height=12mm},
  prep/.style={flow,fill=tclWorkflowVisaBlue!14!white},
  ham/.style={flow,fill=tclWorkflowTeal!18!white},
  diss/.style={flow,fill=tclWorkflowVisaGold!22!white},
  proc/.style={flow,fill=tclWorkflowBerry!14!white},
  result/.style={flow,fill=gray!7},
  arr/.style={-{Latex[length=2mm]},line width=0.75pt,draw=black!70},
  branch/.style={line width=0.7pt,draw=black!70},
  arrAdded/.style={-{Latex[length=2mm]},line width=0.75pt,draw=black!70},
  branchAdded/.style={line width=0.7pt,draw=black!70}
]
\node[prep,text width=15.7cm,minimum height=9mm] (input) at (0,0) {
 \textbf{Model and simulation inputs}\\[-1pt]
 $L=L_xL_y,\;N,\;n_{\max},\;t_{ij},\;g_i,\;\nu_i,\;\Delta,\;
 \kappa_i,\;\gamma_{1,i,r},\;\gamma_{\phi,i,r},\;T_{\max},\;T,\;\varepsilon,\;\rho(0)$};
\node[ham,text width=4.5cm,minimum height=15mm] (hop) at (-5,-1.62) {
 \textbf{Hopping block} $U_{\mathrm{hop}}$\\[-1pt]
 Bond PREP; $O_A,O_B$; selected $S_\ell$\\[-1pt]
 \footnotesize\cref{prop:hop-block-encoding}};
\node[ham,text width=4.5cm,minimum height=15mm] (local) at (0,-1.62) {
 \textbf{Local block} $U_{\mathrm{loc}}$\\[-1pt]
 Pauli diagonals; shared LOBE exchange\\[-1pt]
 \footnotesize\cref{fig:lobe}};
\node[diss,text width=4.5cm,minimum height=15mm] (jumps) at (5,-1.62) {
 \textbf{Dissipative channels}\\[-1pt]
 Photon loss; emitter relaxation\\[-1pt]
 and dephasing\\[-1pt]
 $\{V_j\}$ and $\lambda_{\mathcal D}=\sum_j\lVert V_j\rVert^2$};
\node[ham,text width=9.6cm,minimum height=12mm] (be) at (-2.5,-3.73) {
 \textbf{Complete TCL Hamiltonian block encoding}\\[-1pt]
 $U_{\mathrm{TCL}}$: outer LCU of $U_{\mathrm{hop}}$ and $U_{\mathrm{loc}}$\\[-1pt]
 $(\alpha_{\mathrm{TCL}},a,\delta_{\mathrm{be}})$ block encoding};
\node[diss,text width=4.5cm,minimum height=12mm] (sample) at (5,-3.73) {
 \textbf{Jump sampling}\\[-1pt]
 $p_j=\lVert V_j\rVert^2/\lambda_{\mathcal D}$\\[-1pt]
 $\widetilde V_j=V_j/\sqrt{p_j}$};
\node[proc,text width=6.6cm,minimum height=15mm] (closed) at (-5,-5.66) {
 \textbf{Closed-system time evolution}\\[-1pt]
 Hermitianization and GQSP\\[-1pt]
 $U_{\mathrm{TCL}}\longrightarrow e^{-iH_{\mathrm{TCL}}\tau}$};
\node[proc,text width=6.6cm,minimum height=15mm] (open) at (5,-5.66) {
 \textbf{Open-system time evolution}\\[-1pt]
 Sampled one-qubit dilation; GQSP per step\\[-1pt]
 Repeat $N_{\rm step}=T/\Delta t$ steps};
\node[result,text width=6.6cm,minimum height=14mm] (spec) at (-5,-7.61) {
 \textbf{Retarded spectral response}\\[-1pt]
 $C_{ij}^{\pm}(\tau)\ \longrightarrow\ G_{ij}^{R}(\tau)
 \ \longrightarrow\ A_{ij}(\omega)$\\[-1pt]
 Interferometric correlator estimation};
\node[result,text width=6.6cm,minimum height=14mm] (photo) at (5,-7.61) {
 \textbf{Transient photon statistics}\\[-1pt]
 $\overline n_i(T)$ and $g_i^{(2)}(0;T)$\\[-1pt]
 Computational-basis sampling};
\node[prep,text width=15.7cm,minimum height=7mm] (resource) at (0,-9.00) {
 \textbf{Resource accounting:}\quad $\alpha$, $a$, $\delta_{\rm be}$, $\mathcal C_{\rm Tof}$, $\mathcal C_R$; simulation queries and measurement shots};
\draw[arrAdded] ([xshift=-50mm]input.south) -- (hop.north);
\draw[arrAdded] (input.south) -- (local.north);
\draw[arrAdded] ([xshift=50mm]input.south) -- (jumps.north);
\coordinate (mergeb) at (-2.5,-2.63);
\draw[branchAdded] (hop.south) |- (mergeb);
\draw[branchAdded] (local.south) |- (mergeb);
\draw[arrAdded] (mergeb) -- (be.north);
\draw[arr] (jumps.south) -- (sample.north);
\coordinate (branch2) at (-2.5,-4.65);
\draw[branchAdded] (be.south) -- (branch2);
\draw[branchAdded] (-5,-4.65) -- (3.6,-4.65);
\draw[arrAdded] (-5,-4.65) -- (closed.north);
\draw[arrAdded] (3.6,-4.65) -- ([xshift=-14mm]open.north);
\draw[arrAdded] (sample.south) -- (open.north);
\draw[arr] (closed.south) -- (spec.north);
\draw[arr] (open.south) -- (photo.north);
\end{tikzpicture}}
\caption{End-to-end workflow for the atom-resolved TCL simulation algorithm. The hopping and local operators are block encoded and combined through an outer LCU. The resulting Hamiltonian encoding is used for closed-system GQSP evolution and, together with sampled Lindblad jump operators, for open-system dilation. Spectral correlators and transient photon statistics are estimated using their corresponding measurement routines. Resource accounting applies across all stages.}
\label{fig:tcl-workflow}
\end{figure*}
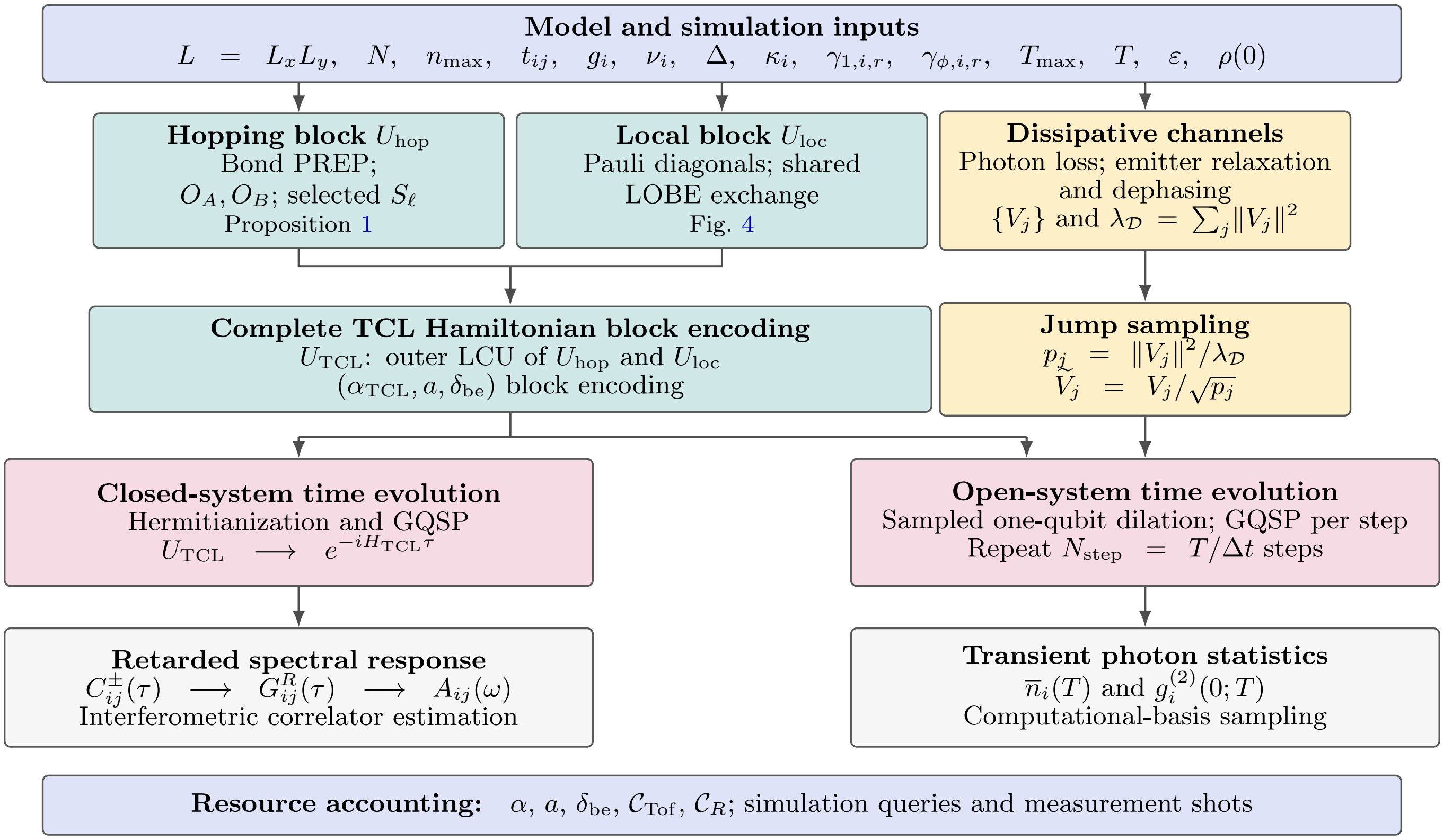

\section{The atom-resolved Tavis--Cummings lattice}
\label{sec:model}

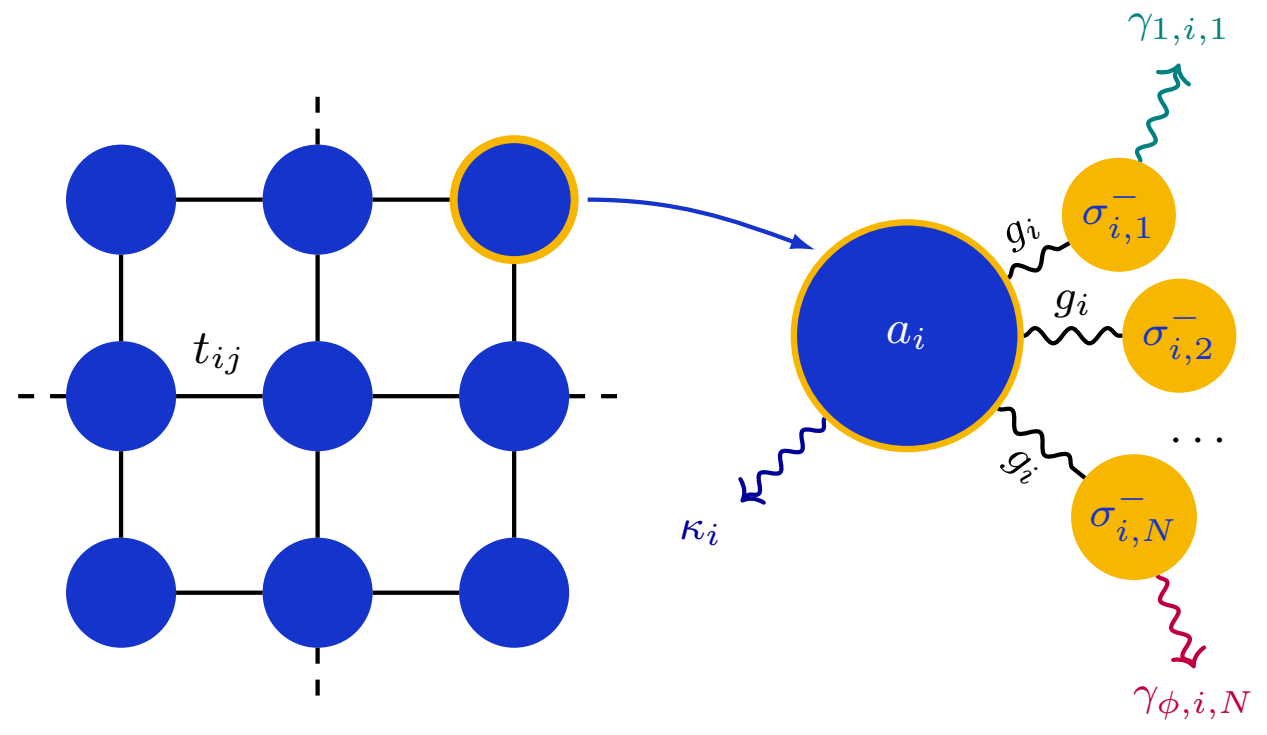
\begin{figure}
\centering
\definecolor{TCLVisaBlue}{HTML}{1434CB}
\definecolor{TCLVisaGold}{HTML}{F7B600}
\begin{tikzpicture}[
  scale=1,every node/.style={font=\small},
  cavity/.style={circle,draw=TCLVisaBlue,fill=TCLVisaBlue,
    line width=0.8pt,minimum size=0.7cm,text=white},
  selected cavity/.style={cavity,draw=TCLVisaGold,line width=1.5pt,
    minimum size=0.8cm},
  emitter/.style={circle,draw=TCLVisaGold,fill=TCLVisaGold,
    line width=0.7pt,inner sep=1pt,minimum size=0.55cm,text=TCLVisaBlue,
    font=\footnotesize}
]

\node[cavity] (c1) at (0,0)     {};
\node[cavity] (c2) at (1.3,0)   {};
\node[cavity] (c3) at (2.6,0)  {};
\node[cavity] (c4) at (0,1.3)   {};
\node[cavity] (c5) at (1.3,1.3) {};
\node[cavity] (c6) at (2.6,1.3) {};
\node[cavity] (c7) at (0,2.6)   {};
\node[cavity] (c8) at (1.3,2.6) {};
\node[selected cavity] (c9) at (2.6,2.6) {};

\draw[thick] (c1) -- (c2);
\draw[thick] (c2) -- (c3);
\draw[thick] (c4) -- (c5);
\draw[thick] (c5) -- (c6);
\draw[thick] (c7) -- (c8);
\draw[thick] (c8) -- (c9);
\draw[thick] (c1) -- (c4);
\draw[thick] (c4) -- (c7);
\draw[thick] (c2) -- (c5);
\draw[thick] (c5) -- (c8);
\draw[thick] (c3) -- (c6);
\draw[thick] (c6) -- (c9);
\node[font=\small] at ($(c4)!0.5!(c5)+(0,0.28)$) {$t_{ij}$};

\draw[thick,dashed] (c4) -- ++(-0.7,0);
\draw[thick,dashed] (c6) -- ++(0.7,0);
\draw[thick,dashed] (c2) -- ++(0,-0.7);
\draw[thick,dashed] (c8) -- ++(0,0.7);

\node[cavity,draw=TCLVisaGold,line width=1.2pt,minimum size=1.5cm] (cav) at (5.2,1.7) {$a_i$};

\draw[draw=TCLVisaBlue,line width=0.8pt,
  -{Latex[length=1.7mm,width=1.1mm]},shorten <=0.6mm,shorten >=0.6mm]
  (c9.east) to[out=0,in=160] (cav.north west);
\node[emitter] (e1) at (6.6,2.5) {$\sigma^-_{i,1}$};
\node[emitter] (e2) at (7.0,1.7) {$\sigma^-_{i,2}$};
\node[font=\small] (edots) at (7.15,1.0) {$\cdots$};
\node[emitter] (e3) at (6.7,0.5) {$\sigma^-_{i,N}$};

\draw[decorate,decoration={snake,amplitude=0.5mm,segment length=2.4mm},thick]
  (cav) -- (e1) node[midway,sloped,above,font=\footnotesize] {$g_i$};
\draw[decorate,decoration={snake,amplitude=0.5mm,segment length=2.4mm},thick]
  (cav) -- (e2) node[midway,sloped,above,font=\footnotesize] {$g_i$};
\draw[decorate,decoration={snake,amplitude=0.5mm,segment length=2.4mm},thick]
  (cav) -- (e3) node[midway,sloped,below,font=\footnotesize] {$g_i$};

\draw[->,decorate,decoration={snake,amplitude=0.45mm,segment length=2mm},thick,blue!60!black]
  (cav) -- ++(-1.1,-1.1) node[below left,font=\footnotesize] {$\kappa_i$};
\draw[->,decorate,decoration={snake,amplitude=0.45mm,segment length=2mm},thick,teal]
  (e1) -- ++(0.4,1.0) node[above,font=\footnotesize] {$\gamma_{1,i,1}$};
\draw[->,decorate,decoration={snake,amplitude=0.45mm,segment length=2mm},thick,purple]
  (e3) -- ++(0.4,-1.0) node[below,font=\footnotesize] {$\gamma_{\phi,i,N}$};

\end{tikzpicture}
\caption{Schematic of the atom-resolved Tavis--Cummings lattice. Left: a $3\times3$
patch of the square lattice of $L=L_xL_y$ cavities, each hosting one photon mode.
Solid lines denote nearest-neighbor hopping $t_{ij}$ (\cref{eq:H-hop}); dashed
segments indicate bonds continuing beyond the displayed patch. The simulated
lattice has periodic boundary conditions. Right: detail of the highlighted cavity
(gold outline in the left panel),
carrying its photon mode $a_i$ and $N$ individually addressed two-level emitters
$\sigma^-_{i,r}$, each exchanging one excitation with the cavity mode with coupling $g_i$
(\cref{eq:H-int}). The three dissipative channels of \cref{sec:dissipative} are shown
schematically: photon loss with rate $\kappa_i$, emitter relaxation with rate
$\gamma_{1,i,r}$, and pure dephasing with rate $\gamma_{\phi,i,r}$ (\cref{eq:jumps}).
Each emitter has its own relaxation and dephasing channels; the repeated channels
are omitted for clarity.}
\label{fig:lattice}
\end{figure}

\subsection{Hamiltonian}

We consider $L=L_xL_y$ cavities on a two-dimensional lattice with periodic boundary conditions, sketched in \cref{fig:lattice}. Cavity $i$ carries one photon mode with untruncated operators $a_i,a_i^\dagger$ satisfying $[a_i,a_j^\dagger]=\delta_{ij}$, and $N$ two-level emitters labeled $r=1,\ldots,N$. The Hilbert space is
\begin{equation}
\mathcal H=\left(\bigotimes_{i=1}^{L}\mathcal H^{(i)}_{\ph}\right)
\otimes\left(\bigotimes_{i=1}^{L}\bigotimes_{r=1}^{N}
\mathcal H^{(i,r)}_{\at}\right),
\label{eq:hilbert}
\end{equation}
where $\mathcal H^{(i)}_{\ph}$ is the Fock space of mode $i$, subsequently truncated to $\nmax+1$ levels, and $\mathcal H^{(i,r)}_{\at}$ is the two-dimensional space of one emitter. After truncation $\dim\mathcal H=(\nmax+1)^{L}2^{LN}$. On the truncated space the ladder operators act as
\begin{equation}
\label{eq:ladder}
\begin{aligned}
a_i\ket{m}&=\sqrt{m}\ket{m-1}, \\
a_i^\dagger\ket{m}&=\sqrt{m+1}\ket{m+1},
\end{aligned}
\end{equation}
with $a_i\ket{0}=0$ and $a_i^\dagger\ket{\nmax}=0$. The finite-dimensional operators therefore obey
\begin{equation}
[a_i,a_j^\dagger]=\delta_{ij}\left(I-(\nmax+1)\proj{\nmax}_i\right),
\label{eq:truncated-commutator}
\end{equation}
where $\proj{\nmax}_i$ acts on photon mode $i$. This boundary term is introduced by the photon cutoff. The hopping block encoding must therefore assign zero amplitude to any transfer that would raise an occupation from $\nmax$ to $\nmax+1$. Throughout we set $\hbar=1$ and measure all frequencies, couplings, and rates in units of a reference light--matter coupling $g$. Adding photon hopping between neighboring cavities to the Tavis--Cummings interaction, while retaining a separate label for every emitter, gives the rotating-wave Hamiltonian \cite{TavisCummings1968,knap2010tavis}
\begin{equation}
\begin{aligned}
H_\TCL&=
\underbrace{H_\hop}_{\text{photon hopping}}
+\underbrace{H_\dg}_{\text{detuning and potential}}
\\
&\quad+\underbrace{H_\intt}_{\text{light--matter exchange}},
\end{aligned}
\label{eq:H-split}
\end{equation}
with
\begin{align}
H_\hop&=-\sum_{\langle i,j\rangle}t_{ij}
\left(a_i^\dagger a_j+a_j^\dagger a_i\right),
\label{eq:H-hop}\\
H_\intt&=\sum_{i=1}^{L}\sum_{r=1}^{N}g_i
\left(a_i\sigma^+_{i,r}+a_i^\dagger\sigma^-_{i,r}\right),
\label{eq:H-int}
\end{align}
and
\begin{equation}
\begin{aligned}
H_\dg={}&-\Delta\sum_{i=1}^{L}\sum_{r=1}^{N}n^{\at}_{i,r}\\
&-\sum_{i=1}^{L}\nu_i\left(n^{\ph}_i
+\sum_{r=1}^{N}n^{\at}_{i,r}\right).
\end{aligned}
\label{eq:H-diag}
\end{equation}
Here $n^{\ph}_i=a_i^\dagger a_i$ and $\Delta=\omega_c-\omega_a$ is the common cavity--emitter detuning. The signed coefficient $\nu_i$ is a site-dependent polariton potential in the rotating frame. The amplitudes $t_{ij}$, potentials $\nu_i$, and couplings $g_i$ are real and may vary across the lattice; the $N$ emitters within a cavity share the coupling $g_i$. We keep the full $LN$-qubit emitter register because independent single-emitter channels can populate emitter-spin sectors outside the maximal-spin manifold. Unequal emitter rates also remove the permutation covariance that can simplify a density-matrix representation.

With the convention $Z\ket0=\ket0$ and $Z\ket1=-\ket1$, where $\ket0$ and $\ket1$ are the emitter ground and excited states, the emitter excitation number is
\begin{equation} 
n^{\at}_{i,r}=\proj{1}_{i,r}=\frac{I-Z^{\at}_{i,r}}{2}. \label{eq:atom-number} 
\end{equation} 
The local polariton number is $n_{p,i}=n^{\ph}_i+\sum_r n^{\at}_{i,r}$ and the total is $N_p=\sum_i n_{p,i}$. Each term of \cref{eq:H-split} either moves a photon between sites, is a function of the number operators, or converts one photon into one emitter excitation, so $[H_\TCL,N_p]=0$. The on-site exchange also conserves $n_{p,i}$, whereas hopping does not.

We use the frame rotating at the common cavity frequency \(\omega_c\). Write the laboratory-frame Hamiltonian as
\begin{equation}
H_{\rm lab}=\omega_c N_p+H_{\TCL},
\label{eq:H-lab}
\end{equation}
and define the rotating-frame transformation and state convention by
\begin{equation*}
\begin{aligned}
R_c(t)&=\exp(i\omega_c tN_p),\\
\lvert\psi_{\rm rot}(t)\rangle
&=R_c(t)\lvert\psi_{\rm lab}(t)\rangle.
\end{aligned}
\end{equation*}
The rotating-frame Hamiltonian is
\begin{equation}
R_c(t)H_{\rm lab}R_c^\dagger(t)
+i\dot R_c(t)R_c^\dagger(t)=H_{\TCL}.
\label{eq:rotating-frame}
\end{equation}
Indeed, \([H_{\TCL},N_p]=0\), so the first term is \(H_{\rm lab}\), while \(i\dot R_c(t)R_c^\dagger(t)=-\omega_cN_p\). The block-encoding normalization for \(H_{\TCL}\) therefore depends on the residual scales \(\Delta,\nu_i,t_{ij}\), and \(g_i\), not on the common carrier \(\omega_c\). This is an exact unitary change of frame for the stated Hamiltonian; no additional approximation is made in this transformation. With the \(e^{+i\omega\tau}\) Fourier convention used here, the laboratory- and rotating-frame spectra satisfy
\begin{equation*}
A^{\rm lab}_{ij}(\omega)=A^{\rm rot}_{ij}(\omega-\omega_c).
\end{equation*}
Thus spectral features shift upward by \(\omega_c\) on the laboratory frequency axis. For the open model,
\begin{align*}
R_c(t)a_iR_c^\dagger(t)&=e^{-i\omega_ct}a_i,\\
R_c(t)\sigma^-_{i,r}R_c^\dagger(t)&=e^{-i\omega_ct}\sigma^-_{i,r}.
\end{align*}
These phases cancel in the photon-loss and emitter-relaxation dissipators because \(\mathcal D[e^{i\phi(t)}V]=\mathcal D[V]\); the dephasing jump, proportional to \(\sigma^z_{i,r}\), is unchanged.

\paragraph{Removing a uniform polariton potential.}
We can also remove a uniform polariton potential by changing the rotating frame. For any real reference $\nu_*$ define
\begin{equation}
\widehat H_\TCL=H_\TCL+\nu_*N_p.
\label{eq:charge-frame-H}
\end{equation}
The transformation replaces $\nu_i$ by $\nu_i-\nu_*$ and leaves the other Hamiltonian terms unchanged. For uniform $\nu_i=\nu$ and $\Delta=0$, choose $\nu_*=\nu$; the full diagonal branch then vanishes. The photon populations and factorial moments are unchanged in this frame, and the stated dissipators remain invariant. With the Fourier convention of this paper,
\begin{equation}
\begin{aligned}
C_{ij,H}^{\pm}(\tau)&=e^{i\nu_*\tau}C_{ij,\widehat H}^{\pm}(\tau),\\
A_{ij,H}(\omega)&=A_{ij,\widehat H}(\omega+\nu_*),\\
E_{H,Q}&=E_{\widehat H,Q}-\nu_*Q .
\end{aligned}
\label{eq:charge-frame-restoration}
\end{equation}
The last identity restores the energy offsets needed to compare different number sectors, for example when locating phase boundaries. For nonuniform potentials, the differences between sites remain in the diagonal branch and must still be encoded.

\subsection{Dissipative model}
\label{sec:dissipative}
Photon loss, emitter relaxation, and emitter dephasing determine the transient observables considered below. We include these processes through the Lindblad generator \cite{gorini1976completely,lindblad1976generators,breuer2002theory}
\begin{equation}
\begin{aligned}
\frac{\partial\rho}{\partial t}=\mathcal L(\rho)={}&-i[H_\TCL,\rho]\\
&+\sum_{j=1}^{J}\left(V_j\rho V_j^\dagger
-\tfrac12\{V_j^\dagger V_j,\rho\}\right),
\end{aligned}
\label{eq:lindblad}
\end{equation}
with jump operators
\begin{equation}
\begin{aligned}
V^{(p)}_i&=\sqrt{\kappa_i}a_i,\\
V^{(1)}_{i,r}&=\sqrt{\gamma_{1,i,r}}\sigma^-_{i,r},\\
V^{(\phi)}_{i,r}&=\sqrt{\frac{\gamma_{\phi,i,r}}{2}}\sigma^z_{i,r}.
\end{aligned}
\label{eq:jumps}
\end{equation}
Here $\kappa_i,\gamma_{1,i,r},\gamma_{\phi,i,r}\ge0$. The factor of $1/2$ in $V^{(\phi)}_{i,r}$ makes an isolated emitter coherence decay at rate $\gamma_{\phi,i,r}$ under the dephasing channel alone. Relaxation contributes a further $\gamma_{1,i,r}/2$ to that decay, so the two rates are fixed by the measured lifetimes of an emitter through
\begin{equation}
\gamma_{1,i,r}=\frac{1}{T_{1,i,r}},
\qquad
\gamma_{\phi,i,r}=\frac{1}{T_{2,i,r}}-\frac{1}{2T_{1,i,r}}.
\label{eq:T1T2}
\end{equation}
Complete positivity requires $\gamma_{\phi,i,r}\ge0$ and therefore $T_{2,i,r}\le2T_{1,i,r}$, which is the usual constraint on the two lifetimes. This parameterization captures variations in $T_1$ and $T_2$ between nominally identical emitters in an array.

Thermal absorption and incoherent pumping are not included. Omitting absorption is justified at optical frequencies, where $\hbar\omega_c$ is of order an electronvolt and therefore far above $k_BT$ at room temperature, so the Bose--Einstein occupation of the cavity mode is negligible. In the microwave regime, this requires $k_BT\ll\hbar\omega_c$, a condition typically achieved at millikelvin temperatures. When thermal excitation cannot be neglected, the corresponding $a_i^\dagger$ and $\sigma^+_{i,r}$ jumps must be added to both the channel count and the dissipative normalization. Counting one photon-loss channel per cavity and one relaxation and one dephasing channel per emitter gives
\begin{equation}
J=L+2LN
\label{eq:J}
\end{equation}
channels in total.

Commutation of the jump operators is not the relevant test for splitting the master equation; the objects being exponentiated are the dissipative superoperators
\begin{equation*}
\mathcal D[V](\rho)=V\rho V^\dagger-\frac12\{V^\dagger V,\rho\}.
\end{equation*}
In the present model, dissipators acting on different subsystems commute. On the same emitter, a direct calculation on the one-qubit operator basis gives
\begin{align*}
[\sigma^z,\sigma^-]&\ne0,\\
[\mathcal D[\sigma^z],\mathcal D[\sigma^-]]&=0.
\end{align*}
The dissipative evolution therefore factors exactly into a product of local loss, relaxation, and dephasing channels. The dissipative generator does not generally commute with the Hamiltonian commutator $-i[H_\TCL,\cdot]$. We use the unitary dilation below to specify circuits for simulating the complete Lindblad dynamics within our oracle model. An alternative is a channel product formula that alternates Hamiltonian evolution with the exact dissipative layer \cite{kliesch2011dissipative,childsli2017lindblad}; we do not compile that approach here.

\subsection{Target observables}
\label{sec:observables}
\paragraph{Retarded cavity spectral function.}
For a prepared eigenstate $\ket{\psi_0}$, we use
\begin{equation*}
\begin{aligned}
H_\TCL\ket{\psi_0}&=E_0\ket{\psi_0},\\
a_i(\tau)&=e^{iH_\TCL\tau}a_i e^{-iH_\TCL\tau},
\end{aligned}
\end{equation*}
and define the correlators by
\begin{align}
C^+_{ij}(\tau)&=\bra{\psi_0}a_i(\tau)a_j^\dagger\ket{\psi_0},\nonumber\\
C^-_{ij}(\tau)&=\bra{\psi_0}a_j^\dagger a_i(\tau)\ket{\psi_0}.
\label{eq:correlators}
\end{align}

If the reference state has total polariton number $Q$, photon addition and removal access the sectors $Q+1$ and $Q-1$, respectively. For the local correlators, insert an energy eigenbasis in each sector. Writing the corresponding energies as $E_{\nu,Q+1}$ and $E_{\mu,Q-1}$ gives
\begin{align*}
C^+_{ii}(\tau)
&=\sum_\nu
\bigl|\langle\nu,Q+1|a_i^\dagger|\psi_0\rangle\bigr|^2\\
&\qquad\times e^{-i(E_{\nu,Q+1}-E_0)\tau},\\
C^-_{ii}(\tau)
&=\sum_\mu
\bigl|\langle\mu,Q-1|a_i|\psi_0\rangle\bigr|^2\\
&\qquad\times e^{-i(E_0-E_{\mu,Q-1})\tau}.
\end{align*}
Thus each oscillation records an energy difference, and its weight is the squared photon-transition matrix element. The retarded response subtracts the removal contribution from the addition contribution. In the vacuum sector, the removal contribution vanishes because $a_i\ket{\psi_0}=0$. The retarded Green function \cite{kubo1957response,Mahan2000ManyParticle} is
\begin{equation}
G^R_{ij}(\tau)=-i\Theta(\tau)
\left[C^+_{ij}(\tau)-C^-_{ij}(\tau)\right].
\label{eq:retarded}
\end{equation}
For an observation window $[0,T_{\max}]$, define its Fourier transform by
\begin{equation}
\widehat G^R_{ij}(\omega)=
\int_0^{T_{\max}}d\tau e^{i\omega\tau}G^R_{ij}(\tau).
\label{eq:frequency-green}
\end{equation}
The corresponding spectral matrix is
\begin{equation}
A_{ij}(\omega)=
-\frac{\widehat G^R_{ij}(\omega)-\widehat G^R_{ji}(\omega)^*}{2\pi i}.
\label{eq:spectral}
\end{equation}
This matrix is Hermitian. Its diagonal entries give the local spectra,
\[
A_{ii}(\omega)=-\frac{1}{\pi}\operatorname{Im}\widehat G^R_{ii}(\omega).
\]
An off-diagonal entry generally requires both index orderings. The elementwise imaginary-part expression also applies when $\widehat G^R_{ij}(\omega)=\widehat G^R_{ji}(\omega)$, as for a reciprocal response. Both particle-addition and particle-removal correlators are required. Retaining only one of them gives an emission- or absorption-type correlation spectrum, not the full retarded spectral function. A finite observation window broadens spectral features on the scale
\begin{equation}
\delta\omega\simeq\frac{2\pi}{T_{\max}},
\label{eq:resolution}
\end{equation}
up to the windowing convention. Consequently, frequency resolution fixes the longest closed-system evolution time.

\paragraph{Transient photon statistics.}
Let $\rho(T)=e^{T\mathcal L}\rho(0)$ for a specified initial state and final time $T$. We consider
\begin{equation}
\label{eq:g2}
\begin{aligned}
\bar n_i(T)&=\operatorname{tr}\left[n^{\ph}_i\rho(T)\right], \\
g^{(2)}_i(0;T)&=
\frac{\operatorname{tr}\left[a_i^{\dagger2}a_i^{2}\rho(T)\right]}
{\bar n_i(T)^{2}}
\end{aligned}
\end{equation}
whenever $\bar n_i(T)>0$. The normalized, normally ordered correlation $g_i^{(2)}(0;T)$ is a standard measure of photon-number fluctuations \cite{glauber1963coherence}. The operator in the numerator is diagonal because
\begin{equation}
a_i^{\dagger2}a_i^{2}\ket{m}=m(m-1)\ket{m}.
\label{eq:g2-diagonal}
\end{equation}

Writing $\mu=\bar n_i(T)>0$, the same factorial moment gives
\begin{equation*}
g_i^{(2)}(0;T)=1+
\frac{\operatorname{Var}(n_i)-\mu}{\mu^2}.
\end{equation*}
Values below or above one therefore indicate number variance below or above the Poisson value $\mu$, respectively. For example, a one-photon number state has $\mu=1$, zero variance, and $g^{(2)}=0$. One computational-basis measurement of all photon registers therefore supplies $n_i$ and $n_i(n_i-1)$ for every site. For independent trajectories, Hoeffding bounds \cite{hoeffding1963bounds} give the sample counts for additive estimates,
\begin{equation*}
\begin{aligned}
O\!\left(\frac{\nmax^2}{\varepsilon_n^2}\right)&\quad\text{for }\bar n_i(T),\\
O\!\left(\frac{\nmax^4}{\varepsilon_f^2}\right)&\quad\text{for the factorial moment}.
\end{aligned}
\end{equation*}
An additive error bound for their ratio also requires a lower bound on $\bar n_i(T)$; no uniform cost for $g^{(2)}$ exists as the state approaches vacuum. \Cref{sec:g2-resources} bounds the ratio error through one correlated combination of the two moments, which relaxes the required occupation accuracy.

By contrast, the correlators in \cref{eq:correlators} are off-diagonal matrix elements between states obtained by applying photon ladder operators. Estimating them requires interferometric measurement, as described in \cref{sec:endtoend}.

\subsection{Truncation and binary encoding}

We take an integer photon cutoff $\nmax\ge1$. The photon cutoff is exact for a useful class of initial states. Suppose $\rho(0)$ is supported on total polariton numbers $N_p\le Q$, with integer $Q\ge0$. The Hamiltonian conserves $N_p$, and every jump in \cref{eq:jumps} preserves it or lowers it by one. The evolution therefore remains in $N_p\le Q$, where each photon occupation satisfies $n_i\le Q$. The sufficient exact cutoffs are therefore
\begin{equation*}
\nmax\ge
\begin{cases}
Q,&\text{closed and open evolution},\\
Q+1,&\text{both photon correlators}.
\end{cases}
\end{equation*}
The extra level in the second sufficient bound accounts for the fact that $a_j^\dagger$ can raise the total number to $Q+1$ in the photon-addition correlator of \cref{eq:correlators}.

A smaller local cutoff is an additional physical approximation. The fact that the dynamics loses total excitation number does not by itself justify a smaller cutoff, because hopping can concentrate the remaining excitations at one cavity. The resource estimates below describe the chosen truncated model; when $\nmax<Q$ for dynamics or $\nmax<Q+1$ for the photon spectrum, comparison with the untruncated model also requires a cutoff-convergence study or a bound on the resulting observable error.

Each mode is stored in
\begin{equation}
\qph=\left\lceil\log_2(\nmax+1)\right\rceil
\label{eq:qph}
\end{equation}
qubits, so the register of one mode has $R=2^{\qph}$ basis states \cite{Sawaya2020ResourceEfficient}. When $\nmax+1$ is not a power of two, the register carries $R-\nmax-1$ labels with no physical meaning. The physical subspace is
\begin{equation}
\mathcal H_{\rm phys}=\operatorname{span}
\left\{\ket{\mathbf n,\mathbf s}:0\le n_i\le\nmax\ \text{for every }i\right\},
\label{eq:phys-subspace}
\end{equation}
with computational basis states
\begin{equation}
\ket{\mathbf n,\mathbf s}=\ket{n_1\cdots n_L}_{\ph}
\otimes\bigotimes_{i=1}^{L}\ket{s_{i,1}\cdots s_{i,N}}_{\at},
\label{eq:basis}
\end{equation}
and $s_{i,r}\in\{0,1\}$. With $\qph$ qubits for each of the $L$ photon registers in \cref{eq:qph} and one qubit for each of the $LN$ emitters, the system register holds
\begin{equation}
n_{\rm sys}=L\qph+LN
\label{eq:nsys}
\end{equation}
qubits. Expanding the occupation in its binary digits as $n_i=\sum_k2^kn_{i,k}$ and using the convention of \cref{eq:atom-number} gives the photon number on $\mathcal H_{\rm phys}$ as
\begin{equation}
n^{\ph}_i=\sum_{k=0}^{\qph-1}2^k\frac{I-Z^{\ph}_{i,k}}{2}.
\label{eq:binary-number}
\end{equation}
For the qubit implementation, we use the right-hand side of \cref{eq:binary-number} as the diagonal extension to the full binary register. It has eigenvalue $n$ on an unused label $\ket n$ with $n>\nmax$. On those labels, this extension differs from $a_i^\dagger a_i$ when the ladder operators are extended by zero. Its restriction to $\mathcal H_{\rm phys}$ remains the physical number operator, so this choice does not change the truncated dynamics.

\subsection{Data access and counting conventions}

Physical cavities are labeled $i=1,\ldots,L$. Their binary addresses are $i-1=0,\ldots,L-1$. Circuit descriptions that use zero-based addresses follow this relabeling.

We take
\begin{equation*}
\begin{gathered}
L_x=2^{\ell_x},\qquad L_y=2^{\ell_y},\\
\ell_x,\ell_y\ge2,\qquad N=2^{n_a}.
\end{gathered}
\end{equation*}
Choosing the lattice dimensions to be powers of two makes the destination update a modular binary increment, so no additional range check is required. The condition $\ell_x,\ell_y\ge2$ also keeps the $+\hat x$ and $-\hat x$ neighbors distinct, and similarly for the $y$ direction. Taking $N=2^{n_a}$ allows the site index $i$ and emitter label $r$ to be stored as contiguous bit fields in a single selection register, so the same unary iteration can address every local branch \cite{Babbush2018}. These assumptions simplify the address arithmetic but do not affect the block-encoding normalization. More general lattice dimensions or emitter counts only require modified address handling.

The circuits below use linear combinations of unitaries (LCU). A \emph{branch} is one labeled term in the linear combination: $\PREP$ prepares the branch amplitudes and $\SELECT$ applies the unitary specified by the branch label. A \emph{signal qubit} is an ancilla whose projected $\ket0$ amplitude provides a generally nonunitary coefficient, as used in \cref{sec:hop,sec:loc}. We use approximate block encodings \cite{gilyen2019quantum} and quote the error in normalized form. Thus, the parameter $\varepsilon$ bounds the error in $H/\alpha$, while the corresponding error in Hamiltonian units is $\alpha\varepsilon$.

\begin{definition}
\label{def:be}
A unitary $U$ acting on $a$ ancilla qubits and the system register is an $(\alpha,a,\varepsilon)$ block encoding of $H$ if
\begin{equation}
\norm{\left(\bra{0^a}\otimes I\right)U\left(\ket{0^a}\otimes I\right)-\frac{H}{\alpha}}\le\varepsilon.
\label{eq:be-def}
\end{equation}
\end{definition}

For an exact block encoding, its action on an initialized input can be written as
\begin{equation*}
\begin{aligned}
&U\bigl(\ket{0^a}\otimes\ket\psi\bigr)\\
&\quad=\ket{0^a}\otimes\frac{H}{\alpha}\ket\psi
+\ket{\Phi_\perp},
\end{aligned}
\end{equation*}
where $(\bra{0^a}\otimes I)\ket{\Phi_\perp}=0$. The ancilla-zero component therefore contains the desired operator, while the remaining component makes the complete circuit unitary. The projection in the definition specifies this component; the simulation circuits keep the ancillas coherent between queries.

For the angle-table rotations, we use a phase-gradient register prepared in the state $\ket{G_{b+1}}$. A phase-gradient state is an eigenstate of modular addition: adding a binary angle produces a phase determined by that angle. The circuit uses this phase kickback to implement the required rotation with reversible arithmetic. This register is an auxiliary catalyst for implementing the required rotations and is not part of the zero-projected ancilla register that defines the block encoding. On the exact state $\ket{G_{b+1}}$, the angle circuit returns the phase-gradient register to the same state, disentangled from all remaining registers, as shown in \cref{eq:angle-catalyst-return}. Accordingly, the parameter $a$ counts only the ancillas projected onto $\ket{0^a}$ in the block encoding. The $b+1$ phase-gradient catalyst qubits, together with the clean arithmetic workspace, are reported separately. The catalyst-state preparation error is included once in the error budget for each experimental run in \cref{sec:endtoend}.

For a circuit $V$, we report $\mathcal C_{\rm Tof}(V)$, the number of Toffoli-family operations, and $\mathcal C_R(V)$, the number of arbitrary single-qubit rotations; Clifford operations are not counted. The primitive costs used throughout, including the convention for a temporary logical AND, are collected in Appendix~\ref{app:primitives}.

\section{Classical simulation cost}
\label{sec:classical}

In this section, we examine the classical computational requirements and limitations of simulating the atom-resolved Tavis--Cummings lattice. The raw Hilbert-space dimension does not establish classical hardness. The Hamiltonian \cref{eq:H-split} conserves total polariton number and is invariant under permutations of the identically coupled emitters at each site. These symmetries reduce the state space when the initial state has the corresponding support. Equilibrium quantum Monte Carlo, tensor networks, and quantum trajectories provide further classical baselines. We describe these reductions and their limitations for the real-time and emitter-resolved open-system targets before comparing circuit costs.

\subsection{State space after symmetry reduction}

The truncated physical Hilbert space contains $(\nmax+1)^L2^{LN}$ basis states, which for a $4\times4$ lattice with $N=4$ and $\nmax=15$ is $2^{128}$. That figure considerably overstates the difficulty of the closed problem, because it takes account of neither the polariton-number symmetry nor the permutation symmetry within a site.

Since $[H_\TCL,N_p]=0$, evolution generated by \cref{eq:H-split} stays in one polariton-number sector when the initial state belongs to that sector. Counting the configurations of $L$ truncated modes and $LN$ emitters that carry $Q$ excitations in total gives the sector dimension as a coefficient of a generating function,
\begin{equation}
D(Q)=[x^{Q}]\left(\sum_{n=0}^{\nmax}x^{n}\right)^{L}(1+x)^{LN},
\label{eq:sector-dim}
\end{equation}
where $[x^{Q}]$ extracts the coefficient of $x^{Q}$, the first factor counts the photon occupation of each cavity and the second counts emitter configurations weighted by the excitations they carry.

The emitter-permutation symmetry gives a second reduction. Because $g_i$ does not depend on $r$, $H_\TCL$ commutes with permutations of the $N$ emitters at a site, so it decomposes into blocks labeled by the total emitter spin at each site. The emitter ground state $\ket{0\cdots0}$ is a maximal-spin state, so a simulation started from it never leaves the maximal-spin manifold, where each site contributes $N+1$ emitter states instead of $2^N$ and \cref{eq:sector-dim} becomes
\begin{equation}
D_{\rm sym}(Q)=[x^{Q}]\left(\sum_{n=0}^{\nmax}x^{n}\right)^{L}
\left(\sum_{k=0}^{N}x^{k}\right)^{L}.
\label{eq:sym-dim}
\end{equation}

\Cref{tab:dims} evaluates both at unit filling $Q=L$. The reductions are large: for the $4\times4$ lattice with $N=4$ and $\nmax=15$ they take $2^{128}$ down to $2^{55.8}$ and then to $2^{40.4}$. A single state vector of $2^{40.4}$ complex double-precision amplitudes occupies about $23$ TB. A Krylov calculation requires several such vectors, in addition to the Hamiltonian action and communication overhead. The symmetry reduction therefore brings the storage requirement within reach of a large distributed-memory calculation, although the computational and communication costs remain substantial. This example provides a classical baseline for the stated lattice, filling, and symmetry sector. This reduction applies to closed dynamics when the emitters start in the maximal-spin sector. More general initial states or independent local dissipation require a larger state space.

\begin{table}[t]
\caption{Base-2 logarithms of the dimensions entering a classical simulation at unit polariton filling $Q=L$. The columns give the full register of \cref{eq:nsys}, the polariton sector of \cref{eq:sector-dim}, the maximal-spin sector of \cref{eq:sym-dim}, and the number $\sum_{q=0}^{Q}D(q)^2$ of complex entries in the block-diagonal density operator reached from sector $Q$, as in \cref{eq:open-sector-storage}. These dimensions refer to the chosen truncated model. For reference, one petabyte holds $2^{45.8}$ complex double-precision amplitudes.}
\label{tab:dims}
\squeezetable
\begin{ruledtabular}
\begin{tabular}{ccc rrrr}
$L$ & $N$ & $\nmax$ & register & sector & max spin & density matrix \\
\colrule
16 & 1 & 7 & 64.0 & 37.1 & 37.1 & 74.5 \\
16 & 2 & 7 & 80.0 & 44.6 & 39.5 & 89.3 \\
16 & 4 & 15 & 128.0 & 55.8 & 40.4 & 111.6 \\
64 & 1 & 7 & 256.0 & 158.1 & 158.1 & 316.5 \\
64 & 2 & 7 & 320.0 & 187.6 & 167.8 & 375.3\\
64 & 4 & 15 & 512.0 & 231.9 & 171.3 & 463.9 \\
\end{tabular}
\end{ruledtabular}
\end{table}

Photon loss and emitter relaxation change the occupied number sector. Starting from sector $Q$, however, the density operator remains block diagonal in total excitation number,
\begin{equation}
\rho(T)=\bigoplus_{q=0}^{Q}\rho_q(T),
\ \
\text{number of entries}=\sum_{q=0}^{Q}D(q)^2.
\label{eq:open-sector-storage}
\end{equation}
To see this, $H_\TCL$ and each $V_j^\dagger V_j$ preserve number, while $V_j\rho_q V_j^\dagger$ belongs either to sector $q$ or to sector $q-1$. Thus the generator cannot create coherence between distinct number sectors from such an initial state. The last column of \cref{tab:dims} gives this storage count, which is $2^{111.6}$ for $L=16$, $N=4$, and $\nmax=15$.

Classical quantum-jump trajectories avoid storing these density matrices. Between jumps a state evolves with $H_{\rm eff}=H_\TCL-\tfrac{i}{2}\sum_jV_j^\dagger V_j$, which also preserves number. A trajectory therefore needs one state vector in its current sector, with a loss or relaxation jump moving it to the next lower sector. The cost then includes repeated trajectory propagation and statistical estimation of the observable \cite{dalibard1992wavefunction,breuer2002theory}.

Independent emitter channels illustrate why permutation invariance of the density operator does not imply support in the maximal-spin manifold. For two emitters with a common nonzero relaxation rate, the jump contribution from $\proj{ee}$ is proportional to
\begin{equation}
\begin{gathered}
\proj{ge}+\proj{eg}
=\proj{S}+\proj{A},\\
\ket{S/A}=\frac{\ket{ge}\mathbin{\pm}\ket{eg}}{\sqrt2}.
\end{gathered}
\label{eq:local-decay-spin}
\end{equation}
The antisymmetric state $\ket A$ lies outside the maximal-spin manifold, although the resulting density operator is permutation invariant \cite{kirton2017suppressing}. Unequal emitter rates further break permutation covariance \cite{shammah2018piqs}. These distinctions motivate retaining one qubit per emitter in the quantum construction.

For closed dynamics, a collective-spin encoding is possible when the reference state lies in the maximal-spin sector at every site. Photon insertions preserve those spin sectors, so each cavity's emitters can be represented by $\lceil\log_2(N+1)\rceil$ qubits instead of $N$ qubits. We keep a separate qubit for each emitter so that the same representation also supports emitter-resolved dissipation. Estimating the resources of a collective-spin implementation would require circuits for its exchange terms and routing, together with corresponding precision bounds.

Direct state-vector storage becomes prohibitive as the lattice grows. A $5\times5$ lattice at unit filling with a single emitter per cavity and $\nmax=7$ already carries $2^{59.7}$ states in its polariton sector, exceeding the $2^{45.8}$ complex amplitudes that fit in a petabyte. For $N=4$, a state vector in the full atom-resolved polariton sector of the $4\times4$ lattice also exceeds this storage capacity.

\subsection{Approximate classical methods}

State-vector storage alone does not give a complete comparison. Other methods can reduce the cost by using special properties of the state, the observable, or the dissipation.

For real nonnegative $t_{ij}$, the Hamiltonian can be made stoquastic in the occupation basis. The hopping matrix elements in \cref{eq:H-hop} are nonpositive. At each site with $g_i\ne0$, multiplying every excited emitter state by the phase factor $-\operatorname{sgn}(g_i)$ makes the exchange matrix elements $-|g_i|$ times the positive ladder amplitude, without changing the diagonal terms. Large-scale quantum Monte Carlo calculations have determined the dynamical critical exponent at the tip of a Mott lobe in the Jaynes--Cummings--Hubbard model~\cite{hohenadler2011jch}. These calculations use stochastic series expansion with directed loops and compare the results with independent worm-algorithm calculations. This supplies an equilibrium baseline. Frustrated or complex hopping can introduce a sign problem, and imaginary-time sampling does not directly provide the real-time observables considered here.

Tensor-network methods are effective for many one-dimensional systems, but two-dimensional simulations are more demanding. Contracting a projected entangled pair state exactly is $\#\mathrm{P}$-hard \cite{schuch2007peps}. A two-dimensional area law still allows entanglement proportional to the length of a spatial cut. Representing such a state with a one-dimensional ordering can require a bond dimension exponential in that length. After a global quench, the entanglement across a cut can also grow linearly in $\tau$, producing exponential-in-time bond-dimension growth for fixed-accuracy representations in hard instances \cite{schuch2008entropy}. The local dimension $(\nmax+1)2^N$ enters the contraction cost as well, and it is $256$ at $\nmax=15$ and $N=4$.

Matrix-product representations also support classical simulation of one-dimensional mixed-state dynamics, with the required bond dimension governed by operator-space correlations \cite{zwolak2004mixed}.

For one-dimensional open dynamics, locally purified tensor networks preserve positivity and provide trace-norm bounds on the errors from time discretization and truncation \cite{werner2016positive}.

For translationally invariant two-dimensional open lattices, infinite projected entangled-pair operator algorithms provide a further approximate dynamical baseline~\cite{mckeever2021stable}.

Permutation-invariant density-matrix methods exploit equal emitter couplings and equal local rates within a site. For a permutation-invariant initial density operator, the covariant Liouvillian preserves this symmetry, allowing the emitter operator space to be represented with dimension polynomial in $N$ rather than $4^N$ \cite{shammah2018piqs}. This representation includes lower-spin sectors and is therefore more general than the maximal-spin Hilbert space in \cref{eq:sym-dim}. Unequal emitter rates remove the full within-site permutation symmetry, although identical groups can retain partial reductions. Site-dependent $\nu_i$, $t_{ij}$, and $g_i$ do not by themselves remove within-site permutation covariance; they instead determine the transport and spatial inhomogeneity of the lattice.

For two-dimensional driven-dissipative lattices, approximate methods include corner-space renormalization \cite{finazzi2015corner} and cluster mean-field \cite{jin2016cluster}, which retain a selected state space or a finite cluster coupled to a mean-field environment. Their accuracy can be assessed by increasing the retained space or cluster size, although a general a priori observable-error bound is unavailable. Our circuit estimates allocate implementation, evolution, and sampling errors for the truncated model; physical cutoff error must be assessed separately when the sufficient conditions above are not met.

\subsection{Worst-case complexity}

Sampling a full photon-number distribution and estimating a selected correlation are different computational tasks. For the untruncated quadratic Hamiltonian $H_\hop=\sum_{ij}h_{ij}a_i^\dagger a_j$, the mode operators satisfy
\begin{equation}
a_i(\tau)=\sum_j U_{ij}(\tau)a_j,
\qquad U(\tau)=e^{-ih\tau}.
\label{eq:free-mode-propagation}
\end{equation}
The retarded commutator is
\begin{equation*}
\langle[a_i(\tau),a_j^\dagger]\rangle=U_{ij}(\tau)
\end{equation*}
for every initial state. For a Fock input with occupations $n_j^{(0)}$, the mean occupation is
\begin{equation*}
\big\langle n_i^{\ph}(\tau)\big\rangle
=\sum_j|U_{ij}(\tau)|^2n_j^{(0)}.
\end{equation*}
Both quantities are computable from an $L\times L$ matrix, and the same results hold with a cutoff sufficient for the relevant number sectors.

Full output sampling from suitable linear-optical networks instead gives the BosonSampling problem, whose approximate-sampling hardness relies on the Aaronson--Arkhipov conjectures \cite{aaronson2011linearoptics}. For spatially local quadratic bosonic Hamiltonians, known bounds distinguish short-time regimes admitting classical sampling from longer-time regimes with conditional sampling hardness \cite{deshpande2018sampling}. These results provide context for many-boson simulation, but they do not establish hardness of the correlators or photon moments in \cref{sec:observables}. The light--matter exchange introduces interactions; the classical cost of the resulting observable calculation depends on the instance and approximation method.

For locally interacting spin systems initialized in a product state, evolution over times polynomial in system size admits efficient classical sampling when constant-rate entanglement-breaking noise exceeds a threshold determined by the interaction strength, range, and lattice dimension. At low noise, some instances on two- and higher-dimensional grids are expected to be hard to sample classically under standard complexity-theoretic assumptions~\cite{trivedi2022transitions}. These results do not determine the classical difficulty of the lattice model with the dissipative channels in \cref{eq:jumps} at the parameters studied here. Our analysis estimates the quantum cost of real-time lattice observables; it does not establish quantum advantage.

\section{Block encoding of photon hopping}
\label{sec:hop}

We now turn to the quantum circuit construction, beginning with a block encoding of the photon-hopping Hamiltonian $H_{\rm hop}$ in \cref{eq:H-hop}. The hopping Hamiltonian is a sum of directed photon transfers. For a fixed directed bond, the construction below is a two-mode instance of the product construction in the LOBE framework \cite{simon2025ladder}; the model-specific work is to route a coherently selected bond, share its occupation-dependent data across all bond labels, and combine the hopping block encoding with those of the remaining Tavis--Cummings terms. For a directed bond $\ell=(j\to i)$, the operator $a_i^\dagger a_j$ removes one photon from cavity $j$, adds one to cavity $i$, and multiplies the state by $\sqrt{n_j(n_i+1)}$. A block encoding must therefore perform two tasks: update the two occupation registers and encode this occupation-dependent coefficient. We implement the update with a reversible modular shift $S_\ell$ and place the coefficient in a diagonal operator $D_\ell$. Setting the appropriate diagonal entries to zero enforces the lower boundary $n_j=0$, the upper boundary $n_i=\nmax$, and all unused binary labels. We then factor the normalized coefficient into one function of $n_j$ and one of $n_i$, which allows it to be loaded from two one-dimensional tables. The following subsections establish the decomposition, construct the corresponding LCU circuit, and evaluate its normalization and cost.

\subsection{Hopping as a modular shift times a diagonal}

We first establish the decomposition of a directed photon transfer into a reversible occupation update and a diagonal coefficient, including its action on the full binary register. Each site has four neighbors, so the number of directed hopping labels $\ell=(j,d)$ with $d\in\{\pm\hat x,\pm\hat y\}$ is
\begin{equation}
S_\hop=4L,
\label{eq:Shop}
\end{equation}
which is a power of two under the assumption on $L_x$ and $L_y$. Writing $i=j+d$ for the destination, define $t_\ell=t_{i(\ell)j(\ell)}$ and $c_\ell=-t_\ell$. Label $\ell$ therefore selects the term $c_\ell a_i^\dagger a_j$. Both orientations of each undirected bond occur among the $4L$ labels, and hence
\begin{equation}
H_\hop=\sum_{\ell}c_\ell\,a_{i(\ell)}^\dagger a_{j(\ell)}
\label{eq:H-hop-labels}
\end{equation}
reproduces \cref{eq:H-hop}. On the truncated register, the photon-transfer operator selected by $\ell$ acts on the source and destination occupations as
\begin{equation}
a_i^\dagger a_j\ket{n_j,n_i}=\sqrt{n_j(n_i+1)}\ket{n_j-1,n_i+1}
\label{eq:hop-action}
\end{equation}
when $1\le n_j\le\nmax$ and $n_i<\nmax$, and gives zero otherwise. The sparse-access construction treats this update as a row map on the complete photon configuration and writes the resulting configuration into a second photon register. LOBE instead applies the increment and decrement to the selected system registers. The next lemma states this construction on the full binary register, including its action at the occupation boundaries and on unused occupation labels.
\begin{lemma}
\label{lem:shift}
Let $S_\ell$ act on the two selected $\qph$-qubit registers by
\begin{equation}
S_\ell\ket{n_j,n_i}=\ket{n_j-1\bmod R,\ n_i+1\bmod R},
\label{eq:shift-def}
\end{equation}
and let $D_\ell$ be diagonal in the occupation basis with
\begin{equation}
D_\ell\ket{n_j,n_i}=
\begin{cases}
\sqrt{n_j(n_i+1)}\ket{n_j,n_i},&1\le n_j\le\nmax\\
&\text{and }n_i<\nmax,\\[2pt]
0,&\text{otherwise}.
\end{cases}
\label{eq:diag-def}
\end{equation}
Then $S_\ell$ is unitary and $a_i^\dagger a_j=S_\ell D_\ell$ on the register space, where $a_i^\dagger a_j$ is understood to annihilate every state carrying an unused label at $j$ or at $i$.
\end{lemma}

\begin{proof}
$S_\ell$ permutes the $R^2$ basis states of the two registers, hence is unitary. Apply $S_\ell D_\ell$ to $\ket{n_j,n_i}$. If the configuration satisfies the two conditions in \cref{eq:diag-def} then $D_\ell$ multiplies by $\sqrt{n_j(n_i+1)}$ and $S_\ell$ sends the state to $\ket{n_j-1,n_i+1}$; no reduction modulo $R$ occurs because $n_j\ge1$ and $n_i<\nmax\le R-1$. This is \cref{eq:hop-action}. If either condition fails then $D_\ell$ annihilates the state and the action of $S_\ell$ on it is irrelevant.
\end{proof}

\cref{eq:shift-def} reversibly updates the two system photon registers. It therefore removes the second $L\qph$-qubit photon register, the register swap, and the reverse row oracle required by the sparse-access construction. The remaining operator $D_\ell$ is diagonal: it does not change either occupation, but supplies the coefficient in \cref{eq:hop-action}. We encode that coefficient in the projected $\ket0$ amplitudes of two signal qubits.

\subsection{Factorizing the occupation-dependent coefficient}

Having established the shift--diagonal decomposition in \cref{lem:shift}, we now consider the implementation of the diagonal factor. A literal table indexed by $(\ell,n_j,n_i)$ has $4LR^2$ entries. The factorization below avoids this joint table in either a direct-register or a sparse-access implementation; the comparison in \cref{tab:compare} gives both implementations the same factorized data. The product construction in LOBE permits the coefficient $\sqrt{n_j(n_i+1)}$ to be encoded as a source factor $\sqrt{n_j}$ and a destination factor $\sqrt{n_i+1}$. For the present lattice, these factors are also independent of the bond label and can therefore be shared across all $4L$ branches. Their validity conditions separate: $1\le n_j\le\nmax$ depends only on the source, while $n_i<\nmax$ depends only on the destination. Each factor can therefore return zero on its own invalid inputs. Dividing each factor by $\sqrt{\nmax}$ places it in $[0,1]$, as required for an amplitude encoded by a single-qubit rotation, and gives the overall normalization factor $1/\nmax$.

\begin{lemma}
\label{lem:factor}
Define, for $n\in\{0,\ldots,R-1\}$,
\begin{align}
A(n)&=\begin{cases}\sqrt{n/\nmax},&n\le\nmax,\\0,&n>\nmax,\end{cases}
\label{eq:A-def}\\[4pt]
B(n)&=\begin{cases}\sqrt{(n+1)/\nmax},&n<\nmax,\\0,&n\ge\nmax.\end{cases}
\label{eq:B-def}
\end{align}
Then $A(n),B(n)\in[0,1]$ for all $n$, and for every pair $(n_j,n_i)\in\{0,\ldots,R-1\}^2$,
\begin{equation}
A(n_j)B(n_i)=\frac{1}{\nmax}\bra{n_j,n_i}D_\ell\ket{n_j,n_i}.
\label{eq:factor}
\end{equation}
\end{lemma}

\begin{proof}
Since $A(n)$ vanishes for $n>\nmax$, its maximum is $A(\nmax)=1$. Similarly, $B(n)$ vanishes for $n\ge\nmax$, and its maximum is $B(\nmax-1)=1$. To prove \cref{eq:factor}, consider the cases in \cref{eq:diag-def}. If $n_j=0$ or $n_j>\nmax$ then $A(n_j)=0$ and the diagonal entry is zero. If $n_i\ge\nmax$ then $B(n_i)=0$ and the diagonal entry is zero. In the remaining case $1\le n_j\le\nmax$ and $n_i<\nmax$, so $A(n_j)B(n_i)=\sqrt{n_j(n_i+1)}/\nmax$, which is the diagonal entry divided by $\nmax$.
\end{proof}

For this lattice, the occupation-dependent factor splits into source and destination functions that are independent of the bond label. This replaces the joint table of $4LR^2$ entries with two tables of $R$ entries. The hopping coefficients $t_{ij}$ are included in the preparation of the bond label, so these two occupation tables do not grow with $L$. Their zero entries enforce $n_j\ge1$, $n_i<\nmax$, and the exclusion of unused binary labels. No separate truncation comparator, validity qubit, or comparator uncomputation is required.

The oracle $O_A$ encodes the source factor $A(n_j)$ on signal qubit $\sigma_1$ by
\begin{equation}
\begin{aligned}
O_A:\ \ket{n_j}\ket{0}_{\sigma_1}\mapsto{}&
\ket{n_j}\Big(A(n_j)\ket{0}_{\sigma_1}\\
&+\sqrt{1-A(n_j)^2}\ket{1}_{\sigma_1}\Big).
\end{aligned}
\label{eq:OA}
\end{equation}
The destination oracle $O_B$ is defined analogously, with $n_j$, $A$, and $\sigma_1$ replaced by $n_i$, $B$, and $\sigma_2$. Each oracle loads a $b$-bit rotation angle from a table of $R$ entries and applies the corresponding $R_y$ rotation using a reusable phase-gradient register \cite{Babbush2018,sunderhauf2024block}.

We use one common integer precision $b\ge1$ for the angle tables of $O_A$, $O_B$, and the exchange rotation $R_\theta$ in \cref{eq:lobe-angle}. The integer $b_{\rm alias}\ge1$ specifies the separate precision of the coefficient preparations. Projection of $\sigma_1$ and $\sigma_2$ onto $\ket{00}$ contributes the product $A(n_j)B(n_i)$. By \cref{lem:factor}, this product equals the corresponding diagonal entry of $D_\ell/\nmax$. This factorized construction uses two signal qubits because successive $R_y$ rotations on one qubit add their angles instead of multiplying their projected amplitudes.

For example, at $\nmax=3$ and with exact rotation angles, the shift and signal amplitudes satisfy
\begin{equation*}
\begin{gathered}
(n_j,n_i)=(2,1)\longmapsto(1,2),\\
A(2)=B(1)=\sqrt{\frac23},\\
A(2)B(1)=\frac23=\frac{\sqrt{2(1+1)}}{\nmax}.
\end{gathered}
\end{equation*}
Projecting the two signal qubits onto $\ket{00}$ therefore contributes the occupation-dependent hopping coefficient divided by $\nmax$. An empty source or a destination at the cutoff makes one factor vanish, giving zero projected hopping amplitude while the modular shift remains unitary.

\subsection{Block-encoding circuit and normalization}
Using the shift--diagonal decomposition of \cref{lem:shift} and the factorized amplitudes of \cref{lem:factor}, we now assemble a block encoding of the hopping Hamiltonian. The directed-bond coefficients determine the LCU preparation, while each selected branch implements the corresponding occupation update and amplitude.
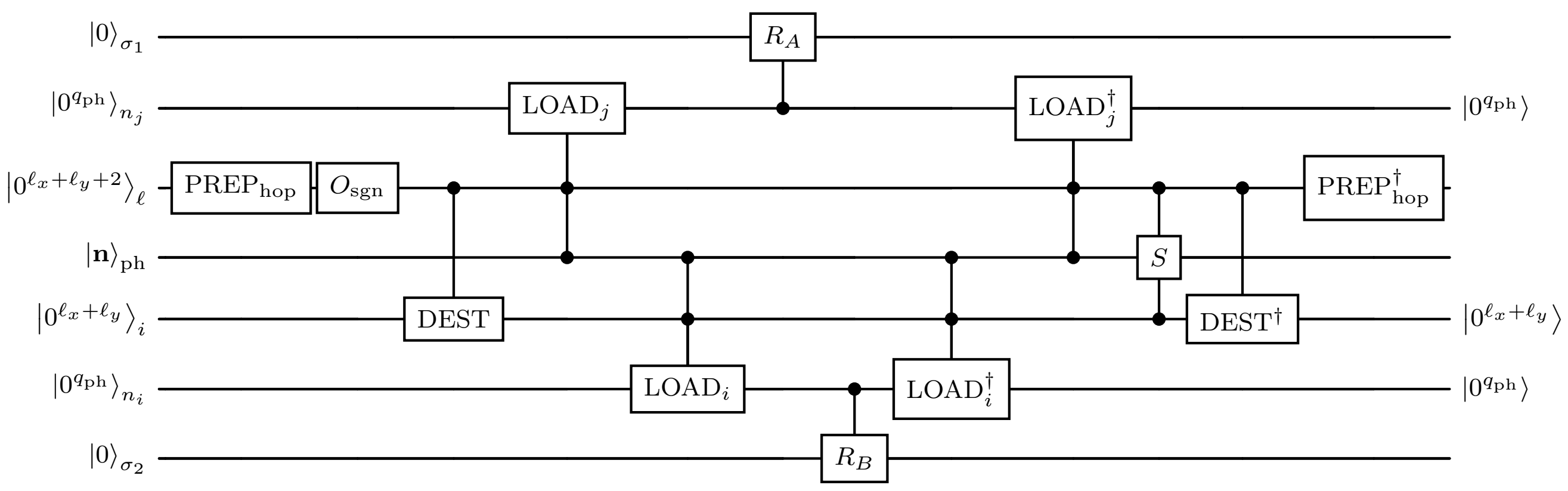
\begin{figure*}[t]
\centering
\CircuitDiagram[\textwidth]{%
\begin{quantikz}[tcl dense circuit]
\lstick{$\ket0_{\sigma_1}$} & \qw & \qw & \qw & \qw & \qw & \qw & \gate{R_A} & \qw & \qw & \qw & \qw & \qw & \qw & \qw \\
\lstick{$\ket{0^{\qph}}_{n_j}$} & \qw & \qw & \qw & \qw & \gate{\LOAD_j} & \qw & \ctrl{-1} & \qw & \qw & \gate{\LOAD_j^\dagger} & \qw & \qw & \qw & \rstick{$\ket{0^{\qph}}$} \\
\lstick{$\ket{0^{\ell_x+\ell_y+2}}_\ell$} & \qw & \gate{\PREP_\hop} & \gate{O_{\sgn}} & \ctrl{2} & \ctrl{-1} & \qw & \qw & \qw & \qw & \ctrl{-1} & \ctrl{1} & \ctrl{2} & \gate{\PREP_\hop^\dagger} & \qw \\
\lstick{$\ket{\mathbf n}_{\ph}$} & \qw & \qw & \qw & \qw & \ctrl{-2} & \ctrl{2} & \qw & \qw & \ctrl{2} & \ctrl{-2} & \gate{S} & \qw & \qw & \qw \\
\lstick{$\ket{0^{\ell_x+\ell_y}}_i$} & \qw & \qw & \qw & \gate{\DEST} & \qw & \ctrl{1} & \qw & \qw & \ctrl{1} & \qw & \ctrl{-1} & \gate{\DEST^\dagger} & \qw & \rstick{$\ket{0^{\ell_x+\ell_y}}$} \\
\lstick{$\ket{0^{\qph}}_{n_i}$} & \qw & \qw & \qw & \qw & \qw & \gate{\LOAD_i} & \qw & \ctrl{1} & \gate{\LOAD_i^\dagger} & \qw & \qw & \qw & \qw & \rstick{$\ket{0^{\qph}}$} \\
\lstick{$\ket0_{\sigma_2}$} & \qw & \qw & \qw & \qw & \qw & \qw & \qw & \gate{R_B} & \qw & \qw & \qw & \qw & \qw & \qw
\end{quantikz}}
\caption{Hopping block encoding of \cref{eq:U-hop}, with the coefficient-sign
oracle included. The occupation words $n_j,n_i$ each contain $\qph$ qubits;
the system photon register contains $L\qph$ qubits. Dots connected to
multi-qubit address wires denote indexed access, rather than a test that
every address bit is one. The two rotations act on separate signal
qubits. Their angle lookup, phase-gradient operation, and cleanup are
expanded in \cref{fig:angle-explicit}. The loaded occupations are erased
before $S$ changes the system data. The enabled increment and decrement
used by $S$ are expanded in \cref{fig:shift-explicit}.
Only the occupation and destination-address work registers are shown
returning to zero. The hopping label, preparation workspace, and two
signal qubits are projected onto zero to obtain
$H_\hop/\alpha_\hop$; the preparation workspace is suppressed in this
diagram. No reset or measurement of these projected registers occurs
inside the block encoding.}
\label{fig:hop}
\end{figure*}

\paragraph{Label preparation.}
Let $\Lambda_\hop=\sum_\ell\abs{t_\ell}$. If $\Lambda_\hop=0$, the hopping contribution vanishes and this branch is omitted. Otherwise, the ideal preparation is
\begin{equation}
\PREP_\hop\ket{0}_{\ell J_\hop}
=\sum_\ell\sqrt{\frac{\abs{c_\ell}}{\Lambda_\hop}}
\ket{\ell}\ket{f_\ell}_{J_\hop}.
\label{eq:PREP-hop}
\end{equation}
Here $J_\hop$ contains the auxiliary outputs of coherent alias sampling, and each $\ket{f_\ell}$ is normalized. Selection preserves the label and acts as the identity on $J_\hop$. These auxiliary outputs need not return to zero after selection, so the projected block includes the full input register $\ell J_\hop$. The finite-precision preparation replaces the desired probabilities by the rounded distribution defined in Appendix~\ref{app:alias-rounding}.

To implement the selected operations, we use two reversible address primitives.  The destination oracle computes the cavity reached by the directed hopping label,
\begin{equation}
\DEST:\ket{j,d}\ket{0}_i
\longmapsto
\ket{j,d}\ket{j+d}_i ,
\end{equation}
where the selected coordinate is updated modulo $L_x$ or $L_y$, as appropriate.  Given a site address $a$, the occupation-load operation coherently reads the selected photon occupation into a clean $q_{\rm ph}$-qubit register,
\begin{equation}
\LOAD_a:
\ket{a}\ket{\mathbf n}_{\rm ph}\ket{0}
\longmapsto
\ket{a}\ket{\mathbf n}_{\rm ph}\ket{n_a}.
\end{equation}
Thus $\LOAD_j$ and $\LOAD_i$ provide the occupations required by $O_A$ and $O_B$, respectively, while $\DEST$ supplies the destination address required by both $O_B$ and the shift $S_\ell$.  These work registers are uncomputed before the hopping branch ends.

The coefficient-sign oracle $O_{\sgn}$ applies $\sgn(c_\ell)$ as a phase. Throughout the selection circuits, we use $\sgn(c)=1$ for $c\ge0$ and $\sgn(c)=-1$ for $c<0$. In particular, a zero coefficient has phase $+1$, so every selected branch remains unitary even when its preparation weight is zero. For an arbitrary sign pattern, this oracle can be implemented by unary iteration over the $4L$ labels at cost at most $4L-1$ temporary ANDs. This term is absorbed into the asymptotic cost below. For uniform hopping, every label has equal weight and $\PREP_\hop$ reduces to Hadamard gates; when all $t_{ij}>0$, the sign $c_\ell<0$ is common to every label and its phase is Clifford. The standard LCU construction \cite{childs2012hamiltonian,berry2015simulating} then gives the complete hopping block encoding
\begin{equation}
\begin{aligned}
U_\hop={}&\PREP_\hop^\dagger\\
&\times\left(\sum_{\ell}\proj{\ell}\otimes I_{J_\hop}\otimes\sgn(c_\ell)S_\ell O_BO_A\right)\\
&\times\PREP_\hop,
\end{aligned}
\label{eq:U-hop}
\end{equation}
drawn in \cref{fig:hop}.

\begin{proposition}[Hopping block encoding]
\label{prop:hop-block-encoding}
Let $U_\hop$ be the circuit of \cref{eq:U-hop}, with $\Lambda_\hop=\sum_\ell |t_\ell|>0$, and let $d_\hop=\log_2(4L)=\ell_x+\ell_y+2$. Assuming the exact phase-gradient catalyst $\ket{G_{b+1}}$, $U_\hop$ is an $(\alpha_\hop,a_\hop,\epsilon_\hop)$ block encoding of $H_\hop$ with
\begin{equation}
\alpha_\hop
=
\nmax\sum_\ell |t_\ell|
\le
4Lt_{\max}\nmax,
\label{eq:alpha-hop}
\end{equation}
\begin{equation}
a_\hop
=
2d_\hop+2b_{\rm alias}+3,
\end{equation}

where $t_{\max}=\max_{\langle i,j\rangle}|t_{ij}|$, and
\begin{equation}
\epsilon_\hop
\le
\left(
\frac{\pi}{2^b}
+
\frac{1}{2^{b_{\rm alias}}}
\right).
\label{eq:hop-block-error}
\end{equation}
Under the measurement-assisted convention of Appendix~\ref{app:primitives},
\begin{equation}
\mathcal C_{\rm Tof}(U_\hop)
=
O\!\left(
L\qph+\nmax+b+b_{\rm alias}
\right).
\label{eq:hop-cost-scaling}
\end{equation}
The phase-gradient catalyst and clean arithmetic workspace are excluded from the projected ancilla count $a_\hop$; both return to their input states before projection.
\end{proposition}

\begin{proof}
Projecting the preparation register $\ell J_\hop$ and the two signal qubits gives, using \cref{lem:factor},
\begin{equation}
\begin{aligned}
&\Big(\bra{0}_{\ell J_\hop}\bra{00}_\sigma\otimes I\Big)
U_\hop
\Big(\ket{0}_{\ell J_\hop}\ket{00}_\sigma\otimes I\Big)\\
&=
\sum_\ell\frac{c_\ell}{\Lambda_\hop}
\frac{S_\ell D_\ell}{\nmax} =
\frac{H_\hop}{\nmax\Lambda_\hop}.
\end{aligned}
\end{equation}
Hence $\alpha_\hop=\nmax\Lambda_\hop =\nmax\sum_\ell|t_\ell| \le4Lt_{\max}\nmax$. Preparation supplies an amplitude $\sqrt{|c_\ell|/\Lambda_\hop}$, and projection through its inverse supplies the same factor again. The sign phase and the unchanged alias output therefore contribute
\begin{equation*}
\sqrt{\frac{|c_\ell|}{\Lambda_\hop}}\,
\sgn(c_\ell)\,
\sqrt{\frac{|c_\ell|}{\Lambda_\hop}}\,
\langle f_\ell|f_\ell\rangle
=\frac{c_\ell}{\Lambda_\hop}.
\end{equation*}
Distinct branch labels are orthogonal, so no cross terms remain. The alias states can depend on $\ell$; selection only needs to leave them unchanged. For $d_\hop=\log_2(4L)$, the projected registers consist of the $d_\hop$-qubit hopping label, the $d_\hop+2b_{\rm alias}+1$ retained alias register, and two signal qubits, giving
\[
a_\hop=2d_\hop+2b_{\rm alias}+3.
\]
The phase-gradient catalyst and clean arithmetic workspace are returned to their input states and are therefore not included in $a_\hop$. The angle- and alias-rounding bounds of Appendix~\ref{app:error} give
\[
\epsilon_\hop
\le
\left(
\frac{\pi}{2^b}
+
\frac{1}{2^{b_{\rm alias}}}
\right).
\]

For the arithmetic cost, Appendices~\ref{app:primitives} and~\ref{app:explicit-circuits} and \cref{tab:ledger} give
\begin{equation}
\begin{aligned}
\mathcal C_{\rm Tof}(O_A)
=
\mathcal C_{\rm Tof}(O_B)
=
2(R-1)+(b-1),
\\
\mathcal C_{\rm Tof}(S_\ell)
=
2\qph(L-1)+2(\qph-1). 
\end{aligned}
\end{equation}

Including destination-address computation, occupation loading and unloading, coefficient preparation, and the sign oracle therefore yields
\begin{equation}
\begin{aligned}
\mathcal C_{\rm Tof}(U_\hop)
={}&2\mathcal C_{\rm Tof}(\DEST)+4\qph(L-1)\\
&+4(R-1)+2(b-1)\\
&+2\qph(L-1)+2(\qph-1)\\
&+2\mathcal C_{\rm Tof}(\PREP_\hop)+\mathcal C_{\rm Tof}(O_{\sgn}).
\end{aligned}
\raisetag{15pt}
\end{equation}
The component scalings are
\begin{align*}
\mathcal C_{\rm Tof}(\DEST)&=O(\ell_x+\ell_y),\\
\mathcal C_{\rm Tof}(\PREP_\hop)&=O(L+b_{\rm alias}),\\
\mathcal C_{\rm Tof}(O_{\sgn})&=O(L),\\
R&=O(\nmax).
\end{align*}
Substituting these scalings gives
\[
\mathcal C_{\rm Tof}(U_\hop)
=
O\!\left(L\qph+\nmax+b+b_{\rm alias}\right).
\]
The two occupation-dependent rotations are implemented with the phase-gradient catalyst, so $\mathcal C_R(U_\hop)=0$.
\end{proof}
The complete Hamiltonian encoding later selects between $U_\hop$ and the on-site encoding. A reversible compute--use--uncompute subroutine need only control the operation between the computation and its inverse. If that subroutine has the form $F^\dagger V F$, then
\begin{equation*}
\begin{aligned}
C_e(V)&=\proj0_e\otimes I+\proj1_e\otimes V,\\
&(I_e\otimes F^\dagger)C_e(V)(I_e\otimes F)\\
&\qquad=\proj0_e\otimes I
+\proj1_e\otimes F^\dagger V F.
\end{aligned}
\end{equation*}
On the inactive branch, $F^\dagger F=I$. The computation and its inverse can therefore run unconditionally, provided that the intervening operation is enabled only on the active branch.

Within $U_\hop$, the destination-address computation, occupation loads, and table lookups are reversed before the branch ends. The direct occupation update $S$, the coefficient-sign phase, and the two signal rotations must obey the outer selection control. The explicit unary traversal in Appendix~\ref{app:shift-explicit} takes this control as its initial enable, so its shift count already includes the control. The rotations in Appendix~\ref{app:angle-explicit} mask the loaded angle with the enable and add $O(b)$ Toffoli-family operations. These constructions preserve the asymptotic scaling. The Pauli and exchange operations in $U_\loc$ must likewise obey the control because their effects remain on the system.

Preparation and reversible address operations may run unconditionally and then be undone. The tables in Appendix~\ref{app:ledger} retain their stated arithmetic subtotals. The explicit shift, rotation, and reflection circuits establish their own gate counts; a complete query count must also specify coefficient preparation, the remaining controls, and the choice of coherent or measurement-assisted erasure. Appendix~\ref{app:walk} combines these costs symbolically.

\section{Block encoding of the diagonal and exchange terms}
\label{sec:loc}

The remaining (on-site) contribution is $H_\loc=H_\dg+H_\intt$. The diagonal part is a linear combination of Pauli operators in the binary representation, and each local exchange operator is encoded using the ladder-operator construction. We assemble both in one linear combination of unitaries.

\subsection{Diagonal coefficients}

Combining the detuning and the local potential through $\delta_i=\Delta+\nu_i$ and substituting \cref{eq:atom-number,eq:binary-number} into \cref{eq:H-diag} gives
\begin{equation}
\begin{aligned}
H_\dg={}&C_{\rm id}I
+\sum_{i=1}^{L}\sum_{k=0}^{\qph-1}\frac{\nu_i2^k}{2}Z^{\ph}_{i,k}\\
&+\sum_{i=1}^{L}\sum_{r=1}^{N}\frac{\delta_i}{2}Z^{\at}_{i,r},
\end{aligned}
\label{eq:Hdiag-LCU}
\end{equation}
with the identity coefficient
\begin{equation}
C_{\rm id}=-\frac12\sum_{i=1}^{L}
\left[\nu_i\left(2^{\qph}-1\right)+N\delta_i\right].
\label{eq:Cid}
\end{equation}
Expanding in Pauli operators is the simplest choice here because $H_\dg$ is linear in the number operators, so the expansion has only $L\qph+LN$ terms; more general diagonal operators can be implemented using dedicated diagonal-circuit constructions \cite{zylberman2025efficient}.

The identity term shifts every eigenvalue equally. For real-time evolution it contributes the global phase $e^{-iC_{\rm id}\tau}$ and may be dropped from the block encoding. For the parameters used in the resource comparison below, doing so reduces $\alpha_\TCL$ by between one quarter and one third. The same shift cancels from energy differences between polariton sectors, including the particle-addition and particle-removal energies used to locate Mott-lobe boundaries. It needs to be retained only when an absolute energy relative to an externally fixed zero is itself the target.

\subsection{Local exchange and its shared angle}

We now focus on the light--matter interaction. For each site--emitter pair, we introduce the exchange operator
\begin{equation}
K_{i,r}=a_i\sigma^+_{i,r}+a_i^\dagger\sigma^-_{i,r},
\qquad
H_\intt=\sum_{i,r}g_iK_{i,r}.
\label{eq:K-def}
\end{equation}
The first entry of $\ket{m,s}$ denotes the photon occupation and $s\in\{0,1\}$ denotes the emitter state.  The exchange operator transfers one excitation between the emitter and the cavity,
\begin{equation}
\begin{aligned}
K_{i,r}\ket{m,1}
=
\sqrt{m+1}\ket{m+1,0},
\\
K_{i,r}\ket{m+1,0}
=
\sqrt{m+1}\ket{m,1}.
\end{aligned}
\end{equation}
Thus, for $m=0,\ldots,\nmax-1$, it preserves the two-dimensional subspace $\operatorname{span}\{\ket{m,1},\ket{m+1,0}\}$ and acts on that subspace as $\sqrt{m+1}\,X$.  The boundary states $\ket{0,0}$ and $\ket{\nmax,1}$ are annihilated by the truncated operator.  The largest singular value is therefore $\sqrt{\nmax}$, attained for $m=\nmax-1$, so $\|K_{i,r}\|=\sqrt{\nmax}.$ The ladder-operator construction of \cite{simon2025ladder} consequently gives a unitary $V_{i,r}$ with signal qubit $\eta$ satisfying
\begin{equation}
\left(\bra0_\eta\otimes I\right)V_{i,r}\left(\ket0_\eta\otimes I\right)
=\frac{K_{i,r}}{\sqrt{\nmax}}.
\label{eq:LOBE-block}
\end{equation}

The circuit implements the two directions of this exchange in one reversible sequence. It first increments the photon occupation when the emitter is excited. The selected photon occupation is then loaded into the work register $m$. This register controls the rotation $R_\theta$ and is subsequently returned to zero. The circuit then conditionally decrements the photon occupation when the emitter is in its ground state and finally flips the emitter. To encode the coefficient $\sqrt{m}$ associated with the loaded photon occupation, the angle satisfies
\begin{equation}
\cos\frac{\theta_m}{2}=\sqrt{\frac{m}{\nmax}},\qquad 0\le m\le\nmax,
\label{eq:lobe-angle}
\end{equation}
and the rotation has zero projected amplitude for $m>\nmax$. Both boundary states therefore have zero projected amplitude without a separate boundary test. The state $\ket{0,0}$ gives $m=0$. For $\ket{\nmax,1}$, the increment gives either the unused label $\nmax+1$ or wraps to $0$; both give zero projected amplitude. For example, take $\nmax=3$ and consider the exchange pair $\ket{1,1}$ and $\ket{2,0}$. Suppressing the signal qubit, the occupation-update and emitter-flip steps give
\begin{align*}
\ket{1,1}&\longrightarrow\ket{2,1}
\longrightarrow\ket{2,0},\\
\ket{2,0}&\longrightarrow\ket{2,0}
\longrightarrow\ket{1,0}
\longrightarrow\ket{1,1}.
\end{align*}
The occupation is loaded after the first arrow in each line, so both inputs use the table entry $m=2$. For the ideal angle, the signal-zero amplitude is therefore $\sqrt{2/3}$ in both directions, reproducing $K_{i,r}/\sqrt{\nmax}$ on this pair. The work register is erased before the remaining occupation update.

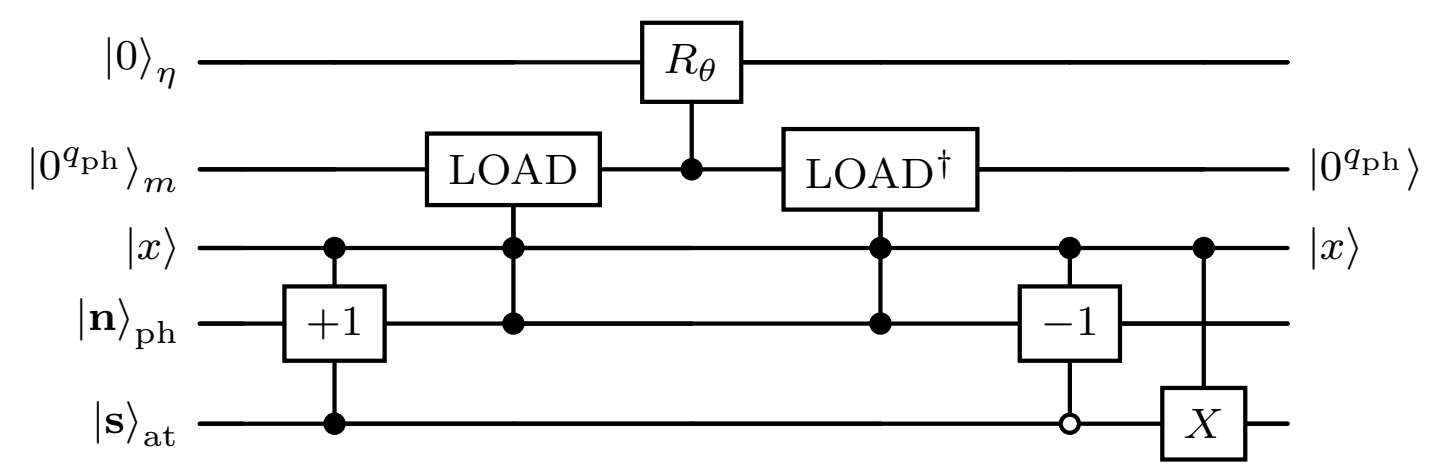
\begin{figure*}
\centering
\CircuitDiagram[\textwidth]{%
\begin{quantikz}[tcl circuit]
\lstick{$\ket0_\eta$} & \qw & \qw & \qw & \gate{R_\theta} & \qw & \qw & \qw & \qw \\
\lstick{$\ket{0^{\qph}}_m$} & \qw & \qw & \gate{\LOAD} & \ctrl{-1} & \gate{\LOAD^\dagger} & \qw & \qw & \rstick{$\ket{0^{\qph}}$} \\
\lstick{$\ket{x}$} & \qw & \ctrl{1} & \ctrl{-1} & \qw & \ctrl{-1} & \ctrl{1} & \ctrl{2} & \rstick{$\ket{x}$} \\
\lstick{$\ket{\mathbf n}_{\ph}$} & \qw & \gate{+1} & \ctrl{-2} & \qw & \ctrl{-2} & \gate{-1} & \qw & \qw \\
\lstick{$\ket{\mathbf s}_{\at}$} & \qw & \ctrl{-1} & \qw & \qw & \qw & \octrl{-1} & \gate{X} & \qw
\end{quantikz}}
\caption{Local exchange selected by $x=(i,r)$ in the $p$-qubit label
register. The photon and emitter registers have $L\qph$ and $LN$ qubits,
respectively. Address connections select the site and emitter named by
$x$. The filled emitter control enables the increment for an excited
emitter; the open control enables the decrement for a ground-state
emitter. The occupation is loaded after the increment and erased before
the decrement. The shared rotation is expanded in
\cref{fig:angle-explicit}; the enabled arithmetic is described in
\cref{app:shift-explicit}. The occupation work register returns to zero,
whereas projecting the signal $\eta$ onto zero gives the exchange block
of \cref{eq:LOBE-block}. The full local selection also conditions this
circuit on the exchange term type and the outer local-branch control.}
\label{fig:lobe}
\end{figure*}

The angle in \cref{eq:lobe-angle} depends only on the selected photon occupation $m$. The site and emitter indices determine which photon and emitter registers are updated, while $g_i$ determines the LCU weight. A separate multiplexed rotation in every site--emitter branch would repeat the same $R$-entry angle table $LN$ times and require $LNR$ arbitrary rotations. Instead, let $x=(i,r)$ denote the selected site--emitter label. We load the occupation of site $i$ once,
\begin{equation}
\ket{x}\ket{\mathbf n}_{\ph}\ket{0}_m
\longmapsto\ket{x}\ket{\mathbf n}_{\ph}\ket{m}_m,
\label{eq:load-m}
\end{equation}
by unary iteration on the site field of $x$. A single $R$-entry table, indexed by $m$, then drives a phase-gradient rotation. \Cref{fig:lobe} shows this branch. The Toffoli costs of its three components are
\begin{equation*}
\begin{aligned}
\text{Occupation load and inverse:}&\quad2\qph(L-1),\\
\text{Table lookup and inverse:}&\quad2(R-1),\\
\text{Phase-gradient rotation:}&\quad b-1.
\end{aligned}
\end{equation*}
These costs follow from Appendices~\ref{app:primitives} and~\ref{app:angle-explicit}. Each cost is incurred once per call to $\SELECT_\loc$, not once per site--emitter branch. Thus the interaction encoding requires no synthesized arbitrary rotations, whereas the multiplexed implementation requires $LNR$.

\subsection{Assembling the encoding}

We now combine the diagonal and exchange contributions into a common LCU representation of \(H_\loc\). We first specify the coefficient list and its normalization, and then construct the corresponding PREP and SELECT operations.

\paragraph{Coefficients.}
The local decomposition has one identity term, $L\qph$ photon-$Z$ terms, $LN$ emitter-$Z$ terms and $LN$ exchange blocks,
\begin{equation}
M_\loc=1+L\left(\qph+2N\right),
\label{eq:Mloc}
\end{equation}
with the following signed coefficients, denoted collectively by $c_x$:
\begin{equation*}
c_x=
\begin{cases}
C_{\rm id},&\text{identity},\\
\nu_i2^{k-1},&\text{photon }Z^{\ph}_{i,k},\\
\delta_i/2,&\text{emitter }Z^{\at}_{i,r},\\
g_i\sqrt{\nmax},&\text{exchange block }(i,r).
\end{cases}
\end{equation*}
The exchange coefficient follows from \cref{eq:LOBE-block}. If all local coefficients vanish, the local branch is omitted. Otherwise, its normalization is their one-norm,
\begin{equation}
\begin{aligned}
\alpha_\loc={}&\abs{C_{\rm id}}
+\frac{2^{\qph}-1}{2}\sum_{i=1}^{L}\abs{\nu_i}\\
&+\frac{N}{2}\sum_{i=1}^{L}\abs{\delta_i}
+N\sqrt{\nmax}\sum_{i=1}^{L}\abs{g_i},
\end{aligned}
\label{eq:alpha-loc}
\end{equation}
where for the second term we used $\sum_{k=0}^{\qph-1}2^{k-1}=(2^{\qph}-1)/2$. The two ways the cutoff enters are already visible here: hopping carries two ladder factors and contributes $\nmax$ through \cref{eq:alpha-hop}, whereas one local exchange carries a single ladder factor and contributes $\sqrt{\nmax}$.

\paragraph{Preparation.}
The selection register has $p=\lceil\log_2M_\loc\rceil$ qubits and supports $\widetilde M=2^p$ labels, with zero weight assigned to unused labels. As in \cref{eq:PREP-hop}, the ideal preparation is
\begin{equation}
\PREP_\loc\ket{0}_{xJ_\loc}
=\sum_x\sqrt{\frac{\abs{c_x}}{\alpha_\loc}}
\ket{x}\ket{f_x}_{J_\loc}.
\label{eq:PREP-loc}
\end{equation}
The states $\ket{f_x}$ are normalized auxiliary outputs. Coherent alias sampling with $b_{\rm alias}$ threshold bits implements the rounded probabilities of Appendix~\ref{app:alias-rounding}, using no arbitrary rotations and $O(M_\loc+b_{\rm alias})$ Toffoli-family operations \cite{Babbush2018}. The same classical coefficients determine the sign phases.

\paragraph{Selection.}
$\SELECT_\loc$ uses unary iteration to select the operation associated with label $x$. It preserves $x$ and acts trivially on $J_\loc$. It applies the coefficient phase $\sgn(c_x)$, with the zero-coefficient convention defined above, followed by $I$, $Z^{\ph}_{i,k}$, $Z^{\at}_{i,r}$, or the exchange branch of \cref{fig:lobe}. Padded labels select the identity. The circuit is unitary for every label, including physical labels with zero weight.

\begin{proposition}[Local block encoding]
\label{prop:local-block-encoding}
Let
\[
U_\loc=\PREP_\loc^\dagger\SELECT_\loc\PREP_\loc,
\]
with $M_\loc$ as in \cref{eq:Mloc}, and $\widetilde M=2^p$. Assuming the exact phase-gradient catalyst $\ket{G_{b+1}}$, $U_\loc$ is an $(\alpha_\loc,a_\loc,\epsilon_\loc)$ block encoding of $H_\loc$, with $\alpha_\loc$ as in \cref{eq:alpha-loc}  and $a_\loc=2p+2b_{\rm alias}+2.$

Writing $\alpha_{\rm int}=N\sqrt{\nmax}\sum_i|g_i|,$ the normalized block-encoding error satisfies
\begin{equation}
\epsilon_\loc
\le
\frac{\pi\alpha_{\rm int}}
     {2^{b+1}\alpha_\loc}
+
\frac{1}{2^{b_{\rm alias}}}.
\end{equation}
Under the measurement-assisted convention of Appendix~\ref{app:primitives},
\begin{equation}
\mathcal C_{\rm Tof}(U_\loc)
=
O\!\left(
LN+L\qph+\nmax+b+b_{\rm alias}
\right).
\end{equation}
The phase-gradient catalyst and clean arithmetic workspace are excluded from the projected ancilla count $a_\loc$; both return to their input states before projection.
\end{proposition}

\begin{proof}
With exact probabilities and angles, projection of the preparation, alias, and signal registers gives
\[
\Big(\bra{0}_{xJ_\loc}\bra{0}_\eta\otimes I\Big)
U_\loc
\Big(\ket{0}_{xJ_\loc}\ket{0}_\eta\otimes I\Big)
=
\frac{H_\loc}{\alpha_\loc}.
\]
The projected registers comprise the $p$-qubit label, the $p+2b_{\rm alias}+1$ retained alias register, and the signal qubit $\eta$, giving $a_\loc=2p+2b_{\rm alias}+2$. The angle- and alias-rounding bounds of Appendix~\ref{app:error} give the stated $\epsilon_\loc$.

The preparation and unary selection contribute $O(M_\loc+b_{\rm alias})$. The exchange branch contributes the selected increment and decrement, one occupation load and its inverse, one lookup in the shared $R$-entry angle table and its inverse, and one phase-gradient rotation. Using $R=O(\nmax)$ therefore gives
\[
\mathcal C_{\rm Tof}(U_\loc)
=
O(LN+L\qph+\nmax+b+b_{\rm alias}).
\]
The occupation-dependent rotation is implemented through the phase-gradient catalyst, so $\mathcal C_R(U_\loc)=0$.
\end{proof}

\section{The complete block-encoding and simulation cost}
\label{sec:full}

We now have separate block encodings of $H_\hop$ and $H_\loc$, with normalizations $\alpha_\hop$ and $\alpha_\loc$, respectively. One additional selection qubit chooses between them. Preparing the selection qubit and undoing that preparation require a single-qubit rotation and its inverse. We first construct the complete encoding and compare the cost of one call with the sparse-access construction. We then derive the number of calls required for real-time evolution and compare the resulting simulation cost with product formulas.

\subsection{LCU combination of the two branches}

\begin{figure*}
\centering
\CircuitDiagram[\textwidth]{\input{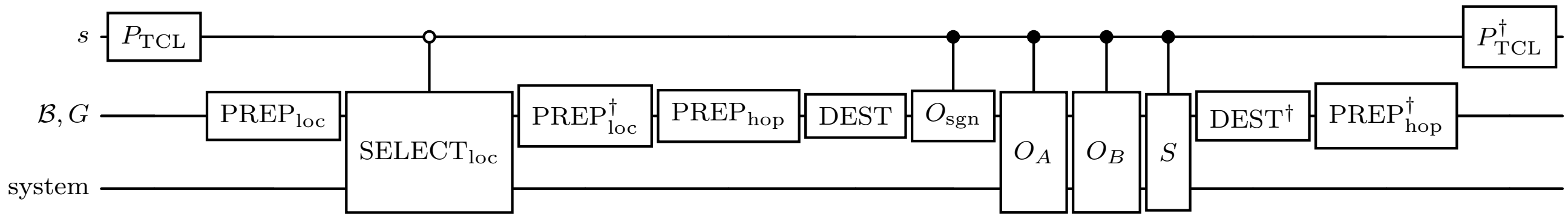}}
\caption{Complete block encoding $U_\TCL$, read from left to right. The outer
selection qubit starts in $\ket0_s$. The register $\mathcal B$ contains the
inner selection, signal, and work qubits, initially zero, while $G$ holds
the reusable phase-gradient state. The two preparations act on their own
label and alias registers within $\mathcal B$ and leave $G$ unchanged.
The open control selects the local branch for $s=0$; its selection includes
the local coefficient signs. Filled controls select hopping for $s=1$.
Here $O_A$ and $O_B$ include the reversible occupation access for their
respective rotations and act on different signal qubits. The destination
address is computed once by $\DEST$ and erased after the shift. The sign
oracle $O_{\sgn}$ is required for general signed hopping coefficients.
The shift $S$ changes occupations directly, without a second photon
configuration register. The preparations are allowed to run without the
outer control because they cancel when their selection is inactive.
All output wires are left unlabeled: projection onto zero defines the
encoded Hamiltonian, and is not a deterministic reset of the selection
and signal qubits. The controlled walk that uses this encoding is shown
separately in \cref{fig:walk-explicit}.}
\label{fig:outer}
\end{figure*}

Write $\alpha_\TCL=\alpha_\loc+\alpha_\hop$. When both branches are nonzero, we combine their block encodings using one selection qubit to encode $H_\TCL=H_\loc+H_\hop$. A single nonzero branch can be used directly. If $\alpha_\TCL=0$, the Hamiltonian vanishes and no evolution circuit is needed. The branch weights must cancel the two normalizations in \cref{eq:alpha-hop,eq:alpha-loc} when the ancillas are projected, so the selection qubit is prepared by
\begin{equation}
P_\TCL\ket{0}_s
=\sqrt{\frac{\alpha_\loc}{\alpha_\TCL}}\ket{0}_s
+\sqrt{\frac{\alpha_\hop}{\alpha_\TCL}}\ket{1}_s,
\label{eq:P-TCL}
\end{equation}
where $P_\TCL$ is implemented by a single arbitrary rotation. The complete encoding is
\begin{equation}
\begin{aligned}
U_\TCL={}&\left(P_\TCL^\dagger\otimes I\right)\\
&\times\left(\proj{0}_s\otimes U_\loc+\proj{1}_s\otimes U_\hop\right)\\
&\times\left(P_\TCL\otimes I\right),
\end{aligned}
\label{eq:U-TCL}
\end{equation}
shown in \cref{fig:outer}. The register $\mathcal B$ includes both inner labels, their signal qubits, and the alias outputs $J_\loc,J_\hop$, together with clean workspace initialized to zero. With exact probabilities and angles, projecting the selection qubit and these inner registers onto zero gives, by \cref{prop:hop-block-encoding} and \cref{prop:local-block-encoding},
\begin{equation}
\begin{aligned}
\Big(\bra{0}\bra{\mathbf 0}_{\mathcal B}\otimes I\Big)&U_\TCL
\Big(\ket{0}\ket{\mathbf 0}_{\mathcal B}\otimes I\Big)\\
&=\frac{H_\loc+H_\hop}{\alpha_\TCL},
\end{aligned}
\label{eq:TCL-projected}
\end{equation}
with
\begin{equation}
\alpha_\TCL=\alpha_\loc+\alpha_\hop
=\alpha_\loc+\nmax\sum_{\ell}\abs{t_\ell}.
\label{eq:alpha-TCL}
\end{equation}
For simulations where a global phase does not affect the observable, we can reduce the normalization by removing the identity term. The diagonal expansion in \cref{eq:Hdiag-LCU} isolates the term $C_{\rm id}I$. Define
\begin{equation}
\begin{aligned}
\overline H_\dg&=H_\dg-C_{\rm id}I,\\
\overline H_\TCL&=H_\TCL-C_{\rm id}I.
\end{aligned}
\label{eq:shifted-H}
\end{equation}
Because the removed term is proportional to the identity,
\begin{equation}
e^{-iH_\TCL\tau}
=e^{-iC_{\rm id}\tau}e^{-i\overline H_\TCL\tau}.
\label{eq:shifted-evolution}
\end{equation}
The first factor is a global phase and does not affect the observables considered here. For controlled evolution, the omitted phase becomes conditional and must be restored unless it cancels in the observable circuit. The correlator constructions in \cref{sec:endtoend} account for this distinction. Thus $\overline H_\TCL$ is not a new physical Hamiltonian; it generates the same real-time dynamics up to that phase. The corresponding block encoding omits the identity coefficient and has normalization
\begin{equation}
\alpha_\TCL^{(0)}=\alpha_\TCL-\abs{C_{\rm id}}.
\label{eq:alpha-shifted}
\end{equation}

For general coefficients, we use the encoding with the identity term removed. In the uniform examples below, we also remove the conserved-charge term using \cref{eq:charge-frame-H}. Then the simulated Hamiltonian is $\widehat H_\TCL=H_\hop+H_\intt$, with normalization
\begin{equation}
\widehat\alpha_\TCL=\alpha_\hop+\alpha_\intt
 =4L|t|\nmax+LN|g|\sqrt{\nmax}.
\label{eq:charge-frame-alpha}
\end{equation}
No identity term remains in these examples. We write $\alpha_{\rm dyn}$ for the normalization of the Hamiltonian used in the simulation: $\widehat\alpha_\TCL$ for these uniform examples, or $\alpha_\TCL^{(0)}$ for the generic identity-shifted construction.

To bound the error from the exchange-angle table, we define
\begin{equation}
\alpha_\intt=N\sqrt{\nmax}\sum_i\abs{g_i}.
\label{eq:alpha-int}
\end{equation}

Combining the bounds in \cref{prop:hop-block-encoding,prop:local-block-encoding} with the overhead for controlling the two branches gives the following lemma.

\begin{lemma}
\label{lem:be}
Let $H_\TCL$ be the truncated Hamiltonian of \cref{eq:H-split}, and assume $\alpha_\TCL=\alpha_\loc+\alpha_\hop>0$. On the exact catalyst state $\ket{G_{b+1}}$, the circuit $U_\TCL$ of \cref{eq:U-TCL} induces an $(\alpha_\TCL,a,\delta_{\be})$ block encoding of $H_\TCL$, with $a$ ancilla qubits projected onto $\ket{0}$, where
\begin{equation}
a=O\left(\log L+\log(N+\qph)+b_{\rm alias}\right).
\label{eq:ancilla-count}
\end{equation}
The catalyst and clean workspace are excluded from the projected ancilla count $a$. Under the arithmetic counting conventions of Appendix~\ref{app:primitives}, its cost satisfies
\begin{equation}
\begin{aligned}
\mathcal C_{\rm Tof}(U_\TCL)
&=O\left(L\qph+LN+\nmax+b+b_{\rm alias}\right),\\
\mathcal C_R(U_\TCL)&\le2.
\end{aligned}
\label{eq:full-cost}
\end{equation}
For angle precision $b$ and coefficient-preparation precision $b_{\rm alias}$,
\begin{equation}
\begin{aligned}
\delta_{\be}
\le\;&
\frac{1}{\alpha_\TCL}
\left[
\frac{\pi(2\alpha_\hop+\alpha_\intt)}{2^{b+1}}
\right.\\
&\left.
+\frac{\alpha_\hop+\alpha_\loc}{2^{b_{\rm alias}}}
\right]
+2\varepsilon_{\rm rot},
\end{aligned}
\label{eq:be-error}
\end{equation}
where $\varepsilon_{\rm rot}$ bounds the operator-norm synthesis error of each outer preparation rotation.
\end{lemma}
With both branches present and the retained padded-label layout, the exact projected width is
\begin{equation*}
\begin{aligned}
p&=\lceil\log_2 M_\loc\rceil,\\
a&=2(p+\ell_x+\ell_y+2)+4b_{\rm alias}+6.
\end{aligned}
\end{equation*}
The $b+1$ catalyst qubits and clean lookup, routing, angle, and carry workspace are excluded from the projected ancilla count $a$; they return to their input states before projection. The total auxiliary space scales as
\begin{equation*}
O\!\left[\log L+\log(LN)+\qph+b+b_{\rm alias}\right]
\end{equation*}
in the stated unary-iteration model. After the Hermitianization in Appendix~\ref{app:walk}, the projected block defines a Hermitian $H_\TCL^\prime$ satisfying
\begin{equation*}
\norm{H_\TCL^\prime-H_\TCL}
\le\alpha_\TCL\delta_{\be}.
\end{equation*}

Let $\epsilon_\hop$ and $\epsilon_\loc$ be the normalized branch block errors stated in the hopping and local propositions. The outer preparation weights the two branch errors by their normalizations:
\begin{equation*}
\delta_{\be}\le
\frac{\alpha_\hop}{\alpha_\TCL}\epsilon_\hop
+\frac{\alpha_\loc}{\alpha_\TCL}\epsilon_\loc
+2\varepsilon_{\rm rot}.
\end{equation*}
Hopping uses two occupation-dependent rotations, while the exchange part of the local branch uses one. Their weighted angle errors therefore give the coefficient $2\alpha_\hop+\alpha_\intt$ in \cref{eq:be-error}. The alias-error weights sum to one because $\alpha_\hop+\alpha_\loc=\alpha_\TCL$. The detailed angle and alias bounds are derived in Appendix~\ref{app:error}.

The two arbitrary rotations are $P_\TCL$ and its inverse, and their synthesis error is included separately. Every other amplitude in the circuit is produced by a table lookup and a phase-gradient rotation. The phase-gradient state is prepared once and reused, so its preparation is an additive setup cost rather than a per-query cost.

\subsection{Real-time evolution}

The preceding bounds concern the cost of one block-encoding call. We now determine the number of calls required to simulate real-time evolution. To this end, we construct a qubitization walk from the Hamiltonian block encoding and employ generalized quantum signal processing (GQSP) to approximate the time-evolution operator $e^{-iH_\TCL\tau}$. GQSP implements this evolution through a polynomial approximation to the exponential function, with the polynomial degree determining the number of walk queries and, consequently, the overall simulation cost. We therefore first establish the degree required for a given evolution time and target precision, and then derive the corresponding resource bounds. We state these bounds for $H_\TCL$; the same analysis applies to the chosen frame Hamiltonian with normalization $\alpha_{\rm dyn}$.

The simulation degree depends jointly on the normalized evolution time and the target precision. For $x>0$ and $0<\epsilon<1/2$, define
\begin{equation}
d_0(x,\epsilon)
=1+x+\frac{\log(1/\epsilon)}
{\log(e+\log(1/\epsilon)/x)},
\label{eq:degree-function}
\end{equation}
and set $d_0(0,\epsilon)=1$. The added constant covers bounded-degree cases; evolution at zero time can simply be omitted. Since $\alpha_\TCL\ge\norm{H_\TCL}$, the spectrum of $H_\TCL/\alpha_\TCL$ lies in $[-1,1]$. The Jacobi--Anger approximation and Hamiltonian simulation by quantum signal processing give
\begin{equation}
d=O \left(d_0(\alpha_\TCL\abs{\tau},\epsilon)\right)
\label{eq:degree}
\end{equation}
for polynomial approximation error $\epsilon$ \cite[Corollary~60]{gilyen2019quantum}. This form is uniform when time and precision vary together. At fixed nonzero $\alpha_\TCL\abs{\tau}$, the precision-dependent term reduces asymptotically to $O[\log(1/\epsilon)/\log\log(1/\epsilon)]$.

The selected exchange and hopping operations are not self-inverse. We first apply the Hermitianization of \cref{lem:hermitian} to the implemented encoding and its exact circuit adjoint. This adds one qubit whose $\ket{+}$ projection averages the implemented block with its adjoint. The enlarged encoding is Hermitian and self-inverse, which permits the walk construction. The projected block then defines a Hermitian $H_\TCL^{\prime}$ with $\norm{H_\TCL^{\prime}-H_\TCL}\le\alpha_\TCL\delta_{\be}$, even when the rounded directed-bond coefficients do not preserve Hermiticity. The resulting qubitized walk supplies the signal unitary \cite{Low2019hamiltonian}. Generalized quantum signal processing implements the complex polynomial using $O(d)$ signal queries and $O(d)$ single-qubit $SU(2)$ rotations \cite{PRXQuantum.5.020368}. Each signal query uses a constant number of controlled calls to the encoding or its adjoint, as detailed in Appendix~\ref{app:walk}.

The walk converts Hamiltonian eigenvalues into unitary phases. For an eigenvalue $\lambda$ of $H_\TCL^\prime$, its corresponding walk eigenvalues satisfy
\begin{equation*}
\begin{aligned}
z_{\lambda,\pm}&=e^{\pm i\theta_\lambda},\\
\cos\theta_\lambda&=\frac{\lambda}{\alpha_\TCL},\\
e^{-i\alpha_\TCL\tau\cos\theta_\lambda}
&=e^{-i\lambda\tau}.
\end{aligned}
\end{equation*}
GQSP approximates this phase function of the walk, producing the required evolution phase for each energy. The construction and its projected-subspace interpretation are given in Appendix~\ref{app:walk}.

\begin{figure*}
\centering
\input{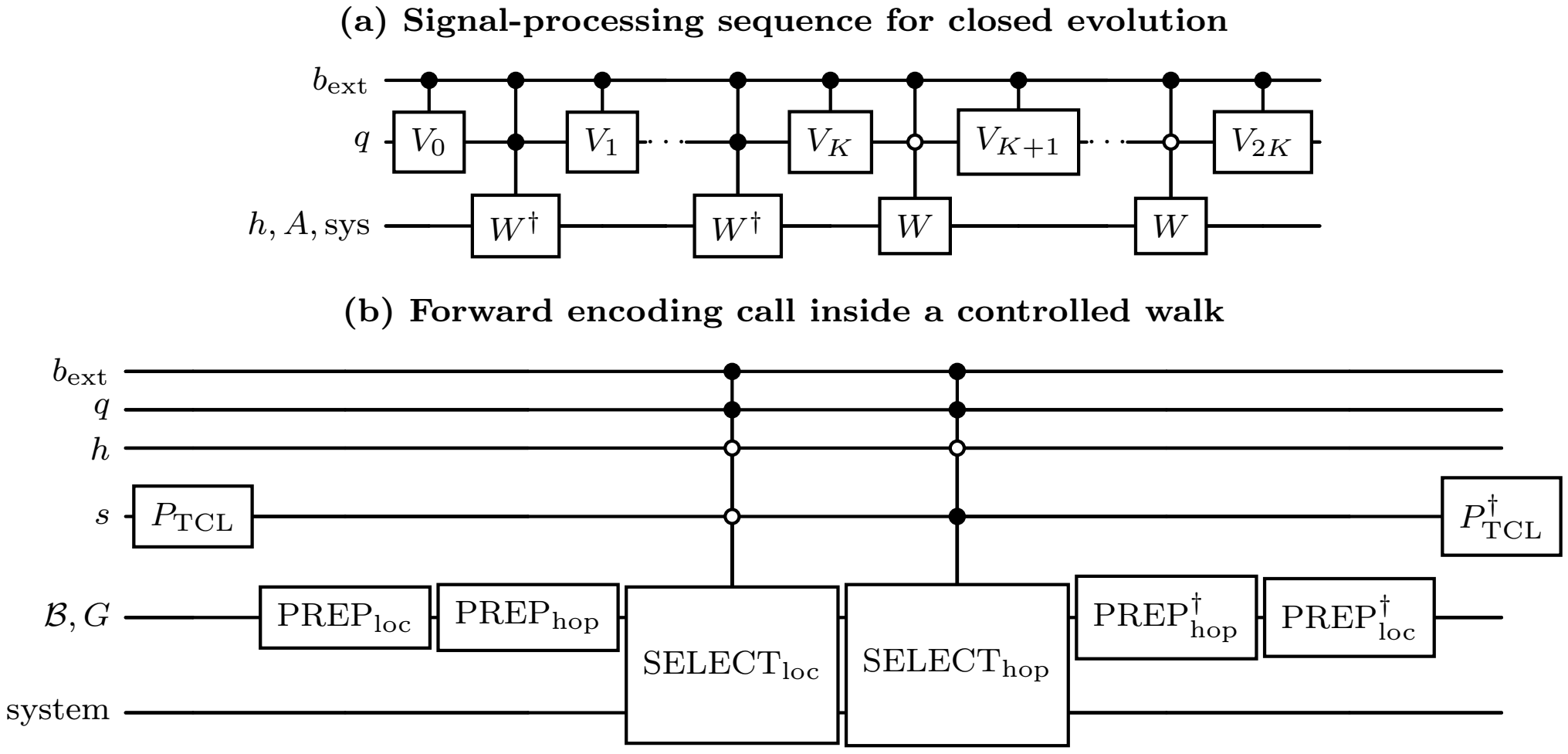}
\caption{ GQSP circuit for time evolution. (a) The signal qubit is $q$, the walk registers are $h,A$, and $b_{\rm ext}$ is an optional interferometer control, distinct from the angle precision $b$. In the common-evolution correlator circuit, $b_{\rm ext}=1$; only the ladder insertions outside this sequence need the interferometer control. The circuit uses $2K$ queries and $2K+1$ signal rotations. (b) The forward encoding call is expanded to show its branch controls. Within each walk, preparation is shared by the forward and adjoint branches, as in \cref{fig:walk-explicit}. Each walk therefore uses one pair of outer preparation rotations. The adjoint reverses the implemented gates exactly.}
\label{fig:dynamics-explicit}
\end{figure*}

\Cref{fig:dynamics-explicit} shows the repeated circuit. Write $K$ for the largest magnitude of a walk power in a bounded Jacobi--Anger approximation. The resulting GQSP circuit uses $K$ walk-adjoint queries controlled on $q=1$, followed by $K$ walk queries controlled on $q=0$, with $2K+1$ single-qubit signal rotations. This is $O(d)$ queries when $d$ in \cref{eq:degree} is evaluated at the chosen polynomial error. The external control $b_{\rm ext}$ in the figure enables the whole sequence when an interferometric measurement needs controlled evolution; it is distinct from the angle-table precision $b$. Prepare $q$ in $\ket0$, $h$ in $\ket{+}$, and $A$ in $\ket{0^a}$; the phase-gradient state is prepared once per run. None of these registers is reset between walk queries. The block obtained by projecting onto $q=0$, $h=+$, and $A=0$ approximates $e^{-iH_\TCL^{\prime}\tau}$. The polynomial and control conventions are specified in Appendix~\ref{app:walk}.

Here the initialized-state error bounds the norm of the difference between the actual joint output and the ideal evolved system with the simulation ancillas in their initialized states, for every system input. This includes leakage into other ancilla states.

To achieve an initialized system--ancilla state error of at most $\varepsilon$, we allocate $\varepsilon/2$ to the Hamiltonian block-encoding error. The remaining budget is divided equally between the GQSP polynomial approximation and signal-rotation synthesis. The sufficient precision requirements are
\begin{equation*}
\begin{aligned}
\abs{\tau}\alpha_\TCL\delta_{\be}
&\le\frac{\varepsilon}{2},\\
\delta_{\rm poly}&\le\frac{\varepsilon^2}{32},\\
\text{per-rotation error}&\le
\frac{\varepsilon}{4N_{\rm rot}},
\end{aligned}
\end{equation*}
where $N_{\rm rot}$ is the number of signal rotations. The polynomial requirement follows because a projected-block error $\delta_{\rm poly}$ can produce an initialized-state error of at most $\sqrt{2\delta_{\rm poly}}$. Importantly, $d_0(x,\varepsilon^2/32)=O[d_0(x,\varepsilon)]$, so this precision requirement does not change the asymptotic polynomial degree. The complete error analysis and corresponding precision choices are provided in Appendix~\ref{app:error}.

We now translate the required GQSP degree into circuit resources. Let $cW$ denote the fully coherent controlled walk of Appendix~\ref{app:walk}. Using the shared preparation in \cref{eq:shared-hermitianization}, its arithmetic cost is
\begin{equation*}
\mathcal C_{\rm Tof}(cW)
=2\mathcal C_{\rm Tof}(cU_\TCL)-C_F+2a,
\end{equation*}
where $C_F=2[P(D_\loc)+P(D_\hop)]$ is the cost of the common alias preparations and their inverses. Sharing these preparations removes one copy of this cost. Each walk also uses one outer preparation rotation and its inverse. The reflection contributes $2a$ coherent Toffolis. The resulting resource bounds for closed-system time evolution are summarized in the following theorem.
\begin{theorem}
\label{thm:closed}
Let $\varepsilon$ be the target initialized-state error for simulating $e^{-iH_\TCL\tau}$. Using the Hermitianization of \cref{lem:hermitian} followed by GQSP, the Toffoli-family arithmetic contribution for one evolution satisfies
\begin{equation}
G_{\rm closed}
=
O\left[
d_0\!\left(
\alpha_\TCL\abs{\tau},
\varepsilon
\right)
\mathcal C_{\rm Tof}(cW)
\right].
\label{eq:G-closed}
\end{equation}
Here $d_0$ is defined in \cref{eq:degree-function} and $\alpha_\TCL$ is given by \cref{eq:alpha-TCL}. The explicit coherent query-cost expressions in \cref{eq:v3-direct-cu,eq:walk-cost} retain the asymptotic scaling of \cref{eq:full-cost} under the stated counting conventions. The measurement-assisted subtotals require different erasure operations and cannot be substituted directly into this coherent count. Synthesis of the $O(d)$ arbitrary signal-processing rotations contributes an additional gate cost.
\end{theorem}

\section{Resource comparisons}

We now compare two distinct resource questions: first, how direct photon-register updates compare with two-register block encodings; second, how GQSP simulation compares with local product formulas at a specified accuracy.
\subsection{Comparison with a sparse-access implementation}
\label{sec:compare}

We compare two implementations of the Hamiltonian block encoding to quantify the savings from updating photon occupations directly. Our construction in \cref{sec:hop} updates the selected occupations in the existing photon registers. The sparse-access implementation writes the updated configuration into a second $L\qph$-qubit photon register through a row oracle $O_F$, then clears the original configuration through a reverse oracle $O_R$. We assign the same cost to these two oracles.

Both implementations use the same factorized occupation tables of \cref{lem:factor} and the same circuits for the on-site terms in \cref{sec:loc}. The comparison therefore isolates the resource savings from the direct photon update. Under the counting assumptions in Appendix \ref{app:ledger}, we denote the arithmetic costs of the direct and sparse-access constructions by $\mathcal C_{\rm Tof}^{\rm dir}$ and $\mathcal C_{\rm Tof}^{\rm sp}$, respectively. \Cref{tab:compare} reports these costs.

The coherent examples use $t_{ij}=0.05>0$, $g_i=\nu_i=1$, and $\Delta=0$. We simulate $\widehat H_\TCL$ from \cref{eq:charge-frame-H}. Every bond has the same hopping sign, so no bond-sign table is needed. All implementations use general alias preparation and the same padded label layout, including diagonal labels with zero weight. Each walk shares its preparation between the encoding and adjoint branches. The counts apply to the full truncated Hamiltonian and do not depend on which number sector is occupied. For the low-density spectrum, the reference state has $Q=1$ and the correlators visit $Q=0,2$. Every cutoff listed here gives exact truncation for those sectors. Applications with an extensive number of excitations require a separate check that the photon cutoff is sufficient.

\begin{figure*}[t]
\centering
\includegraphics[width=\textwidth]{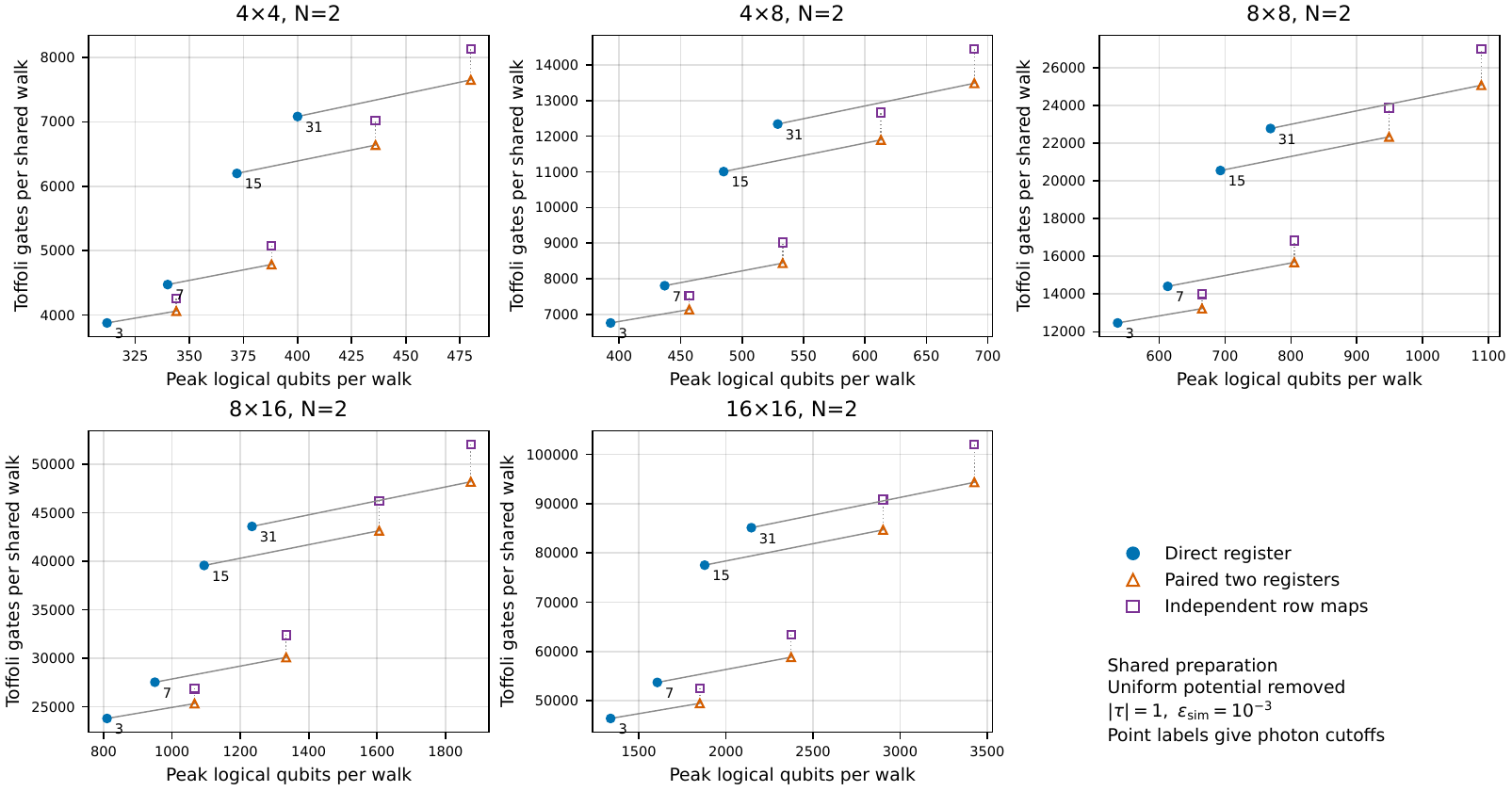}
\caption{ Logical resources for direct updates (filled circles), the paired two-register circuit (open triangles), and independently controlled row maps (open squares). Each panel fixes the lattice geometry at $N=2$; the point labels give the photon cutoff $\nmax$. All instances use $|\tau|=1$, $\varepsilon_{\rm sim}=10^{-3}$, $t=0.05$, $g=\nu=1$, and $\Delta=0$. We remove the uniform polariton potential and use one shared preparation pair per walk. The horizontal axis counts the peak qubits for the stated allocation, including the occupation phase-gradient catalyst and reusable clean workspace. The vertical axis counts coherent Toffoli gates per controlled walk and excludes the outer and GQSP rotations and catalyst preparation.}
\label{fig:qubit-toffoli-resource-tradeoff}
\end{figure*}

\Cref{fig:qubit-toffoli-resource-tradeoff} compares the direct circuit with two alternative two-register constructions. The paired construction uses unconditional CNOT banks around jointly controlled photon shifts on both registers, whereas the independently controlled row-map construction implements separate controlled forward and reverse maps. The figure covers five lattice geometries at $N=2$ and four photon cutoffs. All three constructions share the same preparation and remove the uniform polariton potential.

We compare the number of standard Toffoli (CCX) gates per controlled walk, including coherent uncomputation of temporary workspace. The paired construction is the cheaper of the two-register implementations considered here. It and the row-map circuit use the same occupation coefficients, controls, precision, reflection, and clean workspace as the direct construction. Updating occupations directly removes $L\qph$ qubits and saves the gates needed to update and copy a second photon register. These savings refer to the circuits specified here; other structured encodings may have different costs.

\paragraph{A matched coherent construction.}
We additionally specify a coherent Clifford--Toffoli implementation of both controlled encodings in Appendix \ref{app:matched-query}.  The two implementations share alias preparations, local terms, coefficient tables, controls, the destination address, and the projected reflection register.  The sparse circuit uses separately controlled forward and reverse row-map extensions and a clean $L\qph$-qubit second photon register, returned to zero after each selected update.  At $|\tau|=1$ and $\varepsilon_{\rm sim}=10^{-3}$ for $\widehat H_\TCL$, \cref{tab:v3-coherent-compare} gives the resulting Toffoli counts.  The angle and alias precisions, $b$ and $b_{\rm alias}$, obey \cref{eq:angle-precision,eq:precision-choice}. They therefore differ from the fixed $(20,20)$ precisions used for the arithmetic subtotals in \cref{tab:compare}.

\begin{table*}[t]
\caption{Coherent gate and qubit counts at $|\tau|=1$ and $\varepsilon_{\rm sim}=10^{-3}$ for the uniform charge-removed examples. $B=b_{\rm alias}$. The $cU$ column counts a separate controlled encoding; the $cW$ columns share one preparation pair across the Hermitianization branches. Savings are relative to the paired two-register walk. $P_\ph=L\qph$ is its additional photon register. $\mathsf Q_{\rm walk}$ is the peak qubit count for the stated allocation, including the phase-gradient catalyst for occupation-dependent rotations and the workspace reused by arithmetic and reflection. Appendix~\ref{app:matched-t} adds the outer rotations, GQSP, and preparation of the catalysts.}
\label{tab:v3-coherent-compare}
\squeezetable
\begin{ruledtabular}
\begin{tabular}{c c r r r r r r r}
$(L,N,\nmax)$ & $(b,B)$ & $a$ & $cU$: dir/row & $cW$: dir/row &
$cW$: pair & saving & $P_\ph$ & $\mathsf Q_{\rm walk}$: dir/pair \\
\colrule
$(16,1,3)$ & $(20,19)$ & 108 & $2{,}308/2{,}496$ & $3{,}690/4{,}066$ & $3{,}874$ & $4.75\%$ & 32 & $297/329$ \\
$(64,2,7)$ & $(23,22)$ & 128 & $8{,}834/10{,}046$ & $14{,}410/16{,}834$ & $15{,}682$ & $8.11\%$ & 192 & $613/805$ \\
$(256,4,15)$ & $(27,26)$ & 154 & $52{,}212/58{,}864$ & $83{,}720/97{,}024$ & $90{,}880$ & $7.88\%$ & 1024 & $2{,}399/3{,}423$ \\
\end{tabular}
\end{ruledtabular}
\end{table*}

For the first row, $a=108$ and $C_F=1142$. Using one preparation pair gives
\begin{align*}
\mathcal C_{\rm Tof}(cW_{\rm dir})&=2(2308)-1142+2(108)=3690,\\
\mathcal C_{\rm Tof}(cW_{\rm row})&=2(2496)-1142+2(108)=4066,\\
\mathcal C_{\rm Tof}(cW_{\rm pair})&=3690+184=3874 .
\end{align*}
Thus $1-3690/3874\simeq4.75\%$ is the Toffoli saving relative to the paired circuit. The preparation and reflection costs are common to all columns.

With the second photon register initialized to zero, the paired alternative applies the enabled photon shift to both registers between unconditioned CNOT banks for copying and cleanup. When the enable is zero, the shifts do nothing and the two CNOT banks cancel. This avoids separately controlled copies, but the combined update does not implement separately controlled forward and reverse row maps.

The direct walk saves $4.75\%$, $8.11\%$, and $7.88\%$ relative to the paired circuit. Relative to the separately controlled row maps, the savings are $9.25\%$, $14.40\%$, and $13.71\%$. It also saves $32$, $192$, and $1024$ qubits by removing the second photon register. These counts include alias preparation, indexed reads, local exchange, controlled rotations and shifts, and the reflection. Independent Qualtran checks \cite{harrigan2024qualtran} reproduce the coherent Toffoli counts and peak logical-qubit allocations in \cref{tab:v3-coherent-compare}, using the stated shared
preparation, precision, controls, and workspace conventions. Small-instance gate-level tests additionally check projected operator actions, coherent controls, and clean workspace uncomputation. The measurement-assisted subtotals appear separately in Appendix \ref{app:ledger}, and \cref{tab:v3-T-compare} adds the rotations needed for time evolution.

The normalization bounds the Hamiltonian on the full truncated register, so it can be much larger than the energy range in a low-density sector. For $L=16$, $N=1$, $\nmax=3$, $t=0.05$, and $g=\nu=1$, an identity-only shift gives $\alpha_\TCL^{(0)}=69.31281$ and $\|\overline H_\TCL|_{Q=1}\|=32.10499$. The ratio is $2.16$, but the sector norm is dominated by its common energy offset. After removing $-\nu N_p$, the $Q=1$ spectrum is $[-1.10499,1.10499]$. Adding a photon to a reference state in this sector reaches $Q=2$. This sector has dimension 512 and spectrum $[-2.20204,2.20204]$. Removing a photon reaches the one-dimensional vacuum sector. Comparing the normalization with the centered $Q=2$ spectrum gives
\begin{equation}
\frac{\widehat\alpha_\TCL}
{(E_{\max,\widehat H,Q=2}-E_{\min,\widehat H,Q=2})/2}
=\frac{37.31281}{2.20204}\simeq16.9.
\label{eq:sector-normalization-ratio}
\end{equation}
The denominator is the half-width of the centered spectrum; the full width is $4.40407$. This comparison shows the reduction in the energy scale after removing a common phase. Achieving a corresponding circuit saving would require an encoding restricted to the relevant sectors, including every number and spin sector reached by the observable.

The coefficient tables in \cref{tab:compare} use plain unary iteration. A select--swap lookup can trade additional workspace for fewer Toffoli-family operations, with a dependence on both the number of entries and the output-word width described in Appendix~\ref{app:primitives}. The same lookup method would be used for the shared tables in both constructions. Operations that address the quantum system register require a separate routing analysis. Reducing common table costs can change the ratio of the two totals. We therefore make no numerical prediction for that ratio without specifying the lookup scheme, output widths, and available workspace.

Other ways to generate the ladder coefficients could reduce the costs shared by both constructions. Peng et al.~\cite{peng2025boson} combine in-place cyclic shifts with a sign-LCU approximation and avoid computing square roots explicitly. Their polylogarithmic dependence on the cutoff shows that our $O(R)$ angle-table lookup is one implementation choice. Recent structured block encodings separate the choice of support from the local operation, or route operations through lattice patches ~\cite{summer2026structured,liu2026patch}. Rescaling the patch terms can also accommodate compatible inhomogeneous coupling strengths. Using these constructions for the TCL model would require explicit circuits and gate counts for the binary photon registers.

\subsection{Comparison with product formulas}
\label{sec:trotter}

Ref.~\cite{tudorovskaya2024cavity} gives compiled product-formula circuits for photon hopping in cavity arrays. It therefore addresses a related lattice simulation with a different circuit construction.   Our approach instead uses block encoding to combine the hopping terms through coherent selection.

Having established the cost of GQSP simulation using our block encoding, we compare it with Trotter--Suzuki product formulas. This comparison concerns the total cost of approximating the same evolution at the same accuracy. GQSP has a favorable dependence on the target precision. Its query count depends on the block-encoding normalization, and each query incurs the cost of routing and implementing the block encoding. A product formula avoids PREP/SELECT routing and instead repeats short evolutions under local Hamiltonian terms. Its total cost is set by the number of repetitions needed to suppress Trotter error and by the gate cost of one repetition. \cref{eq:G-closed} alone therefore does not establish which approach is cheaper at finite parameters.

We apply both methods to the shifted Hamiltonian $\overline H_\TCL$ defined in \cref{eq:shifted-H}. For the product formula, we divide this Hamiltonian into commuting layers. On the even periodic square lattices considered here, write
\begin{equation}
\begin{aligned}
\overline H_\TCL
&=\overline H_\dg+\sum_{c=1}^{4}H_{\hop}^{(c)}+\sum_{r=1}^{N}H_{\intt}^{(r)}\\
&=\sum_{\mu=1}^{m}H_\mu,
\qquad m=N+5.
\end{aligned}
\label{eq:pf-layers}
\end{equation}
The four hopping layers are the horizontal-even, horizontal-odd, vertical-even, and vertical-odd edge matchings. All two-cavity terms within one matching have disjoint support and commute. The exchange layer $H_{\intt}^{(r)}=\sum_i g_iK_{i,r}$ also contains commuting terms because it uses one emitter label $r$ across distinct cavities. Finally, all number-operator terms in $\overline H_\dg$ commute. Each layer exponential therefore factorizes into local one- or two-register operations, even though different layers do not commute.

For a step length $\delta$, the first- and second-order formulas are
\begin{align}
S_1(\delta)&=\prod_{\mu=1}^{m}e^{-iH_\mu\delta},
\label{eq:pf-first}\\
S_2(\delta)&=
\left(\prod_{\mu=1}^{m}e^{-iH_\mu\delta/2}\right)
\left(\prod_{\mu=m}^{1}e^{-iH_\mu\delta/2}\right).
\label{eq:pf-second}
\end{align}
Let
\begin{align}
\mathcal K_1&=\sum_{\mu<\nu}\norm{[H_\mu,H_\nu]},
\nonumber\\
\mathcal K_2&=\sum_{\mu,\nu,\xi}^{\prime}
\norm{[H_\mu,[H_\nu,H_\xi]]},
\label{eq:pf-commutators}
\end{align}
where the prime restricts the second sum to nonzero nested commutators. This definition provides a conservative scale for the ordering-dependent second-order error. Commutator-based product-formula bounds give
\begin{align}
\norm{e^{-i\overline H_\TCL\tau}-S_1(\tau/r)^r}
&=O\left(\frac{\mathcal K_1\tau^2}{r}\right),
\nonumber\\
\norm{e^{-i\overline H_\TCL\tau}-S_2(\tau/r)^r}
&=O\left(\frac{\mathcal K_2\abs{\tau}^3}{r^2}\right)
\label{eq:pf-errors}
\end{align}
\cite{childs2021trotter}. A sufficient segment count for an error below $\varepsilon$ is, up to formula- and ordering-dependent constants,
\begin{equation}
\begin{aligned}
r_1&=O\left(1+\frac{\mathcal K_1\tau^2}{\varepsilon}\right),\\
r_2&=O\left(1+\sqrt{\frac{\mathcal K_2\abs{\tau}^3}{\varepsilon}}\right).
\end{aligned}
\label{eq:pf-segments}
\end{equation}
More generally, a formula of fixed order $p$ admits the sufficient segment count
\begin{equation*}
r_p=O\left[1+\left(
\frac{\mathcal K_p\abs{\tau}^{p+1}}{\varepsilon}
\right)^{1/p}\right].
\end{equation*}
Each segment contains an order-dependent number of exponentials. The resulting upper bound has algebraic precision dependence. It need not describe the observed error for a particular state or a Hamiltonian with additional commutation relations.

Locality restricts the commutator sums to overlapping terms. Writing $h_X$ for a local Hamiltonian term supported on $X$ gives
\begin{equation*}
[h_X,h_Y]=0\qquad\text{when }X\cap Y=\varnothing.
\end{equation*}
In a nonzero nested commutator, each additional term must overlap the support accumulated by the inner commutator. For fixed formula order $p$, fixed $N$, and fixed $\nmax$, each starting term has only a bounded number of such overlap sequences. There are $O(L)$ choices of its starting location. Bounding the norm of each sequence by a constant times $\lambda_*^{p+1}$ gives
\begin{equation*}
\mathcal K_p=O\!\left(L\lambda_*^{p+1}\right),
\end{equation*}
where $\lambda_*$ is a local truncated energy scale. For example, $\lambda_*$ may be chosen as a constant multiple of
\begin{equation*}
\begin{aligned}
&t_{\max}\nmax+\nu_{\max}\nmax\\
&\quad+N\delta_{\max}+Ng_{\max}\sqrt{\nmax},
\end{aligned}
\end{equation*}
with each subscript ``max'' denoting the largest coefficient magnitude of that type. If $c_{\rm loc}$ denotes the gate cost of one local exponential and $s_p$ the number of stages in an order-$p$ formula, one segment costs $O(s_pLc_{\rm loc})$. The resulting bound has the form
\begin{equation}
G_{\rm PF}^{(p)}=
O\left[
s_pc_{\rm loc}L\left(1+
\frac{L^{1/p}(\lambda_*\abs{\tau})^{1+1/p}}{\varepsilon^{1/p}}\right)
\right].
\label{eq:pf-local-cost}
\end{equation}
High-order product formulas can consequently approach linear scaling in lattice volume and evolution time for geometrically local Hamiltonians \cite{childs2019nearly}. This locality advantage is important here. With fixed local parameters, the shifted normalization obeys $\alpha_\TCL^{(0)}=O(L\lambda_*)$. Let $cW^{(0)}$ be the fully coherent controlled walk for the shifted encoding of $\overline H_\TCL=H_\TCL-C_{\rm id}I$, and set $C_{\rm qry}^{(0)}=\mathcal C_{\rm Tof}(cW^{(0)})$. Appendix~\ref{app:matched-query} gives explicit coherent query counts for the chosen padded labels. In the frame with the uniform polariton potential removed, use $\alpha_{\rm dyn}$. The corresponding GQSP bound is
\begin{equation}
G_{\rm GQSP}=
O\left[
C_{\rm qry}^{(0)}
d_0(\alpha_\TCL^{(0)}\abs{\tau},\varepsilon)
\right],
\label{eq:gqsp-pf-comparison}
\end{equation}
Let $c_{\rm qry}$ denote a per-site Toffoli-family cost for the same controlled query at fixed local parameters, using common control, erasure, precision, and workspace conventions. If a fully coherent matched compilation of the direct and sparse controlled encodings establishes $C_{\rm qry}^{(0)}=O(Lc_{\rm qry})$, the time-dependent contribution of this bound is $O(L^2c_{\rm qry}\lambda_*\abs{\tau})$. These scalings are conditional: neither $\mathcal C_{\rm Tof}(cU_\TCL^{(0)})$ nor a matched numerical walk-level saving follows from the arithmetic ledger in \cref{tab:compare}.

The two factors of $L$ have different origins. The coefficient sum in the global encoding gives a time-dependent query-count bound $O(L\lambda_*|\tau|)$, while routing and selection give the conditional per-query bound $O(Lc_{\rm qry})$. The direct occupation update changes the routing circuit; it leaves the shared normalization unchanged.

The two methods therefore favor different regimes. At fixed $L$ and $\tau$, the GQSP degree has polylogarithmic precision dependence, whereas the generic fixed-order product-formula bound above scales as $\varepsilon^{-1/p}$. This compares analytic bounds; commuting terms or state-dependent cancellations can reduce the product-formula error. At fixed precision on a growing local lattice, however, a product formula can have the better volume scaling and a smaller step cost. The hopping exponentials $e^{-iH_{\hop}^{(c)}\delta}$ and exchange exponentials $e^{-iH_{\intt}^{(r)}\delta}$ are not free in the binary bosonic encoding; their occupation-dependent rotations must be compiled before the constant $c_{\rm loc}$ is known.

For fixed $N$ and $\nmax$ with bounded, efficiently specified local interactions, Lieb--Robinson-based decompositions also achieve gate costs quasilinear in lattice space--time volume and polylogarithmic in inverse error \cite{haah2018lattice}. Their compilation for the binary bosonic registers is not analyzed here.

Bidirectional GQSP can reduce the number of queries ~\cite{berry2024doubling}. We use the certified $2K$-query sequence here. A numerical comparison would need to include the alternative sequence's controls, reflections, corrections, and phase precision under the same initialized-state error budget.

\section{Extension to open dynamics}
\label{sec:open}

We now extend the Hamiltonian block encoding to the Lindblad generator of \cref{eq:lindblad}. The dissipator contains $J=L+2LN$ jump operators. A direct first-order dilation coherently represents all $J$ jumps, while the construction below samples one jump operator per timestep and rescales it based on its sampling probability to ensure that averaging recovers the target generator.

Each sampled timestep is implemented using a single-ancilla first-order Hamiltonian dilation based on Ref.~\cite{PRXQuantum.5.020332}.

Ref.~\cite{dingchenlin2024singleancilla} also uses a single ancilla to implement dissipative dynamics. We analyze the sampled construction because it chooses the jump label classically, reuses the lattice Hamiltonian encoding, and avoids coherently routing a jump-index register. These features simplify the circuit. Establishing whether it uses fewer gates than a deterministic method would require both methods to be compiled at the same accuracy.

Its $\varepsilon^{-1}$ dependence in the timestep count is not claimed to be optimal. Polylogarithmic precision dependence is achievable for Lindblad simulation under a Pauli-decomposition access model \cite{cleve2017effective}. Higher-order series-expansion methods also achieve polylogarithmic precision dependence when the Hamiltonian and jump operators are provided as block encodings~\cite{liwang2023series}. The exact commuting dissipative layer and symmetric splitting in \cref{eq:exact-dissipative-layer,eq:symmetric-open-step}, as well as randomized and trajectory-based algorithms, provide alternatives under different circuit and access assumptions.

The bounds in this section use the unshifted $H_\TCL$ and $\alpha_\TCL$ as conservative reference values. One may instead construct the dilation with $\overline H_\TCL=H_\TCL-C_{\rm id}I$ and replace the Hamiltonian norm and normalization throughout by $\norm{\overline H_\TCL}$ and $\alpha_\TCL^{(0)}$. The target generator is unchanged because the identity commutes with every state. The finite-step dilation does change, so its error and degree bounds must use the same chosen energy shift.

For uniform $\nu_i=\nu$ and $\Delta=0$, we can also remove the conserved-charge term using \cref{eq:charge-frame-H}. Photon populations and factorial moments are unchanged, so their simulation uses $\widehat H_\TCL$ and $\widehat\alpha_\TCL$ throughout the dilation and error bounds. Applying the known frame unitary recovers the full density matrix and preserves the channel-error norm.

\subsection{Sampled first-order dilation}

Stochastic dilation implements one Lindblad timestep by evolving an enlarged system under a Hermitian Hamiltonian and then tracing out the added register; it requires no postselection \cite{PRXQuantum.5.020332}. Define
\begin{equation*}
\norm{\mathcal L}_{\be}
=1+\norm{H_\TCL}+\sum_j\norm{V_j}^2.
\end{equation*}
The order-$k$ scheme of \cite{PRXQuantum.5.020332} has the resource and error bounds
\begin{equation*}
\begin{aligned}
\text{block count}&\le(J+1)^{k+1},\\
\text{dilation qubits}&\le\lceil(k+1)\log_2(J+1)\rceil,\\
\text{error}&=O\big(T\norm{\mathcal L}_{\be}^{k+1}\Delta t^k\big).
\end{aligned}
\end{equation*}
At first order, one coherent branch is needed for each of the $J$ jumps.

The general order-$k$ theorem restricts the timestep and has constants that depend on $k$ and $J$. For the single-jump construction used here, we derive the remainder explicitly in Appendix~\ref{app:open-cost}, retaining its dependence on the physical channels.

To remove this coherent $J$-way selection, we draw one jump index at each timestep, a sampling strategy also employed in \cite{Chen2025Randomized,watts2026dissipative}. When at least one jump has nonzero norm, define
\begin{equation}
\begin{gathered}
\lambda_{\mathcal D}=\sum_{j=1}^{J}\norm{V_j}^2,
\qquad
p_j=\frac{\norm{V_j}^2}{\lambda_{\mathcal D}},\\
\widetilde{V}_j =\frac{V_j}{\sqrt{p_j}},
\end{gathered}
\label{eq:sampling}
\end{equation}
For this model the truncation gives $\norm{a_i}^2=\nmax$ and $\norm{\sigma^-}=\norm{\sigma^z}=1$, so
\begin{equation}
\begin{aligned}
\lambda_{\mathcal D}={}&\nmax\sum_{i}\kappa_i
+\sum_{i,r}\gamma_{1,i,r}
+\frac12\sum_{i,r}\gamma_{\phi,i,r},
\end{aligned}
\label{eq:lambdaD}
\end{equation}
which is $O[\Gamma_{\max}(L\nmax+LN)]$ for rates bounded by $\Gamma_{\max}$. We sample only jump operators with nonzero dissipation rates. If $\lambda_{\mathcal D}=0$, the evolution is Hamiltonian and we use the closed-system construction without sampling or a dilation qubit. In the remaining case, draw $j$ classically with probability $p_j$. The corresponding rate-rescaled generator is
\begin{equation}
    \mathcal L^{\rm samp}_j=-i[H_\TCL,\cdot]+p_j^{-1}\mathcal D_j, 
\end{equation}
where $\mathcal D_j$ is the $j$th dissipator, hence
\begin{equation}
\mathbb E_{j\sim p}\left[\mathcal L^{\rm samp}_j\right]=\mathcal L,
\label{eq:unbiased}
\end{equation}
as in randomized product formulas \cite{campbell2019random}. Thus the average of the sampled generators equals the target generator, although an individual sample does not.

This equality does not extend to the finite-step exponentials. At fixed cutoff the generators are bounded, so their expansions give
\begin{equation*}
\begin{aligned}
&\sum_jp_j e^{\Delta t\mathcal L_j^{\rm samp}}
-e^{\Delta t\mathcal L}\\
&\quad=\frac{\Delta t^2}{2}
\Big[\sum_jp_j(\mathcal L_j^{\rm samp})^2\\
&\qquad\qquad-\mathcal L^2\Big]+O(\Delta t^3).
\end{aligned}
\end{equation*}
The linear terms agree, while the quadratic terms generally differ. This is the sampling error bounded in Appendix~\ref{app:open-cost}. The probabilities in \cref{eq:sampling} also equalize the rescaled jump norms:
\begin{equation}
\norm{\widetilde{V}_j }^2=\frac{\norm{V_j}^2}{p_j}=\lambda_{\mathcal D},
\label{eq:equal-norm}
\end{equation}
so $\norm{\mathcal L^{\rm samp}_j}_{\be}=\norm{\mathcal L}_{\be}$ for every $j$ and the same timestep may be used whichever channel is drawn. Here, $\norm{\mathcal L}_{\be}\le1+\alpha_\TCL+\lambda_{\mathcal D}$. Substituting the single sampled jump $\widetilde{V}_j $ for the $J$-branch column of the $k=1$ dilation of \cite{PRXQuantum.5.020332} leaves the two-branch matrix
\begin{equation}
\widetilde H_j=
\begin{pmatrix}
\sqrt{\Delta t}H_\TCL & \widetilde{V}_j ^\dagger\\
\widetilde{V}_j  & 0
\end{pmatrix},
\label{eq:dilated}
\end{equation}
that is,
\begin{equation}
\begin{gathered}
\widetilde H_j=\sqrt{\Delta t}\proj{0}\otimes H_\TCL+\mathcal K_j,
\\
\mathcal K_j=\ket{1}\bra{0}\otimes \widetilde{V}_j +\text{h.c.}.
\end{gathered}
\label{eq:dilated-split}
\end{equation}
Note that one timestep initializes the dilation qubit in $\ket0$, applies $e^{-i\sqrt{\Delta t}\widetilde H_j}$, and traces out that qubit.

The two powers of $\sqrt{\Delta t}$ serve different roles:
\begin{equation*}
\sqrt{\Delta t}\,\widetilde H_j
=\Delta t\,\proj0\otimes H_\TCL
+\sqrt{\Delta t}\,\mathcal K_j.
\end{equation*}
Let $K_{a,j}$ be the system operator obtained when the dilation qubit, initially in $\ket0$, ends in $\ket a$. To leading order,
\begin{equation*}
\begin{aligned}
K_{0,j}
&=I-i\Delta t H_\TCL\\
&\quad-\frac{\Delta t}{2}
\widetilde V_j^\dagger\widetilde V_j
+O(\Delta t^2),\\
K_{1,j}
&=-i\sqrt{\Delta t}\,\widetilde V_j
+O(\Delta t^{3/2}).
\end{aligned}
\end{equation*}
The $\ket1$ amplitude is of order $\sqrt{\Delta t}$, so the jump probability is of order $\Delta t$, as required for a dissipative rate. The $\ket0$ branch supplies the accompanying no-jump correction. The remainder constants here are for fixed bounded generators; Appendix~\ref{app:open-cost} gives the norm-dependent remainders and full channel-error bound.

The dilation qubit is reset to $\ket0$ before the next timestep. This sampling is related to averaging over stochastic open-system trajectories \cite{dalibard1992wavefunction}, although each dilation step is implemented as a unitary channel followed by a trace rather than as a postselected jump.

\subsection{Block encoding of the dilated Hamiltonian}

The two terms of \cref{eq:dilated-split} are block encoded separately and combined by a one-qubit linear combination, exactly as in \cref{sec:full}. The resulting normalization is recorded in the following lemma.

\begin{lemma} For each classically sampled jump index $j$, the dilation Hamiltonian $\widetilde H_j$ in \cref{eq:dilated} admits a block encoding $U_{\rm dil}^{(j)}$ with normalization
\begin{equation}
\alpha_{\rm dil}
=
\sqrt{\Delta t}\,\alpha_\TCL
+\sqrt{\lambda_{\mathcal D}}.
\label{eq:alpha-dil}
\end{equation}
The same normalization applies to every sampled realization. Here $\lambda_{\mathcal D}$ is given by \cref{eq:lambdaD}, and $\alpha_\TCL$ is given by \cref{eq:alpha-TCL}.
\label{lem:open_normalization}
\end{lemma}

The construction and normalization are proved in Appendix \ref{app:open-block-encoding}. Finite-precision implementation errors are included in the per-step accuracy budget of \cref{thm:open}.

Sampling changes the register and routing costs, but it does not reduce the block-encoding normalization. The all-jump off-diagonal block and its norm satisfy
\begin{equation*}
\begin{aligned}
M&=\sum_j\big(\ket{j}\bra{0}\otimes V_j+\text{h.c.}\big),\\
\norm M&=\norm{\sum_jV_j^\dagger V_j}^{1/2}\\
&\le\sqrt{\lambda_{\mathcal D}},
\end{aligned}
\end{equation*}
which is the same dissipative bound used in \cref{eq:alpha-dil}. Writing the branch index into the dilation register makes the outputs orthogonal. A preparation weighted by $\sqrt{p_j}$, together with a select operation over block encodings of $V_j/\norm{V_j}$, therefore achieves the normalization $\sqrt{\lambda_{\mathcal D}}$ deterministically as well. The sampled construction therefore provides no improvement in normalization. Its benefit is that the dilation register decreases from $\lceil\log_2(J+1)\rceil$ qubits to one and the $O(J)$ coherent branch routing is removed from every timestep. At fixed photon cutoff and precision, $J=L+2LN$ has the same asymptotic order as $\mathcal C_{\rm Tof}(U_\TCL)$ in \cref{eq:full-cost}. Thus the removal of coherent jump routing is a constant-factor simplification of the per-step circuit rather than a change in its asymptotic scaling.

\subsection{Timestep count and total cost}

We measure channel error in the diamond norm, which bounds the trace-norm error for every input, including an entangled reference. Set $\Lambda=\norm{\mathcal L}_{\be}\le1+\norm{H_\TCL}+\lambda_{\mathcal D}$.

The timestep and arithmetic-cost bounds apply to the channel averaged over independent classical jump samples. For $\lambda_{\mathcal D}>0$, $T>0$, and $0<\varepsilon<1/2$, we use the sampled dilation of \cref{eq:sampling,eq:dilated} with independent jump indices at successive timesteps. A sufficient timestep choice is
\begin{equation}
\begin{gathered}
N_{\rm step}=\max\left\{1,\left\lceil
\frac{cT^2\Lambda^2}{\varepsilon}\right\rceil\right\},
\\
\Delta t=\frac{T}{N_{\rm step}},\\
N_{\rm step}=O\left(1+\frac{T^2\Lambda^2}{\varepsilon}\right),
\end{gathered}
\label{eq:timesteps}
\end{equation}
where a sufficiently large constant $c$ allocates a fixed fraction of $\varepsilon$ to sampling and dilation. This choice also ensures $\Delta t\Lambda\le1$.

Each step simulates $e^{-i\sqrt{\Delta t}\widetilde H_j}$ to initialized-state accuracy $\varepsilon_{\rm step}=\Theta(\varepsilon/N_{\rm step})$, including block-encoding, signal-processing, and gate-synthesis errors. The normalized evolution scale per step is
\begin{equation}
x_{\rm step}=\alpha_{\rm dil}\sqrt{\Delta t}
=\alpha_\TCL\Delta t+\sqrt{\lambda_{\mathcal D}\Delta t}.
\label{eq:x-step}
\end{equation}
The corresponding GQSP degree is
\begin{equation}
\begin{aligned}
d_{\rm step}
&=O\left[d_0\left(x_{\rm step},\varepsilon_{\rm step}^2\right)\right]\\
&=O\left[d_0\left(x_{\rm step},\frac{\varepsilon}{N_{\rm step}}\right)\right].
\label{eq:d-step}
\end{aligned}
\end{equation}
Here $d_0$ is defined in \cref{eq:degree-function}. With these accuracy allocations, the averaged implemented channel approximates $e^{T\mathcal L}$ within diamond-norm error $\varepsilon$.

The averaged-channel timestep scaling follows the general randomized Lindblad analysis of Ref.~\cite{Chen2025Randomized}; the theorem below includes the block-encoding and controlled-query costs of the present model.

Let $C_{\rm qry}^{\rm samp}$ denote the arithmetic cost of one fully coherent controlled signal query built from the sampled dilated block encoding. The resulting total cost is given by the following theorem.

\begin{theorem}[Sampled open-system simulation cost]
\label{thm:open}
With the timestep and accuracy choices in \cref{eq:timesteps,eq:d-step}, the total arithmetic gate complexity of the sampled first-order simulation is
\begin{equation}
G_{\rm open}=O\left[
N_{\rm step}\,
d_0\!\left(x_{\rm step},\frac{\varepsilon}{N_{\rm step}}\right)
C_{\rm qry}^{\rm samp}
\right].
\label{eq:G-open}
\end{equation}
\end{theorem}

The sampling, dilation, and implementation-error bounds, together with the degree and cost derivation, are proved in Appendix~\ref{app:open-cost}. At $T=0$, no evolution step is required.

\paragraph{Sampling step count.}
The sampling error bound contains the term $8T\lambda_{\mathcal D}^2\Delta t$. Allocating $\varepsilon/4$ to this contribution alone gives
\begin{equation}
N_{\rm sampling}
=\left\lceil\frac{32T^2\lambda_{\mathcal D}^2}{\varepsilon}\right\rceil .
\label{eq:explicit-sampling-count}
\end{equation}
For $g=1$, $T=10$, $\varepsilon=0.01$, and uniform $\kappa=\gamma_1=0.1$, $\gamma_\phi=0$, this prescription gives
\begin{center}
\begin{tabular}{c r r}
$(L,N,\nmax)$ & $\lambda_{\mathcal D}$ & $N_{\rm sampling}$\\
$(16,1,3)$ & $6.4$ & $13{,}107{,}200$\\
$(64,2,7)$ & $57.6$ & $1{,}061{,}683{,}200$\\
$(256,4,15)$ & $486.4$ & $75{,}707{,}187{,}200$
\end{tabular}
\end{center}
These step counts ensure that the sampling term fits its error allowance. The dilation remainder, timestep restrictions, and implementation errors may require more steps. The bound gives a conservative sufficient count for this algorithm; it does not set a minimum cost for simulating the model.

The normalized evolution contribution over all timesteps is
\begin{equation*}
\begin{aligned}
N_{\rm step}x_{\rm step}
&=\alpha_\TCL T\\
&\quad+\sqrt{\lambda_{\mathcal D}T N_{\rm step}}.
\end{aligned}
\end{equation*}
The remaining degree contribution accounts for the precision required at each timestep. The sampled jump operator has a local block encoding constructed from ladder or Pauli operators. If the additional controls preserve the lookup scaling, then
\begin{equation*}
\begin{aligned}
C_{\rm qry}^{\rm samp}
&=O\big[\mathcal C_{\rm Tof}(cW)\\
&\qquad+\nmax+b+a_{\rm dil}\big],
\end{aligned}
\end{equation*}
where $a_{\rm dil}$ is the zero-projected width of the dilated encoding. A numerical query cost requires one coherent control convention for both the closed and sampled circuits. Synthesis of the arbitrary signal-processing rotations adds a separate gate cost.

The scheme is first order in $\Delta t$. Replacing the single-jump dilation by a higher-order approximation to $e^{\Delta t\mathcal L^{\rm samp}_j}$ does not remove the sampling error in \cref{eq:samp-error}; improving the $\varepsilon^{-1}$ dependence requires a different sampling or composition scheme. The bound concerns the state averaged over independent classical choices. Let $\Phi_j$ be the reduced timestep channel with an exact phase-gradient catalyst and the timestep ancillas reset. Write
\begin{equation*}
\overline\Phi=\sum_jp_j\Phi_j.
\end{equation*}
Independence of the successive choices gives
\begin{equation*}
\begin{aligned}
\mathbb E_{\mathbf j}\big[
\Phi_{j_{N_{\rm step}}}\circ\cdots\circ\Phi_{j_1}\big]
&=\overline\Phi^{\,N_{\rm step}}.
\end{aligned}
\end{equation*}
A particular sequence $\mathbf j$ generally produces a different conditional state $\rho_{\mathbf j}$. If a measurement outcome $X$ has conditional mean $\mathbb E[X\mid\mathbf j] =\operatorname{Tr}(O\rho_{\mathbf j})$, then
\begin{equation*}
\mathbb E[X]=\operatorname{Tr}\!\left[
O\overline\Phi^{\,N_{\rm step}}(\rho(0))\right].
\end{equation*}
Each independent run samples both the index sequence and the measurement outcome. The same repetitions therefore estimate the observable of the averaged state; a separate multiplicative trajectory count is unnecessary.

Higher-order coherent dilations include products of several jumps and can contain up to $(J+1)^{k+1}$ blocks. Randomized-dissipation methods estimate observables with few ancillas and polylogarithmic dependence on per-circuit precision under a Pauli-decomposition input model ~\cite{kato2026randomized}. To estimate their TCL gate costs, one must also calculate the coefficient one-norms and sampling overhead for the binary photon operators. Trajectory methods give additive jump-query bounds when the total jump rate is scalar, $\sum_jV_j^\dagger V_j=\Gamma I$, and exact block encodings and their inverses are available~\cite{borras2026trajectories}. TCL photon loss and emitter relaxation generally do not satisfy that scalar-rate condition.

Recent query-optimal Lindblad algorithms have additive dependence on normalized time and precision without requiring a scalar jump rate ~\cite{chen2026queryoptimal,wangye2026queryoptimal}. A gate-level TCL comparison must also count the routing and internal operations of their coherent stacked-jump or selected-jump encodings. Geometrically local circuit bounds~\cite{shen2026local} assume fixed local dimension, bounded strength, and rational matrix entries. The square-root ladder coefficients must therefore be rounded, with an error allowance for $T\|\mathcal L_{\rm rounded}-\mathcal L\|_\diamond$. These methods offer alternatives to the first-order construction analyzed here. We leave their compiled TCL costs for a separate comparison.

For the present model, a direct deterministic comparison is also available. Let $\mathcal L_H=-i[H_\TCL,\cdot]$. The commuting dissipators of \cref{sec:dissipative} give the exact local-channel product
\begin{equation}
e^{\Delta t\mathcal D}=\prod_j e^{\Delta t\mathcal D_j}.
\label{eq:exact-dissipative-layer}
\end{equation}
A symmetric step is
\begin{equation}
\mathcal S_{\Delta t}
=e^{\Delta t\mathcal L_H/2}
e^{\Delta t\mathcal D}
e^{\Delta t\mathcal L_H/2}.
\label{eq:symmetric-open-step}
\end{equation}

At fixed photon cutoff, both generators are bounded. Symmetric splitting has $O(\Delta t^3)$ error per step and $O(T\Delta t^2)$ accumulated error, with norm-based constants in the general bound ~\cite{childsli2017lindblad}. An explicit constant follows from contractivity. Each factor of $\mathcal S_{\Delta t}$ and the exact map $e^{\Delta t\mathcal L}$ is completely positive and trace preserving, so each has diamond norm one, and the two maps agree through second order in $\Delta t$. Taylor's theorem with integral remainder bounds each third-order remainder by $(\Delta t\Lambda_2)^3/6$, where
\begin{equation}
\Lambda_2=\norm{\mathcal L_H}_\diamond+\norm{\mathcal D}_\diamond
\le2\bigl(\norm{H_\TCL}+\lambda_{\mathcal D}\bigr)
\label{eq:symmetric-norm}
\end{equation}
because $\norm{\mathcal D[V]}_\diamond\le2\norm{V}^2$ (Appendix~\ref{app:open-cost}). Telescoping over contractive steps bounds the accumulated error by $T\Lambda_2^3\Delta t^2/3$. Hence
\begin{equation}
N_{\rm step}\ge\frac{(T\Lambda_2)^{3/2}}{(3\varepsilon_{\rm split})^{1/2}}
\label{eq:symmetric-steps}
\end{equation}
symmetric steps keep the splitting error below $\varepsilon_{\rm split}$. Expanding the nested commutators in local superoperators could give tighter bounds by retaining only overlapping terms, as in the Hamiltonian analysis above. The dissipative layer itself is exact and contributes no splitting error.

Emitter relaxation and dephasing have elementary exact circuits. Relaxation uses one fresh ancilla, an emitter-controlled rotation of the ancilla by a fixed angle, and a CNOT from the ancilla to the emitter. Dephasing is a mixture of the identity and $Z$, so a classically sampled $Z$ implements it without an ancilla. Exact photon attenuation at finite cutoff has $n_{\max}+1$ Kraus operators for nontrivial attenuation, including amplitudes for losing several photons. A one-qubit exchange dilation approximates this channel only to first order. Implementing the exact local loss channel therefore requires a larger environment or a sequence of operations with corresponding gate counts. One exact construction prepares a $\lceil\log_2(\nmax+1)\rceil$-qubit environment, controlled on the occupation $n$, in the binomial amplitudes for losing $k\le n$ photons. It then subtracts $k$ from the photon word and discards the environment. The occupation-dependent angles can use the lookup circuit of Appendix~\ref{app:angle-explicit}, and the Hamiltonian half steps can use the closed-system walk of Appendix~\ref{app:matched-query} without a dilation qubit. In the two-cavity test of \cref{sec:open-validation}, symmetric splitting gives smaller timestep error at the same number of steps. Comparing total gate costs at a common channel accuracy also requires the local channel circuits and their implementation errors.

\section{Simulating the target observables}
\label{sec:endtoend}

Having established the simulation framework, we now describe how it can be used to evaluate the observables introduced in \cref{sec:model}, namely the retarded cavity spectral function and transient photon statistics. We begin by specifying the initial states required for these calculations.

\subsection{Initial state}
The block encoding does not prepare an input state. For the spectral function we take $\ket{\psi_0}$ to be an eigenstate in a fixed polariton sector. A product of local cavity--emitter eigenstates with definite local polariton numbers summing to $Q$ provides a trial state in the desired sector. It need not be a lattice eigenstate when hopping is present. Its overlap with the target is
\begin{equation}
\chi=\abs{\langle\psi_0|\psi_{\rm trial}\rangle}
\label{eq:overlap}
\end{equation}
where $\chi$ is an amplitude. Eigenstate filtering requires a polynomial whose degree depends on the separation of the target window from unwanted eigenvalues, followed by $O(\chi^{-1})$ amplitude-amplification rounds \cite{brassard2002amplitude,lin2020ground,ge2019faster}. Its cost is therefore not determined by $\chi$ alone. We denote the Toffoli-family arithmetic cost of state preparation by $C_{\rm prep}$ and report it separately from evolution. It depends on the overlap, spectral separation, and required preparation accuracy. Arbitrary-rotation synthesis contributes a separate cost when converting this ledger to a complete logical gate count.

For $L=16$, $N=1$, $t/g=0.05$, $g=\nu=1$, and $\Delta=0$, consider a trial state with one lower polariton localized at a single cavity. Its overlap amplitude with the ground state in the $Q=1$ sector is $\chi=0.2496896$, and the gap to the next distinct energy in this sector is $0.0537383g$. For uniform periodic square lattices at fixed model parameters, the ground state is delocalized and the overlap scales as $L^{-1/2}$. The overlap alone therefore introduces no exponential state-preparation overhead. The cost of eigenstate filtering also depends on the spectral gap, which scales as $L^{-1}$ for large square lattices. A complete resource estimate requires an explicit state-preparation circuit and its precision analysis. Reference states with an extensive number of excitations require a separate analysis of their overlap, gap, and preparation.

To study transient open-system dynamics, $\rho(0)$ must contain an excitation: the vacuum state is unchanged under \cref{eq:lindblad}. A computational-basis photon state, a local polariton product state, or the prepared state used for the spectrum can be chosen according to the relaxation experiment. Its preparation cost is again an explicit input to the end-to-end estimate. For initial emitter states in the maximal-spin sector, global rotations and one-axis twisting also provide a variational preparation route for symmetric spin states~\cite{bond2025global}; this does not by itself prepare a joint cavity--emitter lattice eigenstate.

When a run uses a phase-gradient catalyst, prepare the state $\ket G$ of \cref{eq:gradient-state-explicit}. Let $C_G$ denote its preparation cost under the same counting convention as $C_{\rm prep}$, and let
\begin{equation}
\varepsilon_G=\norm{\rho_G-\proj G}_1
\label{eq:catalyst-input-error}
\end{equation}
for the prepared catalyst $\rho_G$. The complete run, including resets and measurements, is a quantum channel. Contractivity of trace norm therefore bounds the final-state error caused by this initial discrepancy by $\varepsilon_G$, independently of the number of queries. A bounded observable $O$ has an additional bias of at most $\norm O\varepsilon_G$. We prepare independent initial states for different measurement runs and reuse the catalyst within each run. Its preparation cost is paid once per run, and its preparation error is not multiplied by the query count.

The catalyst factors into single-qubit states,
\begin{equation}
\ket{G_{b+1}}=\bigotimes_{k=0}^{b}
\frac{\ket0+e^{2\pi i/2^{b+1-k}}\ket1}{\sqrt2},
\label{eq:gradient-product-preparation}
\end{equation}
where $k=0$ is the least significant bit. Apply a Hadamard to each qubit, followed by the corresponding phase gate. The two largest phase rotations are Clifford gates, so the construction uses $b-1$ arbitrary phase rotations and no Toffoli gates. For $b>1$, let $\delta_G$ bound the operator-norm error of each of the $b-1$ synthesized phase gates. Denote the exact and implemented unitary preparations by $U_G$ and $\widehat U_G$. Telescoping these gates gives
\begin{equation*}
\begin{aligned}
\|\widehat U_G-U_G\|&\le(b-1)\delta_G,\\
\|\rho_G-\proj G\|_1&\le2(b-1)\delta_G.
\end{aligned}
\end{equation*}
Thus $\delta_G\le\varepsilon_G/[2(b-1)]$ is sufficient for the trace-norm budget. At $b=1$, the preparation is entirely Clifford.

Single-qubit Clifford+$T$ synthesis can be used to implement these preparation rotations \cite{ross2016optimal}.

\subsection{Measuring the retarded correlator}

The annihilation operator $a_i$ removes one photon from mode $i$ and multiplies a number state by $\sqrt{n_i}$. The single-register version of \cref{lem:shift} separates these two actions as
\begin{equation}
a=S^{-}\operatorname{diag}\left(\sqrt{n}\right),
\label{eq:a-shift-diag}
\end{equation}
where $S^-$ decrements the occupation modulo $R$ and the diagonal factor returns zero at invalid labels. Thus $a_i$ and $a_i^\dagger$ have block encodings with the tight normalization $\sqrt{\nmax}$ and cost $O(\qph+R+b)$. Let $U_H(\tau)$ denote the full GQSP unitary, acting on the signal, walk, and system registers with the signal and walk ancillas initialized as above. Its projected block approximates $e^{-iH_\TCL\tau}$. The block-encoded products required for the two correlators in \cref{eq:correlators} are
\begin{align}
V^+_{ij}(\tau)&=
U_H^\dagger(\tau)U_{a_i}U_H(\tau)U_{a_j^\dagger},
\nonumber\\
V^-_{ij}(\tau)&=
U_{a_j^\dagger}U_H^\dagger(\tau)U_{a_i}U_H(\tau).
\label{eq:hadamard-target}
\end{align}
Each ladder block encoding uses its own signal ancilla. This is essential: independently projecting the two ancilla spaces reproduces the product of the encoded operators. Reusing one signal ancilla without an intermediate projection would also retain terms from the orthogonal signal subspace and would not isolate the desired product.

With one shared signal, let $P=\proj0\otimes I$. Two block-encoding unitaries then satisfy
\begin{equation*}
\begin{aligned}
P U_AU_B P
&=P U_A P U_B P\\
&\quad+P U_A(I-P)U_B P.
\end{aligned}
\end{equation*}
The second term leaves the zero-signal subspace in the first operation and returns in the second. It need not vanish. Using separate signals prevents this return path from contributing to the joint zero-signal block: if $A$ and $B$ are the normalized operators encoded by $U_A$ and $U_B$, then
\begin{equation*}
\bra{0_a0_b}U_A^{(a)}U_B^{(b)}\ket{0_a0_b}=AB.
\end{equation*}
The bras and kets contract only the signal registers. This identity uses a coherent block projection; it does not require measuring either signal between the ladder insertions.

For an exact eigenstate with known energy $E_0$, one of the two Hamiltonian evolutions in each line of \cref{eq:hadamard-target} can be replaced by a known phase. Stationarity gives~\cite{wecker2015solving}
\begin{align}
C^+_{ij}(\tau)&=e^{iE_0\tau}
\bra{\psi_0}a_i e^{-iH_\TCL\tau}a_j^\dagger\ket{\psi_0},\nonumber\\
C^-_{ij}(\tau)&=e^{-iE_0\tau}
\bra{\psi_0}a_j^\dagger e^{iH_\TCL\tau}a_i\ket{\psi_0}.
\label{eq:stationary-correlators}
\end{align}
The reduced block-encoded products are
\begin{equation*}
\begin{gathered}
U_{a_i}U_H(\tau)U_{a_j^\dagger},\\
U_{a_j^\dagger}U_H^\dagger(\tau)U_{a_i},
\end{gathered}
\end{equation*}
and each contains one evolution. In the ideal circuit, multiplying their measured complex amplitudes by $\nmax$ and the indicated phase gives the correlators in \cref{eq:stationary-correlators}. When the circuit simulates $\overline H_\TCL=H_\TCL-C_{\rm id}I$, these phases use the shifted eigenvalue $E_0-C_{\rm id}$. The two-evolution products in \cref{eq:hadamard-target} remain available without a known $E_0$. If an estimate $\widetilde E_0$ is used, its additional correlator bias satisfies
\begin{equation*}
\text{additional bias}
\le\nmax T_{\max}\abs{\widetilde E_0-E_0}.
\end{equation*}
Any cost of obtaining or certifying that estimate must be included in the workload.

A Hadamard test of $V^\pm_{ij}$ returns the real part of $\bra{\mathbf0,\psi_0}V^\pm_{ij}\ket{\mathbf0,\psi_0}$~\cite{somma2002simulating}; an $S^\dagger$ on the control before measurement returns the imaginary part. No postselection is needed. For exact evolution and ladder blocks, the two ladder normalizations $\sqrt{\nmax}$ give
\begin{equation}
\bra{\mathbf0,\psi_0}V^\pm_{ij}(\tau)
\ket{\mathbf0,\psi_0}=\frac{C^\pm_{ij}(\tau)}{\nmax}.
\label{eq:hadamard-value}
\end{equation}

\paragraph{A common-evolution interferometer.}
Both interferometer branches can share one time evolution, as in Ref.~\cite[Sec. III C 3 and Fig. 11]{somma2002simulating}. Write $\mathbf0$ for the initialized ancillas, including two separate ladder signal qubits. The Hermitianizing qubit is initialized in $\ket+$, as above. The addition and removal circuits produce the branch states {
\begin{equation}
\begin{aligned}
\ket{\phi_0^+}&=U_H\ket{\mathbf0,\psi_0},\\
\ket{\phi_1^+}&=U_{a_i}U_HU_{a_j^\dagger}\ket{\mathbf0,\psi_0},\\
\ket{\phi_0^-}&=U_HU_{a_j}\ket{\mathbf0,\psi_0},\\
\ket{\phi_1^-}&=U_{a_i}U_H\ket{\mathbf0,\psi_0}.
\end{aligned}
\label{eq:common-evolution-branches}
\end{equation}
} The ideal overlaps satisfy $\langle\phi_0^\pm|\phi_1^\pm\rangle=C_{ij}^{\pm}/\nmax$. Apply the first ladder insertion on the appropriate control branch, then apply $U_H$ to both branches and the second ladder insertion on its branch. Measurements of $X$ and $Y$ on the Hadamard control give the real and imaginary parts without postselection. Using separate ladder signals prevents the unwanted return path described above. The circuit works for general input states and requires no eigenenergy estimate. Only the ladder insertions need the additional interferometer control. The evolution retains its internal GQSP signal controls.

Let $\varepsilon_C>0$ be the target total additive error of either complex correlator. A statistical error of at most $\varepsilon_C/2$ is achieved with
\begin{equation}
N_{\rm shot}=O\left(\frac{\nmax^{2}}{\varepsilon_C^{2}}\right)
\label{eq:shots}
\end{equation}
independent shots per quadrature and time point at fixed constant failure probability. The two quadratures and the two correlators require four Hadamard-test circuits at each time. The common-evolution circuit uses two controlled ladder blocks and one evolution shared by the two branches. \Cref{eq:shots} gives a sufficient shot count when the correlators themselves are requested. For the spectrum, we can reduce this count by bounding the error of their weighted sum directly.

Every valid full-space ladder normalization obeys $\alpha_a\ge\norm{a_i}=\sqrt{\nmax}$. Thus $\nmax$ is the smallest product normalization for separately encoded ladder insertions. \cref{eq:shots} gives a worst-case sufficient shot count for these independent Hadamard tests.

The remaining error budget controls state preparation and circuit implementation. Let $\widetilde\rho_0$ be the prepared state and define its trace-norm error by
\[
\varepsilon_{\rm prep}=\norm{\widetilde\rho_0-\proj{\psi_0}}_1.
\]
Let $\varepsilon_{\rm sim}$ bound the norm error of each implemented evolution on inputs with its signal and walk ancillas initialized as above, uniformly over the sampled times, and let $\varepsilon_a$ bound the operator-norm error of each normalized ladder block relative to $a_i/\sqrt{\nmax}$ or $a_j^\dagger/\sqrt{\nmax}$. All normalized blocks are contractions. Replacing the state and the four operator factors one at a time therefore bounds the systematic error of either correlator by
\begin{equation}
\varepsilon_{C,{\rm sys}}
\le\nmax\left(\varepsilon_{\rm prep}+\varepsilon_G+2\varepsilon_{\rm sim}+2\varepsilon_a\right).
\label{eq:correlator-sys-error}
\end{equation}
Choosing the right-hand side to be at most $\varepsilon_C/2$ and assigning at most $\varepsilon_C/2$ to statistical error gives total correlator error at most $\varepsilon_C$. A sufficient choice is
\begin{equation*}
\begin{aligned}
\varepsilon_{\rm sim}
&=\min\left\{\frac14,\frac{\varepsilon_C}{12\nmax}\right\},\\
\varepsilon_{\rm prep},\,\varepsilon_G,\,\varepsilon_a
&\le\varepsilon_{\rm sim}.
\end{aligned}
\end{equation*} The evolution-error bound includes errors from finite-precision block encoding, polynomial approximation, and gate synthesis, assuming an exact catalyst. The catalyst's initial preparation error is included once through $\varepsilon_G$. The costs $C_{\rm prep}$ and $C_G$ and the circuit precisions must satisfy this allocation.

For the stationary-state circuit with known energy, the evolution error enters once, reducing $2\varepsilon_{\rm sim}$ to $\varepsilon_{\rm sim}$. In the common-evolution interferometer, both branches depend on the evolution, so we retain the conservative $2\varepsilon_{\rm sim}$ bound. Estimating the eigenenergy contributes $\nmax T_{\max}\abs{\widetilde E_0-E_0}$ only in the stationary-state phase reduction. When fixed-angle rotations use a second catalyst, $\varepsilon_G$ includes the preparation errors of both catalysts.

\subsection{Time grid}

The observation window fixes the nominal resolution through \cref{eq:resolution}. Choose $W$ to bound the transition frequencies contributing to the correlators, rather than only the desired plotting range. The Nyquist condition and number of time points are
\begin{equation*}
\begin{aligned}
\delta\tau&<\frac{\pi}{W},\\
N_t&=\left\lceil\frac{T_{\max}}{\delta\tau}\right\rceil+1.
\end{aligned}
\end{equation*}
If transitions outside this range are omitted, their total absolute spectral weight needs a separate bound: an omitted weight $\zeta$ in either correlator changes its value by at most $\zeta$ at every time and can be propagated through \cref{eq:spectral-stat-error}. An unproved bandwidth estimate does not provide an aliasing guarantee.

We estimate the local spectral density $A_{ii}(\omega)$ using the common-evolution interferometer and normalization $\alpha_{\rm dyn}$ for the chosen frame. The walk retains its GQSP signal control, while the evolution needs no additional Hadamard control. A uniform degree bound is
\begin{equation}
d_{\max}=O\!\left[d_0(\alpha_{\rm dyn}T_{\max},
\varepsilon_{\rm sim}^{2})\right]
=O\!\left[d_0(\alpha_{\rm dyn}T_{\max},\varepsilon_{\rm sim})\right].
\label{eq:max-degree}
\end{equation}
Frame phases are restored using \cref{eq:charge-frame-restoration}.

Let $A_h(\omega)$ denote the finite-window quadrature estimate with real weights $w_k$. Arbitrary correlator errors obey the deterministic propagation bound
\begin{equation}
|\delta A_h(\omega)|\leq
\frac{2\varepsilon_C}{\pi}\sum_k|w_k|.
\label{eq:spectral-stat-error}
\end{equation}
This bound includes systematic bias. For statistical error, we can obtain a tighter guarantee by treating the weighted spectrum directly. Write $C_k^\pm=\nmax(s_{k,R}^\pm+i s_{k,I}^\pm)$ and measure each $s$ with $N_{\rm shot}$ independent outcomes in $\{-1,1\}$. Then
\begin{equation}
\begin{aligned}
\widehat A_h(\omega)=\frac{\nmax}{\pi}\sum_k w_k
\bigl[&\cos(\omega\tau_k)(\widehat s_{k,R}^+-\widehat s_{k,R}^-)\\
&-\sin(\omega\tau_k)(\widehat s_{k,I}^+-\widehat s_{k,I}^-)\bigr].
\end{aligned}
\label{eq:weighted-spectrum-estimator}
\end{equation}
The sum of squared individual outcome ranges is $8\nmax^2\sum_k w_k^2/(\pi^2N_{\rm shot})$. Hoeffding's inequality and a union bound over $N_\omega$ specified frequencies give the sufficient count
\begin{equation}
N_{\rm shot}=\left\lceil
\frac{4\nmax^2\sum_k w_k^2}{\pi^2 e_{\rm stat}^2}
\ln\frac{2N_\omega}{\delta}\right\rceil .
\label{eq:weighted-spectral-shots}
\end{equation}
The same measurement samples can be reused at every frequency. The measurement runs must be independent, but the union bound does not require the frequency estimates to be independent.

Allocate $e_{\rm stat}=\varepsilon_A/2$ and the other half to systematic spectral error. For a positive-weight rule with $\sum_kw_k=T_{\max}$, take $\varepsilon_C=\pi\varepsilon_A/(2T_{\max})$ and retain the systematic correlator allowance $\varepsilon_{C,{\rm sys}}\leq \varepsilon_C/2$ from \cref{eq:correlator-sys-error}. For a tapered window, use the modified weights in both bounds. Errors from time quadrature, omitted frequencies, finite-window broadening, and the frequency grid must be allowed for separately. This concentration bound applies to the specified frequency grid. Extending it to a continuous interval requires an interpolation bound.

The arithmetic workload for one ordered local spectrum is
\begin{equation}
\begin{aligned}
\mathcal C_A=O\!\Bigl[4N_{\rm shot}N_t\Bigl(
&C_{\rm prep}+C_G\\
&+d_{\max}\mathcal C_{\rm Tof}(cW)+\qph+R+b\Bigr)\Bigr].
\end{aligned}
\label{eq:cost-spectral}
\end{equation}
\Cref{eq:v3-spectrum-T} gives the corresponding logical-$T$ cost, using the degree and precision chosen at each time and including ladder operations and catalyst preparation. For $N_t=90$, $N_\omega=1000$, and $\delta=0.01$, trapezoidal weights reduce the shot count by about $328$ relative to requiring all pointwise correlator estimates to meet the same allocated tolerance. The deterministic bias bound is unchanged.

At fixed time spacing, $\sum_kw_k^2=O(T_{\max}\delta\tau)$. The total number of runs, $4N_tN_{\rm shot}$, therefore scales approximately as $T_{\max}^2/\varepsilon_A^2$ for fixed absolute accuracy, apart from confidence logarithms. Doubling the observation window requires about four times as many runs. If accuracy is instead fixed as a fraction of the height of an isolated unit-weight pole under a hard window, $T_{\max}/\pi$, the leading run count is constant. The longer evolutions still cost more. Overlapping peaks and peaks with small weight require an accuracy criterion suited to the spectrum of interest.

Amplitude estimation can reduce measurement repetitions at the cost of greater coherent depth~\cite{brassard2002amplitude}. Response sampling using phase estimation~\cite{roggero2019response} avoids the time-grid Fourier estimator, but has different preparation and resolution costs. Determining whether either method is cheaper for TCL spectra requires a separate circuit comparison. For off-diagonal spectra, estimate both ordered pairs as required by the spectral-matrix definition; this changes the circuit count by at most a factor of two.

\subsection{Exact-diagonalization check of the spectral calculation}
\label{sec:validation}

\begin{figure*}
\includegraphics[width=\textwidth]{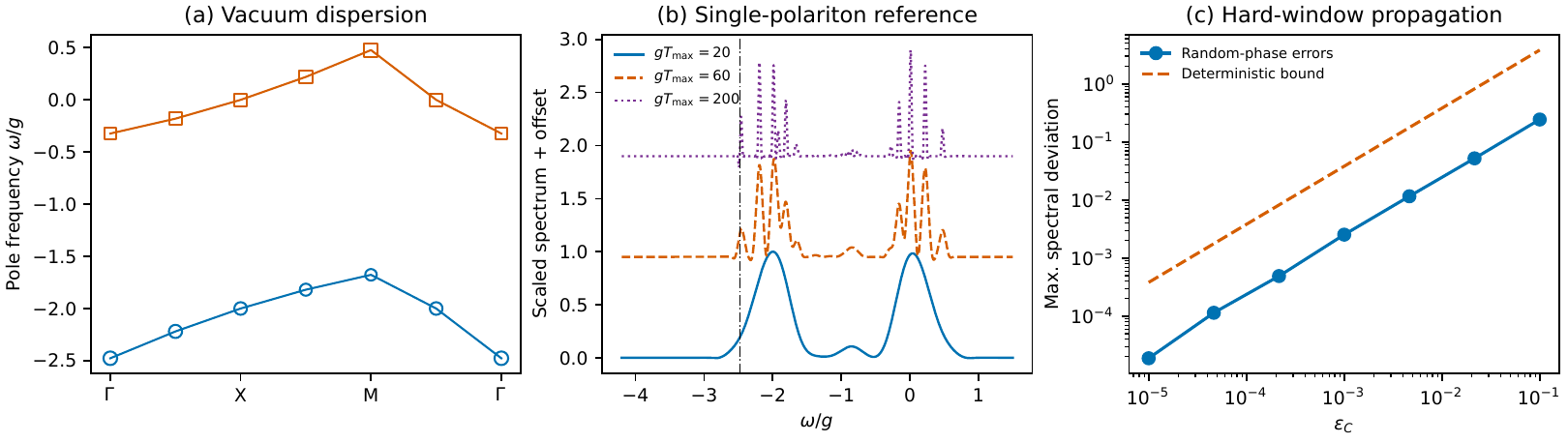}
\caption{Exact diagonalization of the $4\times4$ lattice with $N=1$, $\nmax=2$, $t_{ij}=0.2$, $g_i=1$, $\Delta=0$ and $\nu_i=1$. (a) Poles of the momentum-resolved particle-addition correlator evaluated in the vacuum state, overlaid on the closed-form polariton dispersion of \cref{eq:polariton}; marker size indicates the photon weight $\abs{u_\pm(\mathbf k)}^2$ of the corresponding eigenvector. (b) Spectral density at one site, computed from \cref{eq:retarded,eq:spectral} with the reference state taken as the lowest single-polariton eigenstate, so that both correlators of \cref{eq:correlators} contribute. Traces are offset and scaled to a common height, with the one-sided Hann taper $w_{T_{\max}}(\tau)=[1+\cos(\pi\tau/T_{\max})]/2$ for $0\le\tau\le T_{\max}$ in place of the hard window. Tapering changes sidelobes and linewidths and can move the maxima of overlapping features. It leaves the underlying pole frequencies unchanged; the resolution still scales as $1/T_{\max}$, with a taper-dependent constant \cite{harris1978windows}. The dash-dotted line marks the particle-removal pole. (c) Largest deviation of the quadrature spectrum at $T_{\max}=60$, compared with the bound in \cref{eq:spectral-stat-error}. Both sampled correlators are perturbed by independent complex errors of magnitude $\varepsilon_C$ with uniformly random phases. Panel (c) uses the hard window of \cref{eq:spectral}, for which the quadrature weights sum to $T_{\max}$. }
\label{fig:spectral}
\end{figure*}

We compute the observables defined above by exact diagonalization on a small lattice to check the correlator definitions, the resolution estimate of \cref{eq:resolution}, and the error propagation in \cref{eq:spectral-stat-error}. We take the $4\times4$ lattice with one emitter per cavity and $\nmax=2$. The reference state is the lowest eigenstate of the single-polariton sector, so the particle-addition correlator evolves in the two-polariton sector and the particle-removal correlator in the vacuum sector; since the reference sector has $Q=1$ and $\nmax\ge Q+1$, the truncation is exact for every state either correlator reaches.

Taking the vacuum as the reference state instead makes the addition correlator solvable in closed form, which supplies an independent check on the Hamiltonian and correlator implementation. In the single-excitation sector the lattice decouples by momentum, and at each $\mathbf k$ the photon and bright-emitter mode energies are
\begin{equation*}
\begin{aligned}
\epsilon_{\mathbf k}&=-2t(\cos k_x+\cos k_y)-\nu,\\
\epsilon_a&=-\Delta-\nu.
\end{aligned}
\end{equation*}
The bright mode is the normalized symmetric combination of the $N$ emitter excitations at momentum $\mathbf k$. Its coupling to the photon is $g\sqrt N$, while the $N-1$ orthogonal emitter combinations do not
couple to the photon. The two coupled modes therefore have energies \begin{equation} \omega_\pm(\mathbf k)=\frac{\epsilon_{\mathbf k}+\epsilon_a}{2} \pm\sqrt{\left(\frac{\epsilon_{\mathbf k}-\epsilon_a}{2}\right)^{2}+g^{2}N}. \label{eq:polariton} \end{equation} \Cref{fig:spectral}(a) compares the poles and weights extracted from the correlator with \cref{eq:polariton} along $\Gamma\to X\to M\to\Gamma$. The largest differences between the numerical and analytic momentum-resolved frequencies and weights are $4.45\times10^{-16}$ and $1.67\times10^{-16}$, respectively. Comparing the full $Q=1$ eigenvalue lists gives a maximum difference of $2.67\times10^{-15}$. The addition/removal weight sum rule is satisfied to $1.34\times10^{-15}$. These checks test the Hamiltonian matrix, the number-sector restrictions, and the photon correlators. They do not test the reversible circuit implementation.

Returning to the single-polariton reference state, the addition and removal branches carry spectral weights $1.0429$ and $0.0429$, so their difference is $1$ to ten decimal places, as the bosonic commutator requires below the cutoff. The removal weight is $4.1\%$ of the addition weight. Its pole lies at $\omega=-2.477033$, while the nearest addition pole lies at $\omega=-2.464380$. Their separation, $0.01265$, is smaller than the finest nominal resolution $2\pi/200=0.03142$ used in \cref{fig:spectral}(b). The removal term therefore changes the nearby line shape; the plotted windows do not resolve it as an isolated feature. Longer windows reveal additional structure around the two polariton branches, with nominal resolutions $0.31$, $0.10$, and $0.03$ for the three windows shown and a taper-dependent linewidth constant.

\Cref{fig:spectral}(c) illustrates the deterministic bound in \cref{eq:spectral-stat-error} using the hard window, for which $\sum_k\abs{w_k}=T_{\max}$. Both sampled correlators are perturbed by independent complex errors of magnitude $\varepsilon_C$ and uniformly random phase, and the largest resulting deviation of the quadrature spectrum is recorded. For the random realization shown (seed 917), the observed deviations are approximately $15$--$20$ times smaller than the deterministic bound. For the independent, zero-mean perturbations used here, the root-mean-square error at a fixed frequency scales with $(\sum_k\abs{w_k}^2)^{1/2}$, whereas the deterministic bound scales with $\sum_k\abs{w_k}$. This explains cancellation in the random-phase experiment. The statistical guarantee for measurement samples follows from the Bernoulli concentration bound in \cref{eq:weighted-spectral-shots}. The random-phase test only illustrates cancellation of independent errors. \Cref{eq:spectral-stat-error} bounds arbitrary errors, including systematic bias.

\subsection{Accuracy and cost of transient photon statistics}
\label{sec:g2-resources}
We now turn to the transient photon statistics introduced in \cref{sec:model}, which characterize the dissipative evolution of the TCL system. Unlike the spectral function, these are equal-time observables that can be estimated directly from computational-basis measurements after the sampled open-system evolution. In particular, the second-order correlation $g_i^{(2)}(0;T)$ is a ratio of occupation moments, so its accuracy depends on both the errors in these moments and the mean photon occupation.

For each reported site, define
\begin{equation*}
\mu=\bar n_i(T),\qquad
f=\langle n_i(n_i-1)\rangle.
\end{equation*}

Assume a stated bound $\mu\ge\mu_{\min}>0$. The estimates $\widehat\mu$ and $\widehat f$ come from the same samples, so their errors are correlated. Write $\widehat\mu=\mu+\Delta_\mu$, $\widehat f=f+\Delta_f$, and $g=f/\mu^2$. Then, exactly,
\begin{equation}
\begin{aligned}
\frac{\widehat f}{\widehat\mu^2}-\frac{f}{\mu^2}
&=\frac{\Delta_\varphi-g\Delta_\mu^2}{\widehat\mu^{2}}, \\
\Delta_\varphi&=\Delta_f-2g\mu\,\Delta_\mu.
\end{aligned}
\label{eq:g2-ratio-identity}
\end{equation}
The leading error is therefore the error in the mean of the single-shot variable
\begin{equation*}
\varphi(n)=n(n-1)-c\,n,\qquad c=2g\mu=\frac{2f}{\mu}.
\end{equation*}
The coefficient $c$ is unknown but fixed, and $f\le(\nmax-1)\mu$ gives $0\le c\le2(\nmax-1)$. For integers $0\le j<k\le\nmax$ with $d=k-j$,
\begin{align*}
|\varphi(k)-\varphi(j)| &= d\,|k+j-1-c| \\
&\le d(2\nmax-1-d) \\
&\le \nmax(\nmax-1).
\end{align*}
{Thus $\varphi$ has a range width no larger than that of the factorial moment, and} $|\varphi(n)|\le\nmax(\nmax-1)$ because $\varphi(0)=0$. The estimator $\widehat f/\widehat\mu^2$ is unchanged; $\varphi$ enters only its error analysis.

Suppose $|\Delta_\mu|\le\varepsilon_n<\mu_{\min}$ and $|\Delta_\varphi|\le\varepsilon_\varphi$. Since $g\le(\nmax-1)/\mu$, \cref{eq:g2-ratio-identity} gives
\begin{equation}
\left|\frac{\widehat f}{\widehat\mu^2}-\frac{f}{\mu^2}\right|
\le\frac{\varepsilon_\varphi+(\nmax-1)\varepsilon_n^2/\mu_{\min}}
{(\mu_{\min}-\varepsilon_n)^2}.
\label{eq:g2-ratio-error}
\end{equation}
{The linear contribution of the occupation error is included in $\Delta_\varphi$. Once this combination is controlled, the remaining term in the numerator is quadratic in $\Delta_\mu$, so the occupation tolerance can be much larger than the tolerance for $\Delta_\varphi$.} For $\nmax\ge2$ and a target ratio error $\varepsilon_g$ with $\varepsilon_g\mu_{\min}\le(\nmax-1)/8$, one sufficient allocation is
\begin{equation}
\begin{aligned}
\varepsilon_n&=\mu_{\min}\left[\frac{\varepsilon_g\mu_{\min}}{16(\nmax-1)}\right]^{1/2},\\
\varepsilon_\varphi&=\frac34\,\varepsilon_g\mu_{\min}^2.
\end{aligned}
\label{eq:g2-moment-allocation}
\end{equation}
The condition gives $\varepsilon_n\le\mu_{\min}/11$, so the right side of \cref{eq:g2-ratio-error} is at most $(13/16)(11/10)^2\varepsilon_g<\varepsilon_g$. Bounding $\widehat f$ and $\widehat\mu$ separately would instead amplify the occupation error linearly through the denominator and require $\varepsilon_n=O(\varepsilon_g\mu_{\min}^2/\nmax)$. At $\nmax=1$, the factorial moment vanishes exactly, so $g^{(2)}=0$ wherever $\mu>0$.

For $\nmax\ge2$, use half of each tolerance in \cref{eq:g2-moment-allocation} for statistics and half for preparation and evolution. To keep the measured quantities bounded even if an imperfect circuit produces an unused binary label, cap the recorded occupation at $\nmax$ before computing its factorial moment. This leaves the ideal truncated observables unchanged. If $e$ bounds the final-state trace-norm error, the systematic errors in $\mu$ and in the mean of $\varphi$ are at most $\nmax e$ and $\nmax(\nmax-1)e$. It is therefore sufficient to require
\begin{equation}
e\le\min\left\{\frac{\varepsilon_n}{2\nmax},
\frac{{\varepsilon_\varphi}}{2\nmax(\nmax-1)}\right\}.
\label{eq:g2-state-allocation}
\end{equation}
For example, assign at most $e/3$ each to system-state preparation, catalyst preparation, and the averaged-channel error of \cref{sec:open}. The last allocation includes sampling, dilation, and circuit implementation.

Both moments at every site are estimated from the same computational-basis samples. Hoeffding bounds, applied to $n$ and to $\varphi(n)$, and a union bound give a sufficient number of samples for all these estimates:
\begin{equation}
\begin{aligned}
N_{\rm shot}=O\Bigg[\max\Bigg\{&\frac{\nmax^2}{\varepsilon_n^2},\\
&\frac{\nmax^2(\nmax-1)^2}{\varepsilon_\varphi^2}\Bigg\}
\log\frac{2L}{\delta}\Bigg]
\end{aligned}
\label{eq:g2-shared-shots}
\end{equation}
The simultaneous estimates have failure probability at most $\delta$, where $0<\delta<1$. The arithmetic workload is $N_{\rm shot}(C_{\rm prep}+C_G+G_{\rm open})$, with $G_{\rm open}$ evaluated at the allocated channel accuracy. Rotation synthesis is added in the same way as for the spectral workload. This bound combines the number of measurement repetitions with the cost of one open-system simulation. It depends on the specified initial-state preparation cost and the lower bound $\mu\ge\mu_{\min}>0$.

For measurements reported at $M$ times, a sufficient shot count at each time is
\begin{equation}
\begin{aligned}
N_{\rm shot}\geq 2\max\Bigg\{&\frac{\nmax^2}{\varepsilon_n^2},\\
&\frac{\nmax^2(\nmax-1)^2}{\varepsilon_\varphi^2}\Bigg\}
\ln\frac{4LM}{\delta}.
\end{aligned}
\label{eq:g2-explicit-shots}
\end{equation}
We allocate statistical tolerances $\varepsilon_n/2,$  $\varepsilon_\varphi/2$. The same occupation measurements give both moments at every site. For $L=16,\nmax=3,M=1,\mu_{\min}=0.5,\varepsilon_g=0.1,\delta=0.01$, \cref{eq:g2-moment-allocation,eq:g2-state-allocation} give $\varepsilon_n=1.97642\times10^{-2}$, $\varepsilon_\varphi=1.875\times10^{-2}$, and final-state tolerance $e=1.5625\times10^{-3}$. The explicit count is $1{,}794{,}879$ independent runs. Bounding the two moments separately would instead give $\varepsilon_n=1.04167\times10^{-3}$, $e=1.73611\times10^{-4}$, and $145{,}385{,}127$ runs. The correlated bound therefore reduces the runs by a factor of $81$ and relaxes the state tolerance ninefold. If $e/3$ is assigned to the sampled channel, the rates and time in \cref{eq:explicit-sampling-count} require  $251{,}658{,}240$ steps per run to satisfy that sampling allowance alone. Together, these bounds give $4.517\times10^{14}$ sampled steps across all runs. Dilation and implementation errors may require more steps. Appendix~\ref{app:open-circuits} gives the gate cost of each sampled step. Use the uniform $(16,1,3)$ coefficients of \cref{tab:v3-coherent-compare} in the charge-removed frame of \cref{sec:open}. To evaluate the cost at the timestep count above, take exact input-state and catalyst preparation and assign $e/3$ to circuit implementation. Then $\varepsilon_{\rm step}=e/(6N_{\rm step})$ limits the accumulated implementation error to $e/3$. \Cref{eq:sampled-data-precision,eq:sampled-fixed-precision} give {$b=33$, $B=24$, $K=6$, and $m=48$}, and \cref{eq:sampled-step-T} gives about  {$4.2\times10^{5}$} logical $T$ gates per step on average over the sampled channels; Appendix~\ref{app:sampled-gates} lists the per-channel counts. At these precision widths, the evolution circuit has a conservative allocation of $434$ logical qubits (\cref{eq:sampled-qubit-allocation}).  The mean evolution cost over the stated runs and timesteps is therefore about  {$1.9\times10^{20}$} $T$ gates, excluding input-state and catalyst preparation. This tally does not yet bound the full cost at the requested $g^{(2)}$ accuracy: the dilation remainder must also fit the remaining error budget. The remaining cost is set mainly by the first-order timestep count. With $\norm{\widehat H_\TCL}\le\widehat\alpha_\TCL$ and $\varepsilon_{\rm split}=e/3$, \cref{eq:symmetric-norm,eq:symmetric-steps} give a sufficient count of $653{,}956$ symmetric steps per run, $385$ times fewer than the sampling allowance above, although each symmetric step applies a full layer of local channels. The large estimate results from the conservative moment-error and sampling bounds used here; it does not set a minimum cost for estimating $g^{(2)}$.

\subsection{Numerical check of open dynamics}
\label{sec:open-validation}

We test the open-system construction on two coupled cavities with two emitters per cavity. We set $g_i=g=1$, $t_{12}=0.2g$, $\Delta=\nu_i=0$, and $\kappa_i=0.1g$. The two relaxation rates in each cavity are $\gamma_{1,i,1}=0.02g$ and $\gamma_{1,i,2}=0.20g$, with no dephasing. The numerical benchmark in \cref{sec:open-validation} tests photon loss and emitter relaxation, but not dephasing. The dephasing jump circuit is specified separately in Appendix~\ref{app:sampled-jumps}. Initially the photons are in vacuum, both emitters in the first cavity are excited, and both emitters in the second cavity are in their ground state. This state has $Q=2$, so $\nmax=2$ is exact for its subsequent dynamics and the photon moments. The reachable number sectors have dimensions $1$, $6$, and $17$. This two-cavity graph provides a small check of the channels and observables; it does not use the periodic-lattice routing assumptions of the resource ledger.

\begin{figure*}
\centering
\includegraphics[width=\textwidth]{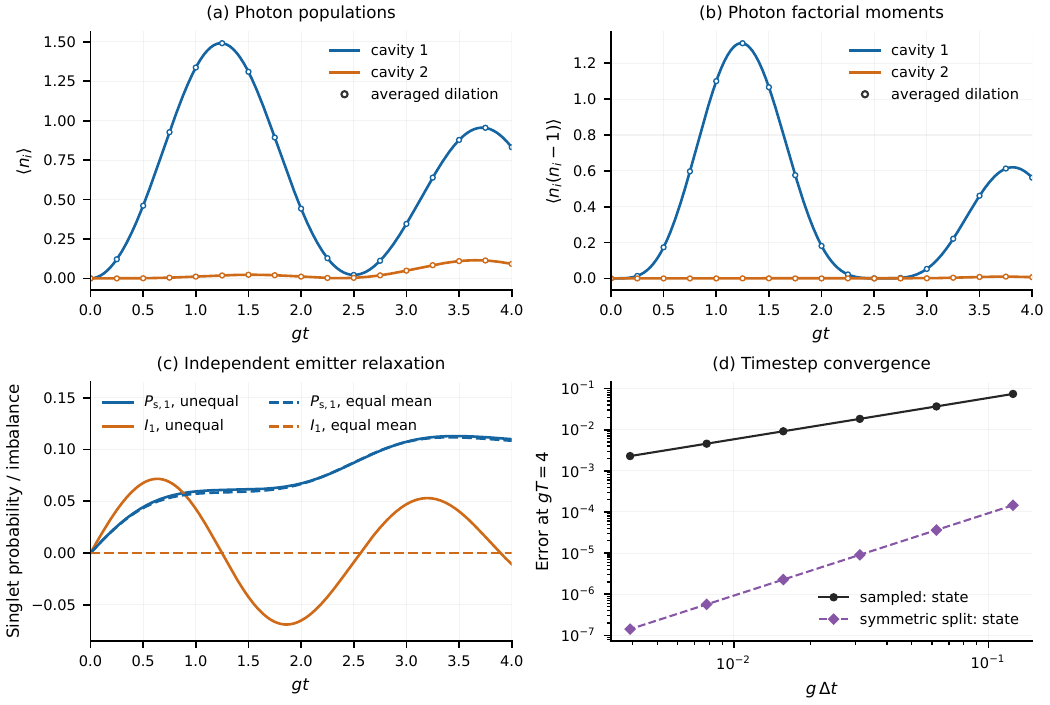}
\caption{Open dynamics of two coupled cavities with two emitters per cavity, for the parameters and initial state in \cref{sec:open-validation}. (a) Photon populations and (b) photon factorial moments. Solid curves show the exact master-equation evolution; circles show the averaged sampled dilation with $g\Delta t=1/256$. (c) The singlet probability $P_{\mathrm{s},1}$ and emitter population imbalance $I_1$ in the first cavity under unequal relaxation rates $(0.02g,0.20g)$ and equal rates $(0.11g,0.11g)$. Both choices populate the singlet sector, whereas only unequal rates generate an imbalance. (d) Trace-norm state error at $gT=4$. Timestep halving gives asymptotic orders $1.001$ for sampled dilation and $2.000$ for the symmetric Hamiltonian/local-channel split, as estimated from the two smallest timesteps. The channels and Hamiltonian evolutions use exact matrix exponentials or exact local Kraus maps, so the comparison isolates timestep error.}
\label{fig:open-validation}
\end{figure*}

For each timestep we compute the six nonzero jump-channel dilations in \cref{eq:dilated} by matrix exponentiation and average their reduced channels with the probabilities in \cref{eq:sampling}. Repeating this averaged map gives the expected evolution under independent channel sampling, without Monte Carlo error. We compare it with the exact master-equation evolution. The photon populations and factorial moments agree as the timestep decreases, as shown in \cref{fig:open-validation}. An independent calculation in the full $144$-dimensional truncated Hilbert space agrees with the number-sector calculation to $1.7\times10^{-15}$ in the Frobenius norm of the final density matrix.

To distinguish the effects of independent relaxation and unequal rates, we also solve the master equation with both emitter rates set to their mean, $0.11g$. Let $P_{\mathrm{s},1}$ be the emitter singlet probability in the first cavity and $I_1$ the emitter population imbalance there, with
\begin{equation*}
\begin{aligned}
\text{emitter singlet}&=\frac{\ket{eg}-\ket{ge}}{\sqrt{2}},\\
I_1&=\langle n^{\at}_{1,1}-n^{\at}_{1,2}\rangle.
\end{aligned}
\end{equation*}
Both rate choices populate the singlet sector, showing that independent local relaxation can drive the state out of the maximal-spin sector even when the rates are equal. Unequal rates also create an emitter imbalance, which remains zero for equal rates. Both rate choices give the same initial singlet slope, while the imbalance slope depends on the rates:
\begin{equation*}
\begin{aligned}
\left.\frac{dP_{\mathrm{s},1}}{dt}\right|_{t=0}&=0.11g,\\
\left.\frac{dI_1}{dt}\right|_{t=0}&=\begin{cases}
0.18g,&\text{unequal rates},\\
0,&\text{equal rates}.
\end{cases}
\end{aligned}
\end{equation*}

At $gT=4$, increasing $N_{\rm step}$ from $512$ to $1024$ reduces the sampled-dilation state error $\norm{\rho_{\rm approx}(T)-\rho_{\rm exact}(T)}_1$ from $4.5684\times10^{-3}$ to $2.2832\times10^{-3}$, consistent with first-order convergence. At $1024$ steps the largest final-time errors over the two cavities are $6.18\times10^{-4}$ for the photon population and $2.55\times10^{-5}$ for the factorial moment. A symmetric product formula using exact Hamiltonian half steps and the exact local dissipative channels gives state errors $5.6811\times10^{-7}$ and $1.4203\times10^{-7}$ at the same step counts, consistent with second-order convergence. This comparison concerns timestep error; it does not compare compiled gate costs. All evolutions preserve trace and positivity within numerical precision. These tests concern ideal finite-dimensional channels for the stated initial state, without gate-synthesis or measurement error; they do not provide a numerical diamond-norm certificate.

\section{Discussion and conclusions}
\label{sec:discussion}
We have specified a logical circuit construction for the atom-resolved Tavis--Cummings lattice and connected it to a finite-window retarded photon spectrum and transient photon statistics under local loss, relaxation, and dephasing. The hopping construction specializes ladder-operator block encoding to a coherently selected bond: it updates source and destination photon occupations in place and uses two occupation tables shared across directed bonds. This avoids the second photon-configuration register of the sparse-access implementation compared here while retaining the same coefficient factorization and Hamiltonian normalization.

With the same shared preparation, precision, and controls, the direct walk uses $4.75$--$8.11\%$ fewer Toffoli gates than the paired two-register circuit for the three examples. It also removes $L\qph$ qubits. Removing the uniform polariton potential reduces the normalization from $69.3,716.3,7165.9$ to $37.3,428.3,4733.9$, respectively. These savings apply to the specified encodings on the full truncated register. Low-density sectors can have much smaller energy ranges, and product formulas can exploit locality without routing a global selection register. Comparing those approaches would require their own circuit counts at the same accuracy. These counts alone do not establish optimality or quantum advantage.

Sharing the evolution between interferometer branches and bounding the weighted spectral error directly reduce the illustrative spectral cost. For the specified construction, the estimated cost of the
evolution and ladder operations is $4.36\times10^{15}$ $T$ gates for the stated $1\%$ accuracy relative to an isolated unit-weight pole, before preparation. The conservative bounds for jump sampling and photon-moment measurements also give large counts. At matched precision, each sampled walk costs $11+C_{J,j}$ Toffolis more than the closed-system walk (\cref{eq:sampled-walk-cost}). At the timestep count determined by sampling alone, the transient-statistics example of \cref{sec:g2-resources} has a mean evolution cost of about $1.9\times10^{20}$ $T$ gates, before preparation and any increase needed to control the dilation error. For a particular physical application, the evolution counts must be combined with input-state and catalyst preparation costs, accounted for separately above. The chosen photon cutoff must also be justified by excitation constraints or a cutoff-convergence study.

For spectral estimation, finer frequency resolution requires a longer observation window and more accurate correlator estimates, increasing both evolution and measurement costs. For transient photon statistics, a small mean occupation amplifies moment-estimation errors in $g_i^{(2)}(0;T)$. Our ratio-error bounds therefore use a stated positive lower bound on the photon population. When such a bound is unavailable, photon populations and factorial moments can still be estimated directly.

The sampled open-system dilation uses one dilation qubit and removes coherent selection among all $J$ jump channels, but its accumulated error is first order in the timestep size. The two-cavity validation in \cref{sec:open-validation} tests the method with unequal emitter rates and checks the predicted timestep convergence against the exact master-equation solution; symmetric splitting converges at second order for the same timestep sequence, without establishing a gate-cost comparison. For the transient-statistics example, the sufficient symmetric-splitting count from \cref{eq:symmetric-steps} is about $385$ times smaller than the count obtained from
the first-order sampling allowance alone.Each symmetric step requires Hamiltonian evolution and a full layer of local dissipative channels, rather than one sampled jump. Determining the net resource saving therefore requires compiled circuits for these channels, particularly exact photon loss, together with matched implementation-error allocations. This offers a promising route to reducing the open-system cost.

Measurement repetitions remain a large factor in both examples. Amplitude estimation~\cite{brassard2002amplitude} replaces the $1/\varepsilon^2$ repetition scaling of the Hadamard tests by $1/\varepsilon$ at greater coherent depth. For the $1.96\times10^{5}$ shots per time, correlator, and quadrature of the spectral example, of order $10^{2}$--$10^{3}$ coherent repetitions would replace them. The method applies directly to the unitary spectral circuits, but not to the sampled open-system evolution, whose timesteps discard ancillas. For $g^{(2)}$, the correlated bound of \cref{sec:g2-resources} still uses worst-case ranges. Variance-aware or adaptive estimators could reduce its $1.8\times10^{6}$ runs further: for near-coherent statistics at $\mu\approx0.5$, an asymptotic delta-method estimate needs a few times $10^{4}$ runs.

Future work should test the complete evolution circuits with numerically generated GQSP phases on interacting closed-system instances, compare optimized constructions at matched observable accuracy, and translate the logical resource estimates into physical-qubit and runtime costs for a specified fault-tolerant architecture, including
error correction and magic-state distillation.

\section*{Data and code availability}
The accompanying repository contains the validation notebooks, consolidated circuit implementations, resource-estimation notebooks, and supporting data. It includes classical spectral
and open-system benchmarks, small-instance gate-level tests, and executable Qualtran decompositions for the coherent resource estimates. Generic GQSP resource schedules are distinguished from numerical phase-specific circuits. Repository access details are provided at {\url{https://github.com/quanta-sim/TCL_simulation-and-resource-estimation}}.

\section*{AI Usage Disclosure}
The authors developed the core research ideas, model choices, overall simulation approach, and initial error bounds, and designed the initial circuit primitives and their operation sequences. AI tools assisted with drafting and refining code for numerical checks, preparing figures, and editing the manuscript. During later revisions, AI tools also helped build additional circuits from these primitives, lower the gate counts of existing circuits, and refine the analytical arguments and the resource and error calculations. The authors independently checked the derivations, circuit constructions, figures, and resource calculations, and tested all code. The authors reviewed and approved all AI-assisted material and take full responsibility for the manuscript.

\section*{Disclaimer}
All contributions to this paper were made while the author (Nam Nguyen) was an employee/researcher at George Mason University. Visa does not own, sponsor, or endorse this work and the views expressed are the author's own.

\section*{Acknowledgments}
We thank Michael Bremner, Sam Elman, and Dominic Berry for helpful discussions.

Soumya Sarkar and Rakshit Gharat thank Abhishek Rawat and Gargi Tyagi for insightful discussions 

We also extend our thanks to Matthew Otten and Hao Zhang for helping us with the classical numerical simulations of these models.

Nam Nguyen thanks Sunpreet Arora of Visa Research for his personal support and encouragement during the preparation of this manuscript.

\clearpage
\onecolumngrid

\appendix

\section{Primitive costs}
\label{app:primitives}

This appendix specifies the primitive counting conventions used in the resource ledger. We count Toffoli-family operations and omit Clifford gates. Computing a temporary logical AND counts as one operation, while its corrected measurement-assisted uncomputation adds no further Toffoli cost \cite{gidney2018halving}. This free erasure applies to temporary AND workspace; reversing a lookup or another subroutine to erase its output is counted separately. Where explicit coherent implementations are available, we also give their ordinary Toffoli counts. These costs describe individual primitives, with catalyst preparation and the remaining controlled-query overheads accounted for separately.

\emph{Table lookup and indexed access.} Unary iteration implements a QROM over $D$ entries using $D-1$ temporary ANDs, with the same cost for its inverse \cite{Babbush2018}. For indexed access to the quantum photon register, one operation over $L$ addresses costs $L-1$ temporary ANDs and $O(\log L)$ reusable workspace \cite{Babbush2018}. Reading a $\qph$-bit occupation at a selected address $a$ therefore costs
\begin{equation*}
\mathcal C_{\rm Tof}(\LOAD_a)
=\mathcal C_{\rm Tof}(\LOAD_a^\dagger)
=\qph(L-1).
\end{equation*}
Loading and subsequently erasing one selected occupation thus costs $2\qph(L-1)$. The same indexed-access construction, without the Hadamard conjugation, implements the selected bit flips used by the shift $S$ of \cref{eq:shift-def} and by the increment and decrement of \cref{fig:lobe}.

\emph{Lookup with additional workspace.} Select--swap offers an alternative to plain unary iteration. For a $D$-entry table of $w$-bit words, it loads a group of $k$ words and selects the requested word from that group. Its Toffoli-family cost has the form
\begin{equation*}
O(\lceil D/k\rceil+wk)
\end{equation*}
and its additional workspace is $O(wk)$ qubits, with constants and clean- or dirty-ancilla requirements set by the chosen implementation~\cite{low2024trading,sunderhauf2024block}. When $1\le w\le D$, choosing $k$ of order $\sqrt{D/w}$ gives cost and additional workspace of order $\sqrt{Dw}$. The commonly quoted $O(\sqrt D)$ dependence assumes fixed word width. Here $w$ grows with angle precision or alias-record size. These trades are not used in Appendix~\ref{app:ledger}, which retains the plain unary-iteration convention.

\emph{Coordinate update.} The destination address $i=j+d$ is computed by a direction-controlled modular increment or decrement of an $\ell$-bit coordinate. The ripple carry chain computes
\begin{equation*}
c_1=x_0,\qquad c_{k+1}=c_k\wedge x_k,
\end{equation*}
using $\ell-1$ temporary ANDs. Including the two direction controls adds constant overhead. One controlled shift therefore costs $2(\ell-1)+O(1)$, compared with $O(\ell^2)$ for a decomposition into successive multi-controlled NOT gates. If a multi-controlled construction is preferred, conditionally clean ancillae give an $m$-controlled NOT in $O(m)$ Toffoli operations with two clean ancillas \cite{Khattar:2024pqa}. For the pair of coordinates, we use the bound
\begin{equation*}
\mathcal C_{\rm Tof}(\DEST)\le2(\ell_x+\ell_y)+4.
\end{equation*}

\emph{Phase-gradient rotation.} For a loaded $b$-bit angle register, an $R_y$ rotation by the stored angle requires $b-1$ temporary ANDs and no synthesized arbitrary rotations when implemented with a phase-gradient state \cite{sunderhauf2024block,low2024trading}. The explicit construction in Appendix~\ref{app:angle-explicit} uses $2(b-1)$ ordinary Toffoli gates with coherent carry erasure, and its enable mask adds $b$ temporary ANDs or $2b$ coherent Toffolis. Catalyst preparation is counted separately per run.

\emph{Coherent alias sampling.} The weighted LCU preparations $\PREP_\hop$ and $\PREP_\loc$ are implemented using coherent alias sampling \cite{Babbush2018}. For a preparation over $\widetilde M$ padded labels with $b_{\rm alias}$-bit thresholds, we assign the cost
\begin{equation*}
\widetilde M-1+b_{\rm alias}+\left\lceil\log_2\widetilde M\right\rceil
\end{equation*}
Toffoli-family operations, with no arbitrary rotations. This is the counting convention used in the resource ledger rather than a complete gate-level compilation of the controlled preparation. The following subsection specifies the rounded thresholds and retained registers, and derives the coefficient error independently of this cost convention.

\subsection{Rounding the alias thresholds}
\label{app:alias-rounding}

Let $D$ be the padded number of labels and let $\rho_x$ be the desired probability of label $x$. An exact alias table assigns each uniformly selected column $z$ a threshold $u_z\in[0,1]$ and an alternative label $a_z$. It outputs $z$ with probability $u_z$ and $a_z$ otherwise. Round each threshold to the nearest multiple of $2^{-b_{\rm alias}}$ by
\begin{equation}
\begin{aligned}
T_z&=\left\lfloor 2^{b_{\rm alias}}u_z+\frac12\right\rfloor,\\
\widetilde u_z&=\frac{T_z}{2^{b_{\rm alias}}}.
\end{aligned}
\label{eq:alias-rounding}
\end{equation}
Use a uniform $b_{\rm alias}$-bit integer $r$. The comparison $r<T_z$ chooses $z$; otherwise it chooses $a_z$. A threshold $T_z=2^{b_{\rm alias}}$ is encoded by storing zero and choosing $z$ as its own alternative label. This column then always outputs $z$, without an extra threshold bit. Thus both endpoint probabilities are exact.

For $D=2^d$, the reversible preparation starts with a $d$-qubit uniform column register and a $b_{\rm alias}$-qubit uniform integer register $r$. A lookup writes the alternative label into $d$ clean qubits and the stored threshold into $b_{\rm alias}$ clean qubits. A reversible comparison computes one flag $v$ indicating that the alternative should be selected; its temporary workspace is erased while retaining $v$. A swap controlled by $v$ exchanges the column and alternative registers, leaving the selected label in the column register. The random integer, alternative register, threshold, and flag are retained in $J$. Their width is
\begin{equation}
w=d+2b_{\rm alias}+1.
\label{eq:alias-work-width}
\end{equation}
The unary lookup workspace is also returned to zero before the swap. The inverse preparation first reverses the swap, restoring the original column before reversing the comparison and lookup. Selection acts only on the resulting label and system, leaving every register in $J$ unchanged. This specifies the retained output width, without assigning a complete controlled gate count to the preparation.

For each column, moving probability between its two possible outputs changes the total absolute probability difference by at most $2|\widetilde u_z-u_z|/D$. Therefore the rounded distribution satisfies
\begin{equation}
\begin{aligned}
\sum_x|\widetilde\rho_x-\rho_x|
&\le\frac{2}{D}\sum_z|\widetilde u_z-u_z|\\
&\le 2^{-b_{\rm alias}}.
\end{aligned}
\label{eq:alias-distribution-error}
\end{equation}
The result applies to both weighted LCU preparations, $\PREP_\hop$ and $\PREP_\loc$. An exact zero-weight label receives no probability in any column; the endpoint convention preserves that property after rounding. Write $B_x$ for the normalized block selected by label $x$. Since selection preserves $x$ and leaves $J$ unchanged, preparation and its inverse weight this block by the label probability:
\[
B_{\rm exact}=\sum_x\rho_x B_x,
\qquad
B_{\rm rounded}=\sum_x\widetilde\rho_x B_x.
\]
Each $B_x$ is a contraction, so
\[
\|B_{\rm rounded}-B_{\rm exact}\|
\le\sum_x|\widetilde\rho_x-\rho_x|\,\|B_x\|
\le2^{-b_{\rm alias}}.
\]
Thus the probability error enters linearly; a distance bound between the complete preparation states is unnecessary here.

\section{Explicit reversible circuits}
\label{app:explicit-circuits}

\subsection{Selected photon updates}
\label{app:shift-explicit}
\begin{figure*}
\centering
\input{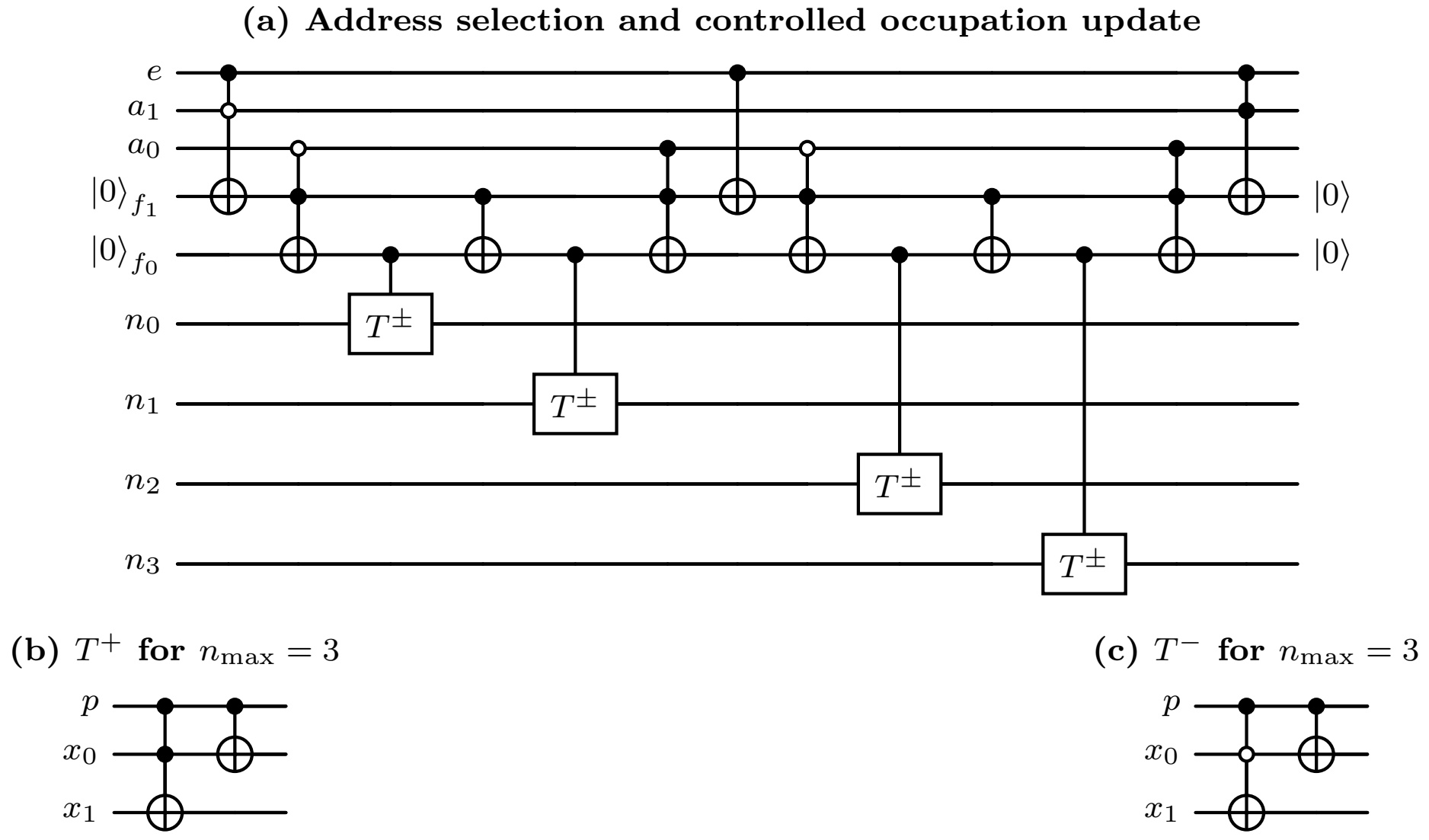}
\caption{Explicit selected photon update. (a) A four-register example of the binary address traversal, with $a=2a_1+a_0$. Each occupation wire contains $\qph$ qubits. The two clean flags $f_1,f_0$ select one occupation register when $e=1$ and return to zero. The update is the identity when $e=0$. Filled and open controls require their qubit to be one and zero, respectively. The same traversal extends to $L$ registers by adding address levels. (b,c) Expansion of the enabled increment and decrement for $\nmax=3$, with $n=x_0+2x_1$ and leaf flag $p$. Each uses one Toffoli and one CNOT. Apply panel (a) with $T^-$ at address $j$, followed by $T^+$ at address $i$, to implement the selected hopping shift. All gates shown have a coherent inverse. The measurement-assisted alternative described in the text replaces only clean-flag and carry erasures.}
\label{fig:shift-explicit}
\end{figure*}

We implement the selected occupation update using a binary-tree unary-iteration traversal \cite{Babbush2018, Khattar:2024pqa}. Let $a$ be an address and let $e$ enable the update. The required operation is
\begin{equation}
\begin{aligned}
\ket{e,a,\mathbf n}&\longmapsto\ket{e,a,\mathbf n'},\\
n'_k&=\begin{cases}
(n_k\pm e)\bmod R,&k=a,\\
n_k,&k\ne a.
\end{cases}
\end{aligned}
\label{eq:shift-enabled}
\end{equation}
Here the sites are numbered from $0$ to $L-1$ for binary addressing. The hopping branch uses the decrement at $a=j$ and the increment at $a=i$. Its outer selector activates this branch at $s=1$, so the enable is simply $e=s$. The four-register example in \cref{fig:shift-explicit} illustrates the addressing circuit independently of the lattice geometry.

\paragraph{Address selection.}
The purpose of the address-selection circuit is to activate exactly one occupation register, namely the register whose site index is specified by the binary address $a$. The address bits are processed from most to least significant. At each level of the binary address tree, a temporary flag records whether the current address prefix matches the branch being followed.

Let $p$ denote the flag associated with the current prefix and let $a_r$ be the next address bit. A clean flag $f$ is first computed as
\begin{equation*}
f=p(1-a_r),
\end{equation*}
so that $f=1$ only when the previously processed address bits match the current branch and the next address bit is zero. The circuit then processes the subtree whose addresses extend the current prefix with $a_r=0$. Afterwards, a CNOT from $p$ to $f$ changes the flag from $p(1-a_r)$ to
\begin{equation*}
f=pa_r,
\end{equation*}
so the same qubit can be reused to process the subtree corresponding to $a_r=1$. Once both subtrees have been visited, $f$ is returned to zero with a Toffoli controlled by $p$ and $a_r$.

This procedure is repeated recursively until a leaf is reached. At a leaf, the corresponding occupation update is applied, controlled by the leaf flag. Thus, for $e=1$, exactly one leaf is active and only the occupation register selected by the address $a$ is modified. For $e=0$, no leaf is active and the update is the identity. The initial parent flag is $e$. Since the same flag is reused for the two branches at each level, only $\log_2 L$ temporary flags are required. All address bits and parent flags remain unchanged during the traversal, so the temporary flags can be uncomputed even though the photon occupation data may change.

\paragraph{Carry and borrow gates.}
Write
\begin{equation*}
q=\qph\quad\text{and}\quad n=\sum_{r=0}^{q-1}2^r x_r,
\end{equation*}
so $x_0$ is the least significant bit. The two-bit circuits in \cref{fig:shift-explicit} first flip $x_1$ when a carry or borrow is required and then flip $x_0$. For general $q\ge2$, take the leaf flag as $c_0=p$. Compute $q-2$ clean prefix bits from
\begin{equation}
\begin{aligned}
c_r&=c_{r-1}u_{r-1},\\
u_r&=
\begin{cases}
x_r,&\text{increment},\\
1-x_r,&\text{decrement},
\end{cases}
\end{aligned}
\label{eq:shift-prefix}
\end{equation}
for $r=1,\ldots,q-2$. Each prefix uses one Toffoli, with an open occupation control for the decrement. Flip $x_{q-1}$ with a Toffoli controlled by $c_{q-2}$ and $u_{q-2}$. Next, for $r=q-2,\ldots,1$, apply the following updates in order:
\begin{equation*}
\begin{aligned}
x_r&\leftarrow x_r\mathbin{\oplus}c_r
&&\text{(CNOT)},\\
c_r&\leftarrow c_r\mathbin{\oplus}c_{r-1}u_{r-1}
&&\text{(Toffoli)}.
\end{aligned}
\end{equation*}
The second update erases $c_r$ using its original controls $c_{r-1},u_{r-1}$. Those lower bits have not yet changed. Finish with a CNOT from $p$ to $x_0$. This order returns every prefix bit to zero.

For $q=2$ there are no prefix bits, and for $q=1$ only the final CNOT is needed. The arithmetic is modular, including at $n=0$ and $n=R-1$; the coefficient tables, rather than this unitary shift, enforce the physical boundaries.

\paragraph{Coherent and measurement-assisted costs.}
The fully coherent circuit has $L-1$ internal address nodes. Each node uses two Toffolis for flag computation and erasure. At each leaf, the $q$-bit update uses $2q-3$ Toffolis when $q\ge2$. The source decrement and destination increment therefore use
\begin{equation}
2\left[2(L-1)+L\max(2q-3,0)\right]
\end{equation}
ordinary Toffoli gates. Reversing this gate sequence gives the exact adjoint. The extra clean workspace is $\log_2 L+\max(q-2,0)$ qubits. Both the address traversal and the arithmetic already include the outer enable.

The temporary-AND convention of Appendix~\ref{app:primitives} gives a smaller count. To erase a clean target holding $f=uv$, measure it in the $X$ basis. If the outcome is one, apply $\mathrm{CZ}$ to its two controls, then reset the target to zero. An open control is handled by conjugating this correction with $X$. The outcome-dependent phase is thereby removed.

In the address traversal, the flag has the value $pa_r$ at erasure, although it was initially computed as $p(1-a_r)$; the correction uses its value at erasure. In the carry chain, erasure occurs before either lower control changes. Thus the measurement and correction have the same action on the retained registers as coherent erasure, including for superpositions of addresses and enable states.

Each internal address node now costs one temporary AND, and each occupation update costs $q-1$ Toffoli-family operations. The pair of shifts costs
\begin{equation}
2\left[(L-1)+L(q-1)\right]
=2q(L-1)+2(q-1).
\label{eq:shift-explicit-cost}
\end{equation}
This reproduces the shift and carry entries of \cref{tab:ledger} under its measurement-assisted convention. This subtotal covers the selected shifts and carry chains; it is only part of the complete Hamiltonian-query cost.

For the measurement-assisted construction, implement the adjoint by compiling the inverse logical update and its associated measurement-assisted uncomputation. The measurements cannot simply be reversed as unitary gates.

\begin{figure*}
\centering
\input{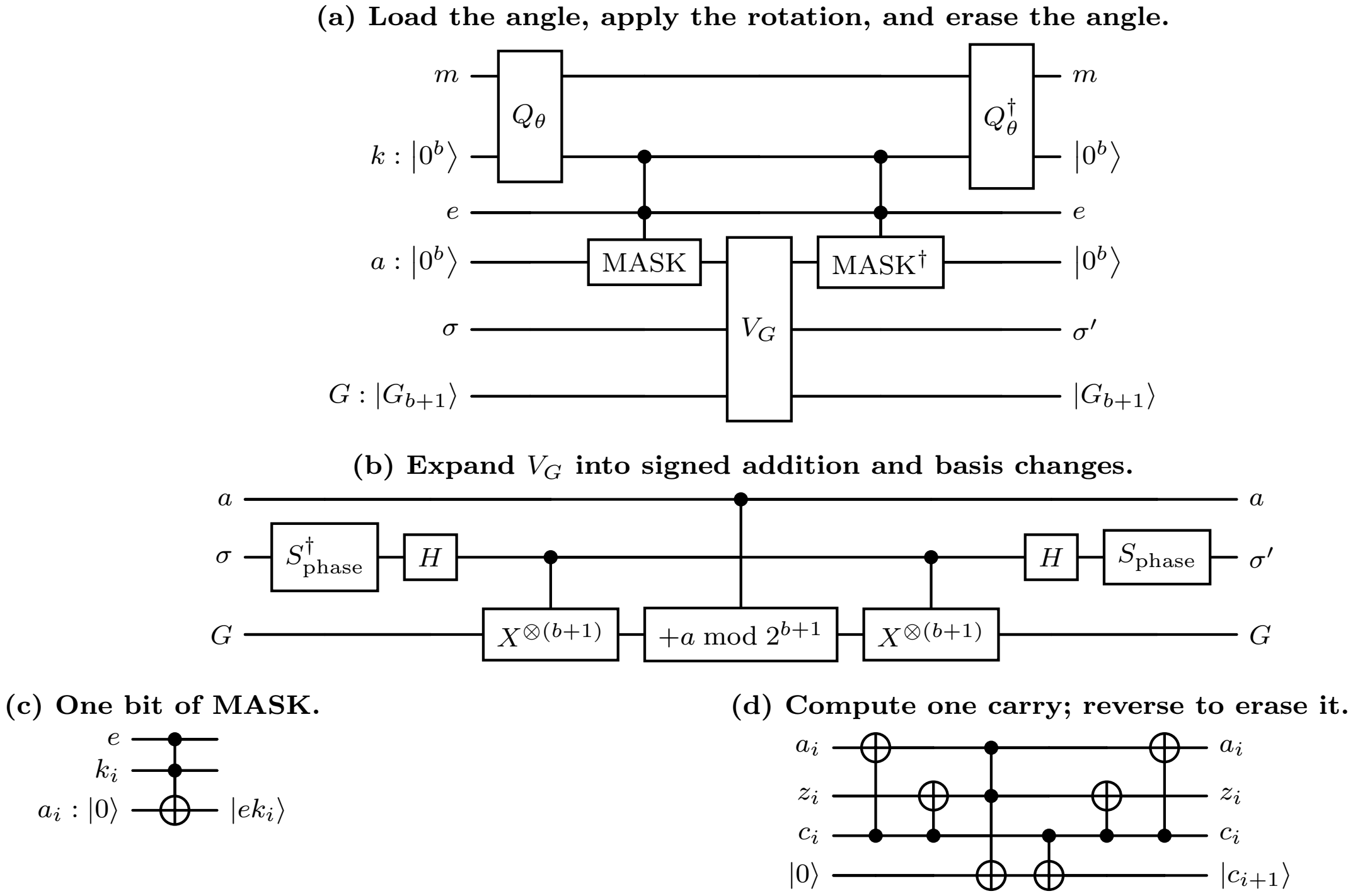}
\caption{Shared angle lookup and explicitly controlled phase-gradient rotation.
The address $m$ loads the $b$-bit integer $k_m$ through $Q_\theta$, after
which the enable bit $e$ forms the masked integer $a=e k_m$. Panel (a)
shows the complete lookup--rotate--uncompute sequence. Panel (b) expands
$V_G$: phase-gradient addition implements
$R_y(2\pi a/2^b)$ on the signal qubit $\sigma$ while returning the catalyst
$\ket{G_{b+1}}$ exactly. The lines from $k$ to MASK and from $a$ to the
addition represent whole-register operands rather than multi-qubit
controls. Panel (c) shows the mask circuit for one bit. This circuit is repeated for all
$b$ angle bits. Panel (d) shows one carry computation used in the modular
addition of \cref{eq:angle-carry}; when $c_0=0$, only the Toffoli gate is
required. The $b-1$ carry qubits are omitted from the drawing and are
initialized and returned to zero. The table, mask, and carry registers are
uncomputed, while the signal qubit need not return to its input state. For
an unconditional rotation, MASK is omitted and the loaded angle is used
directly. The same circuit realizes the angle rotations required by $O_A$,
$O_B$, and the local exchange block by changing the lookup table.}
\label{fig:angle-explicit}
\end{figure*}

\subsection{Shared angle lookup and phase-gradient rotation}
\label{app:angle-explicit}
A phase-gradient state is an eigenstate of modular addition, allowing an angle-dependent phase to be generated through phase kickback. This enables programmable rotations using reversible arithmetic rather than separately synthesized rotations for each angle. Throughout this work, we use reusable phase-gradient states to implement rotations in both the Hamiltonian block encoding and GQSP, with shared angle lookup for occupation-dependent amplitudes.

The two hopping rotations and the local exchange rotation use the same circuit in \cref{fig:angle-explicit}, with different shared angle tables. For an address $m$ and enable qubit $e$, the circuit loads a rounded angle, forms an enabled copy of its binary representation, and applies the corresponding $R_y$ rotation using a reusable phase-gradient state. It then uncomputes the angle and arithmetic workspace. We first establish this circuit action and then describe the reversible addition and its cost.

For each table address $m$, the desired angle $\theta_m\in[0,\pi]$ is rounded to
\begin{equation}
\begin{gathered}
\widetilde\theta_m=\frac{2\pi k_m}{2^b},
\\
0\le k_m\le 2^{b-1}.
\end{gathered}
\label{eq:angle-grid-explicit}
\end{equation}
The integer $k_m$ fits in $b$ bits, and nearest rounding gives
\begin{equation*}
|\widetilde\theta_m-\theta_m|\le\pi/2^b,
\end{equation*}
as used in Appendix~\ref{app:error}. Both endpoints, $0$ and $\pi$, are represented exactly, including when $b=1$. Assigning every forbidden occupation the exact value $k_m=2^{b-1}$ therefore preserves its zero projected amplitude.

The rounded angle is loaded by the reversible XOR lookup
\begin{equation}
Q_\theta:\ket{m}\ket{k}\longmapsto
\ket{m}\ket{k\mathbin{\oplus}k_m}.
\end{equation}
Unary iteration decodes each address in turn; the decoded flag controls a NOT on each output bit that is one in the corresponding classical table entry. To enable the rotation, write
\begin{equation*}
a_i=e k_i
\end{equation*}
into $b$ clean qubits, as shown in Fig.~\ref{fig:angle-explicit} panel (c). The resulting word is
\begin{equation*}
a=0\quad\text{when}\quad e=0\quad\text{and}\quad a=k_m\quad\text{when}\quad e=1.
\end{equation*}
After the rotation, reverse the mask and lookup to return both words to zero.

The rotation is implemented by adding the masked angle to the $b+1$-qubit phase-gradient catalyst
\begin{equation}
\ket{G_{b+1}}=
\frac{1}{\sqrt{2^{b+1}}}
\sum_{z=0}^{2^{b+1}-1}
e^{2\pi i z/2^{b+1}}\ket{z}.
\label{eq:gradient-state-explicit}
\end{equation}
Let $A_a$ denote addition of $a$ modulo $2^{b+1}$ on this register, and let $F=X^{\otimes(b+1)}$ complement all of its bits. The identities
\begin{equation}
\begin{aligned}
A_a\ket{G_{b+1}}&=e^{-2\pi i a/2^{b+1}}\ket{G_{b+1}},\\
F A_a F&=A_{-a}.
\end{aligned}
\label{eq:signed-gradient-add}
\end{equation}
show that addition and subtraction produce opposite phases while returning the catalyst to its original state. The controlled complements in panel (b) give the two signal states the phases
\begin{equation*}
\begin{gathered}
e^{-i\widetilde\theta/2}\quad\text{and}\quad e^{i\widetilde\theta/2},\quad\text{respectively},\\
\text{where}\quad\widetilde\theta=2\pi a/2^b.
\end{gathered}
\end{equation*}
These are the two phases of $R_z(\widetilde\theta)$, with no additional phase depending on the stored angle. The surrounding Clifford gates then convert this rotation into $R_y$: with
\begin{equation*}
\begin{gathered}
S_{\rm phase}=\operatorname{diag}(1,i),\\
S_{\rm phase} H R_z(\widetilde\theta)H S_{\rm phase}^\dagger
=R_y(\widetilde\theta).
\end{gathered}
\end{equation*}
Thus the complete circuit acts as
\begin{equation}
\begin{aligned}
&\ket{m}\ket{e}\ket{\psi}_\sigma\ket{G_{b+1}}\ket{0}_{\rm work}\\
&\longmapsto
\ket{m}\ket{e}R_y(e\widetilde\theta_m)\ket{\psi}_\sigma
\ket{G_{b+1}}\ket{0}_{\rm work}.
\end{aligned}
\label{eq:angle-catalyst-return}
\end{equation}
This action holds for arbitrary superpositions of the address, enable, and signal registers, including inputs entangled with additional registers. The catalyst returns to its original state and remains unentangled with the other registers, while the inactive branch $e=0$ is the identity with no relative phase. The walk reflection can therefore omit the catalyst on the invariant subspace in which it begins in the state of \cref{eq:gradient-state-explicit}.

\paragraph{Addition and carry gates.}
The identities above reduce the rotation to implementing $A_a$ on the catalyst register. We use ripple-carry arithmetic~\cite{gidney2018halving} with the explicit updates below. Write $a_i$ and $z_i$ for the bits of $a$ and of the catalyst, ordered from least to most significant. With $c_0=0$, compute $b-1$ carry bits into clean qubits according to
\begin{equation}
c_{i+1}=c_i\mathbin{\oplus}
[(a_i\mathbin{\oplus}c_i)(z_i\mathbin{\oplus}c_i)],
\label{eq:angle-carry}
\end{equation}
for $i=0,\ldots,b-2$. For $i\ge1$, the reversible updates are applied in the following order:
\begin{equation*}
\begin{aligned}
a_i&\leftarrow a_i\mathbin{\oplus}c_i,
&z_i&\leftarrow z_i\mathbin{\oplus}c_i,\\
c_{i+1}&\leftarrow c_{i+1}\mathbin{\oplus}a_i z_i,\\
c_{i+1}&\leftarrow c_{i+1}\mathbin{\oplus}c_i,\\
a_i&\leftarrow a_i\mathbin{\oplus}c_i,
&z_i&\leftarrow z_i\mathbin{\oplus}c_i.
\end{aligned}
\end{equation*}
The first and last lines use CNOTs, the second uses a Toffoli, and the third uses a CNOT. This is the carry circuit in Fig.~\ref{fig:angle-explicit} panel (d). For $i=0$, where $c_0=0$, it reduces to a single Toffoli from $a_0,z_0$ to $c_1$.

The highest catalyst qubit is $\ket{-}$ in \cref{eq:gradient-state-explicit}, so a carry into this qubit contributes only a minus sign. With
\begin{equation*}
u=a_{b-1},\qquad v=z_{b-1},\quad\text{and}\quad w=c_{b-1},
\end{equation*}
the corresponding carry value is
\begin{equation*}
uv\mathbin{\oplus}uw\mathbin{\oplus}vw.
\end{equation*}
Its action is implemented by the three Clifford gates
\begin{equation*}
\operatorname{CZ}(u,v),\qquad
\operatorname{CZ}(u,w),\quad\text{and}\quad\operatorname{CZ}(v,w);
\end{equation*}
the gates involving $c_0$ are omitted when $b=1$.

After computing the carries, update the catalyst bits from most to least significant:
\begin{equation*}
z_i\leftarrow z_i\mathbin{\oplus}a_i\mathbin{\oplus}c_i,
\qquad i=b-1,b-2,\ldots,0.
\end{equation*}
Use CNOTs for this update and erase $c_i$ by reversing its computation whenever $i>0$. The lower bits required to erase $c_i$ have not yet changed, so each inverse carry computation is valid. This implements modular addition on the invariant subspace in which the highest catalyst qubit is $\ket{-}$ and returns all carry qubits to zero.

Measurement-assisted erasure provides an alternative to reversing the carry computation. First expose its two XOR controls and undo the CNOT from $c_i$ to $c_{i+1}$, then measure the resulting AND bit in the $X$ basis. For the minus outcome, apply CZ to its two controls; reset the measured qubit and restore the controls. Under the temporary-AND convention of Appendix~\ref{app:primitives}, each carry is computed once and its corrected erasure adds no Toffoli cost. The addition costs are therefore
\begin{equation*}
\begin{aligned}
\text{Measurement-assisted:}&\quad b-1
\text{ Toffoli-family operations},\\
\text{Fully coherent:}&\quad 2(b-1)
\text{ ordinary Toffoli gates}.
\end{aligned}
\end{equation*}
The coherent count includes the inverse carry computations. This implementation remains available whenever a literal gate-by-gate adjoint is required.

In addition to these carry costs, the enable mask and lookup require
\begin{equation*}
\begin{aligned}
\text{Enable-mask workspace:}&\quad b\text{ clean qubits},\\
\text{Enable mask, measurement-assisted:}&\quad b\text{ temporary ANDs},\\
\text{Enable mask, fully coherent:}&\quad 2b\text{ coherent Toffolis},\\
\text{Table lookup and its inverse:}&\quad 2(R-1)\text{ lookup operations}.
\end{aligned}
\end{equation*}
The measurement-assisted mask count uses corrected erasure. Its overhead is therefore $O(b)$ for the controlled construction shown here, while the lookup operations are the separate costs used in the ledger.

The circuit identities assume an exact finite phase-gradient state. The catalyst is prepared once per run, with its preparation error included separately in the run's error budget; subsequent calls require no synthesized arbitrary rotations. Catalyst preparation and the choice between coherent and measurement-assisted erasure must be specified before these line items are converted into a compiled logical gate count.

\subsection{Circuit verification}
The accompanying gate-level tests check address comparison, destination arithmetic, alias preparation, indexed occupation access, and projected block-encoding actions. They also test
coherent controls, adjoints, catalyst return, and clean-workspace uncomputation on small instances. Complete coherent circuit resource counts are checked separately using executable Qualtran decompositions.

\section{Cost ledgers}
\label{app:ledger}

\Cref{tab:compare} gives fixed-precision arithmetic subtotals under the measurement-assisted convention. They describe the Hamiltonian with its identity term removed and exclude the walk controls and reflection. The main-text savings instead use the complete coherent walk counts.

\begin{table*}[t]
\caption{Arithmetic-cost subtotals for the sparse-access and direct-register constructions under the stated primitive counting conventions of Appendix~\ref{app:ledger}, with $b=b_{\rm alias}=20$. The common local branch is included in both totals, while the controlled-query and reflection overheads of Appendix~\ref{app:walk} are excluded. The normalization is evaluated at $g_i=1$, $t_{ij}=0.05$, $\Delta=0$, $\nu_i=1$ and rounded to the nearest integer; $\alpha_\TCL^{(0)}$ omits the identity coefficient. The counts use general alias preparation and padded selection labels. They give the arithmetic cost of one call at fixed precision; a comparison of total simulation costs also requires a common evolution time and target accuracy.}
\label{tab:compare}
\squeezetable
\begin{ruledtabular}
\begin{tabular}{ccc rr r rr}
$L$ & $N$ & $\nmax$ & $\mathcal C_{\rm Tof}^{\rm sp}$ & $\mathcal C_{\rm Tof}^{\rm dir}$ & sp./dir. & $\alpha_\TCL$ & $\alpha_\TCL^{(0)}$ \\
\colrule
16  & 1 & 3  & 1{,}222  & 1{,}016  & 1.20 & 101   & 69   \\
64  & 2 & 7  & 5{,}356  & 4{,}186  & 1.28 & 1{,}004 & 716  \\
256 & 4 & 15 & 31{,}020 & 24{,}854 & 1.25 & 9{,}598 & 7{,}166 \\
\end{tabular}
\end{ruledtabular}
\end{table*}

\Cref{tab:compare} reports arithmetic-cost subtotals for the Hamiltonian block encoding under the following counting assumptions. The breakdown identifies the circuit primitives included in the resource estimates and makes the reported counts reproducible. The explicit circuits in Appendix~\ref{app:explicit-circuits} and~\ref{app:walk} establish the selected occupation update, shared rotation, and controlled reflection. A complete query count must also include the angle masks, coefficient preparation and its controls, and the final register allocation. The ledger retains the stated primitive convention. Write
\begin{equation}
\begin{aligned}
\mathcal A(D)&=D-1+b_{\rm alias}+\left\lceil\log_2D\right\rceil,\\
\mathcal I&=\qph(L-1),\\
\mathcal E&=2(\ell_x+\ell_y)+4,
\end{aligned}
\label{eq:ledger-shorthand}
\end{equation}
for one coherent alias preparation over $D$ padded labels, one indexed read of a $\qph$-bit occupation out of the photon register, and one destination-address computation. The quantities $\mathcal C_{\rm Tof}^{\rm dir}$ and $\mathcal C_{\rm Tof}^{\rm sp}$ are the sums of the common lines and the corresponding hopping lines in \cref{tab:ledger}. The numerical comparison uses the sign assumptions and unoptimized shared preparations stated in \cref{sec:compare}. For general bond signs, the cost of $O_{\sgn}$ in \cref{sec:hop} must be added in both implementations.

The two constructions of \cref{tab:compare} share $\PREP_\hop$, $\PREP_\loc$, the unary routing of $\SELECT_\loc$, the exchange increment and decrement, and the shared angle table; they differ only in the hopping branch. Given the input configuration $\ket{\mathbf n}$ and a label $\ell=(j,d)$, the sparse-access forward oracle $O_F$ computes the destination address $i=j+d$ and loads $n_j$ and $n_i$ into two $\qph$-qubit work registers by indexed access. Conditioned on those loads, it decrements the output register at $j$ and increments it at $i$, then erases the destination address and the two loaded occupations. The reverse oracle $O_R$ clears the residual input configuration at the same cost. Because the two occupation tables of \cref{lem:factor} already return zero outside $1\le n_j\le\nmax$ and $n_i<\nmax$, neither oracle needs a separate validity comparator.

\begin{table*}[t]
\caption{Conditional Toffoli-family line items for the two hopping-branch implementations compared in \cref{tab:compare}, in the shorthand of \cref{eq:ledger-shorthand}. A temporary logical AND counts as one Toffoli-family operation with free measurement-assisted uncomputation \cite{gidney2018halving}; reapplying a subroutine to erase its output is counted, which is why most lines carry a factor of two. Both branches have $\mathcal C_R=0$, since every occupation-dependent amplitude is produced by a phase-gradient rotation rather than a synthesized one; the two rotations of \cref{lem:be} are common to both constructions and are not listed here.}
\label{tab:ledger}
\begin{ruledtabular}
\begin{tabular}{lr}
Line item & $\mathcal C_{\rm Tof}$ \\
\colrule
\multicolumn{2}{l}{\textit{Common to both constructions}}\\
$\PREP_\hop$ and inverse & $2\mathcal A(4L)$\\
$\PREP_\loc$ and inverse & $2\mathcal A(\widetilde M)$\\
$\SELECT_\loc$ unary routing & $\widetilde M-1$\\
Exchange increment and decrement & $2\mathcal I+2(\qph-1)$\\
Shared angle table: load, table, rotation & $2\mathcal I+2(R-1)+(b-1)$\\
\colrule
\multicolumn{2}{l}{\textit{Hopping branch, direct-register shift (\cref{sec:hop})}}\\
$\DEST$ and inverse & $2\mathcal E$\\
Two occupation loads and inverses & $4\mathcal I$\\
Tables $A$, $B$ and inverses & $4(R-1)$\\
Two phase-gradient rotations & $2(b-1)$\\
Direct occupation shift $S$ & $2\mathcal I$\\
Carry and borrow chains & $2(\qph-1)$\\
\colrule
\multicolumn{2}{l}{\textit{Hopping branch, sparse access}}\\
$O_F$: address, loads, update, cleanup & $2\mathcal E+6\mathcal I+2(\qph-1)$\\
Reverse oracle $O_R$ & as $O_F$\\
Tables $A$, $B$ and inverses & $4(R-1)$\\
Two phase-gradient rotations & $2(b-1)$\\
\end{tabular}
\end{ruledtabular}
\end{table*}

\paragraph{A worked column.} For $L=16$, $N=1$, $\nmax=3$, and
$b=b_{\rm alias}=20$, the register and table sizes are
\begin{equation*}
\begin{aligned}
\qph&=2,& R&=4,\\
M_\loc&=65,&\widetilde M&=128.
\end{aligned}
\end{equation*}
The shared lines total
\begin{equation*}
\begin{aligned}
&2\mathcal A(64)+2\mathcal A(128)+127+62+85\\
&\qquad=178+308+127+62+85=760.
\end{aligned}
\end{equation*}
The direct-register hopping branch adds
\begin{equation*}
\begin{aligned}
&2\mathcal E+4\mathcal I+4(R-1)+2(b-1)\\
&\qquad+2\mathcal I+2(\qph-1)=256,
\end{aligned}
\end{equation*}
giving $1{,}016$, the ledger subtotal in \cref{tab:compare}. The sparse-access branch instead adds
\begin{equation*}
\begin{aligned}
&2\left[2\mathcal E+6\mathcal I+2(\qph-1)\right]\\
&\qquad+4(R-1)+2(b-1)=412+50=462,
\end{aligned}
\end{equation*}
giving $1{,}222$, matching \cref{tab:compare}.

\paragraph{Independent Qualtran checks.}
We implemented the ledger primitives and forward and reverse XOR
row oracles, to reproduce \cref{tab:compare} using executable Qualtran circuits
\cite{harrigan2024qualtran}. Counting their decomposed operations under the ledger's temporary-AND and measurement-assisted uncomputation convention gives arithmetic subtotals below the stated bounds. The difference arises solely from a cheaper reversible destination-address circuit; all other checked arithmetic line items agree. This primitive-level audit supports the ledger bounds but does not establish the cost of a complete Hamiltonian block encoding or controlled walk at those arithmetic subtotals.

\section{Error budget}
\label{app:error}

We now quantify how finite-precision circuit operations affect the accuracy of the TCL block encoding. The construction introduces errors through the discretization of occupation-dependent rotation angles, finite-precision coefficient preparation, and the synthesis of the remaining arbitrary rotations. Specifically, these contributions arise from the $b$-bit angle grids used in $O_A$ and $O_B$ of \cref{eq:OA} and $R_\theta$ of \cref{eq:lobe-angle}, the $b_{\rm alias}$-bit thresholds in $\PREP_\hop$ and $\PREP_\loc$, and the synthesis of $P_\TCL$ and $P_\TCL^\dagger$. We bound these contributions separately and combine them to obtain an error bound for $U_\TCL$.

We first consider the angle discretization. Representing $\theta\in[0,\pi]$ on a uniform $b$-bit grid and rounding to the nearest representable angle gives
\begin{equation*}
|\widetilde\theta-\theta|\le\pi/2^b.
\end{equation*}
Since
\begin{equation*}
|\cos x-\cos y|\le|x-y|,
\end{equation*}
each corresponding rotation amplitude is perturbed by at most
\begin{equation}
\varepsilon_{\rm amp}=\frac{\pi}{2^{b+1}}.
\label{eq:eta}
\end{equation}
The hopping branch applies two such rotations, and since both amplitudes lie in $[0,1]$ by \cref{lem:factor},
\begin{equation}
\begin{aligned}
\lvert\widetilde A\widetilde B-AB\rvert
&\le\lvert\widetilde A-A\rvert\lvert\widetilde B\rvert
+\lvert A\rvert\lvert\widetilde B-B\rvert\\
&\le2\varepsilon_{\rm amp}.
\end{aligned}
\label{eq:product-error}
\end{equation}
Writing
$\widetilde H_\hop=\sum_\ell c_\ell S_\ell\widetilde D_\ell$
and using
$\norm{\widetilde D_\ell-D_\ell}\le2\varepsilon_{\rm amp}\nmax$
together with \cref{eq:alpha-hop},
\begin{equation}
\norm{\widetilde H_\hop-H_\hop}
\le\sum_\ell\abs{c_\ell}\norm{\widetilde D_\ell-D_\ell}
\le2\varepsilon_{\rm amp}\alpha_\hop.
\label{eq:hop-error}
\end{equation}
The exchange branch applies one rotation, so its contribution is $\varepsilon_{\rm amp}\alpha_\intt$, with $\alpha_\intt$ defined in \cref{eq:alpha-int}. For the coherent alias samplers, \cref{eq:alias-distribution-error} establishes that nearest-threshold rounding introduces a distribution error bounded by
\begin{equation*}
\sum_x|\widetilde\rho_x-\rho_x|\le 2^{-b_{\rm alias}}.
\end{equation*}
Since the alias outputs do not affect the selected operations and each selected block has operator norm at most one, the error in the resulting projected block is bounded by the same quantity. Multiplying by the corresponding block-encoding normalizations, the contributions from $\PREP_\hop$ and $\PREP_\loc$ are bounded by $\alpha_\hop 2^{-b_{\rm alias}}$ and $\alpha_\loc 2^{-b_{\rm alias}}$, respectively. Combining these with the angle-discretization errors gives the total perturbation of the encoded Hamiltonian,
\begin{equation}
\varepsilon_H^{\rm(data)}
\le
2\varepsilon_{\rm amp}\alpha_\hop
+
\varepsilon_{\rm amp}\alpha_\intt
+
\frac{\alpha_\hop+\alpha_\loc}{2^{b_{\rm alias}}}.
\label{eq:be-sum}
\end{equation}

Finally, we account for the synthesis error of the outer branch-selection rotation $P_\TCL$, which prepares the relative amplitudes of the local and hopping contributions to the block encoding [\cref{eq:P-TCL}]. As shown in \cref{eq:U-TCL}, the complete construction contains both $P_\TCL$ and $P_\TCL^\dagger$. If each is synthesized with an operator-norm error at most $\varepsilon_{\rm rot}$, the triangle inequality gives a total contribution of at most $2\varepsilon_{\rm rot}$ to the normalized projected-block error. Combining this with the Hamiltonian perturbation from finite-precision data loading, we obtain
\begin{equation}
\delta_{\be}
\le
\frac{\varepsilon_H^{\rm(data)}}{\alpha_\TCL}
+
2\varepsilon_{\rm rot}.
\label{eq:normalized-be-error}
\end{equation}
Using \cref{eq:be-sum}, this gives
\begin{equation}
\delta_{\be}
\le
\frac{1}{\alpha_\TCL}
\left[
\frac{\pi(2\alpha_\hop+\alpha_\intt)}
{2^{b+1}}
+
\frac{\alpha_\hop+\alpha_\loc}
{2^{b_{\rm alias}}}
\right]
+
2\varepsilon_{\rm rot},
\end{equation}
consistent with the block-encoding error stated in \cref{lem:be}. Let $B$ be the normalized projected block of the implemented $U_\TCL$, so that
\begin{equation*}
\norm{B-H_\TCL/\alpha_\TCL}\le\delta_{\be}.
\end{equation*}
Rounding does not necessarily preserve the Hermiticity of the projected block $B$. Apply \cref{lem:hermitian} using the exact adjoint of the implemented circuit and define
\begin{equation}
H_\TCL^{\prime}=\frac{\alpha_\TCL}{2}(B+B^\dagger).
\label{eq:effective-hermitian-H}
\end{equation}
Then $H_\TCL^{\prime}$ is Hermitian and
\begin{equation*}
\norm{H_\TCL^{\prime}-H_\TCL}\le\alpha_\TCL\delta_{\be}.
\end{equation*}

We now translate this Hamiltonian error into an error in the simulated dynamics. Since the qubitized walk encodes $H_\TCL^{\prime}$ rather than $H_\TCL$, the corresponding time evolution differs from the target evolution. The Duhamel identity bounds this difference by
\begin{equation}
\norm{e^{-iH_\TCL\tau}
-e^{-iH_\TCL^{\prime}\tau}}
\le
\abs{\tau}\norm{H_\TCL-H_\TCL^{\prime}}
\le
\abs{\tau}\alpha_\TCL\delta_{\be}.
\label{eq:time-error}
\end{equation}

We now allocate the error budget to ensure that the initialized-state simulation error does not exceed $\varepsilon>0$. As discussed in \cref{sec:full}, we assign half of this budget to errors arising from the approximate block encoding and the remaining half to the GQSP implementation. For the block encoding, we allocate $\varepsilon/8$ each to angle discretization and alias-sampling errors, and $\varepsilon/4$ to the synthesis of the two outer branch-selection rotations $P_\TCL$ and $P_\TCL^\dagger$. These contributions are bounded after propagation over the evolution time $\tau$, ensuring that $|\tau|\alpha_\TCL\delta_{\be}\le\varepsilon/2$.

The remaining budget accounts for the GQSP polynomial approximation and the synthesis of its signal-processing rotations. Since the polynomial approximation controls only the projected block, we require $\delta_{\rm poly}\le\varepsilon^2/32$, which bounds the corresponding initialized-state error by $\sqrt{2\delta_{\rm poly}}\le\varepsilon/4$. In contrast, the synthesis error is bounded in operator norm for the full GQSP circuit and therefore contributes directly to the state error. We allocate at most $\varepsilon/4$ to this contribution, so that the combined simulation error is at most $\varepsilon$.

If $\tau=0$ or $\alpha_\TCL=0$, no Hamiltonian evolution circuit is needed. Otherwise, for $2\alpha_\hop+\alpha_\intt>0$, a sufficient common angle precision is
\begin{equation}
b\ge\max\left\{1,\left\lceil
\log_2\frac{4\pi\abs{\tau}(2\alpha_\hop+\alpha_\intt)}{\varepsilon}
\right\rceil\right\}.
\label{eq:angle-precision}
\end{equation}
If $2\alpha_\hop+\alpha_\intt=0$, the ladder-angle tables are absent and this constraint is unnecessary. For the coefficient preparations, choose
\begin{equation}
b_{\rm alias}\ge\max\left\{1,\left\lceil
\log_2\frac{8\abs{\tau}(\alpha_\hop+\alpha_\loc)}{\varepsilon}
\right\rceil\right\}.
\label{eq:precision-choice}
\end{equation}
A zero-weight branch is omitted and requires no coefficient preparation. If both branches are present, require the synthesis error of each outer rotation to satisfy
\begin{equation}
\varepsilon_{\rm rot}\le
\frac{\varepsilon}{8\abs{\tau}\alpha_\TCL}.
\label{eq:rotation-precision}
\end{equation}
When only one branch is present, the outer selection and these rotations are omitted.

Choose the polynomial degree such that $\delta_{\rm poly}\le\varepsilon^2/32$. For a GQSP sequence with $N_{\rm rot}>0$ signal-processing rotations, synthesizing each to operator-norm error at most $\varepsilon/(4N_{\rm rot})$ bounds the total synthesis error by $\varepsilon/4$ through a telescoping sum. Let $U_{\rm poly}$ denote the ideal GQSP circuit and $U_{\rm sim}$ its synthesized implementation. Define $J$ as the isometry that appends $\ket0_q\ket{+}_h\ket{0^a}_A$, the exact phase-gradient state, and zero-initialized workspace. The projected-block bound
\begin{equation*}
\norm{J^\dagger U_{\rm poly}J-e^{-iH_\TCL^{\prime}\tau}}
\le\delta_{\rm poly}
\end{equation*}
then implies, by unitarity,
\begin{equation*}
\begin{aligned}
\norm{U_{\rm poly}J-Je^{-iH_\TCL^{\prime}\tau}}
&\le\sqrt{2\delta_{\rm poly}}\\
&\le\varepsilon/4,
\end{aligned}
\end{equation*}
including ancilla leakage. Together with
\begin{equation*}
\norm{U_{\rm sim}-U_{\rm poly}}\le\varepsilon/4,
\end{equation*}
this bounds the initialized-state error by $\varepsilon/2$. Combining with \cref{eq:time-error} yields the target accuracy $\varepsilon$. Finally, although $b$ and $b_{\rm alias}$ enter additively in \cref{eq:full-cost}, a complete logical gate count must also include synthesis of the outer branch-selection and GQSP signal-processing rotations.

The square-root conversion used above follows directly from the unitarity of the polynomial circuit and its target. Set $V=e^{-iH_\TCL^{\prime}\tau}$. For every normalized system state $\ket\psi$, the isometry $J$ defined above satisfies
\begin{equation*}
\begin{aligned}
\|U_{\rm poly}J\ket\psi-JV\ket\psi\|^2
&=2-2\operatorname{Re}
\langle\psi|V^\dagger J^\dagger U_{\rm poly}J|\psi\rangle\\
&\le2\delta_{\rm poly}.
\end{aligned}
\end{equation*}
Taking the supremum over $\ket\psi$ gives $\|U_{\rm poly}J-JV\|\le\sqrt{2\delta_{\rm poly}}$. This bound includes leakage from the projected ancilla subspace, which explains why the polynomial tolerance is quadratic in the target initialized-state error.

\section{Qubitized walk and controlled queries}
\label{app:walk}

We now describe how the TCL block encoding is converted into the qubitized walk required for GQSP simulation. The standard walk construction $W=(2\Pi-I)\SELECT$ \cite{Low2019hamiltonian} assumes a Hermitian, self-inverse selection operator. However, the selected operations in \cref{sec:full} do not generally satisfy this condition. To obtain a suitable walk operator, we first Hermitianize $U_\TCL$ using one additional ancilla, yielding a Hermitian and unitary encoding. The following lemma establishes this construction and specifies the corresponding projection.

\begin{lemma}
\label{lem:hermitian}
Following the Hermitianization construction of Ref.~\cite[Lemma 10]{Low2019hamiltonian}, let $U$ be an $(\alpha,a,0)$ block encoding of a Hermitian $H$, let $h$ be one further qubit, and set
\begin{equation}
\widetilde U=\ket0\bra1_h\otimes U+\ket1\bra0_h\otimes U^\dagger.
\label{eq:hermitianized}
\end{equation}
Then $\widetilde U$ is both unitary and Hermitian, hence self-inverse, and
\begin{equation}
\left(\bra{+}_h\bra{0^a}\otimes I\right)\widetilde U
\left(\ket{+}_h\ket{0^a}\otimes I\right)=\frac{H}{\alpha}.
\label{eq:hermitianized-block}
\end{equation}
\end{lemma}

\begin{proof}
Taking the adjoint of \cref{eq:hermitianized} exchanges the two terms, so $\widetilde U^\dagger=\widetilde U$. In the product $\widetilde U\widetilde U^\dagger$, the two cross terms carry $\bra1\ket0$ or $\bra0\ket1$ on $h$ and vanish. The remaining terms give
\begin{equation*}
\begin{aligned}
\widetilde U\widetilde U^\dagger
&=\proj0_h\otimes UU^\dagger
+\proj1_h\otimes U^\dagger U\\
&=I.
\end{aligned}
\end{equation*}
Hence $\widetilde U$ is unitary, and being also Hermitian it satisfies $\widetilde U^2=I$. For the projected block, apply \cref{eq:hermitianized} to $\ket{+}_h$: the first term sends it to $\ket0_h\otimes U/\sqrt2$ and the second to $\ket1_h\otimes U^\dagger/\sqrt2$. Thus
\begin{equation*}
\bra{+}_h\widetilde U\ket{+}_h=\frac{U+U^\dagger}{2}.
\end{equation*}
Projecting the remaining $a$ ancillas onto $\ket{0^a}$ and using the block-encoding property of $U$ gives
\begin{equation*}
\frac{H+H^\dagger}{2\alpha}=\frac{H}{\alpha},
\end{equation*}
because $H$ is Hermitian.
\end{proof}

The same argument applies to an approximate block encoding. If the projected block of $U$ is
\begin{equation*}
H/\alpha+E\quad\text{with}\quad\norm{E}\le\delta_{\be},
\end{equation*}
then the block of $U^\dagger$ is $H/\alpha+E^\dagger$, and the argument above averages the two rather than adding them, so the Hermitianized encoding has error
\begin{equation*}
\norm{(E+E^\dagger)/2}\le\delta_{\be}.
\end{equation*}
The bound of \cref{lem:be} therefore applies to $\widetilde U$ with the same constant.

Since $\widetilde U$ is self-inverse, the walk operator,
\begin{equation*}
\begin{aligned}
W&=(2\Pi-I)\widetilde U,\\
\text{with }\Pi&=\proj{+}_h\otimes\proj{0^a}\otimes I,
\end{aligned}
\end{equation*}
satisfies $\cos\theta_\lambda=\lambda/\alpha_\TCL$ for eigenvalues $\lambda$ of the Hermitian effective Hamiltonian in \cref{eq:effective-hermitian-H}. For an exact encoding this is $H_\TCL$.

To see how the Hamiltonian eigenvalue becomes a walk phase, let $\ket{\psi_\lambda}$ be an eigenstate of $H_\TCL^{\prime}$ and define
\begin{equation*}
\ket{v_\lambda}=\ket{+}_h\ket{0^a}\ket{\psi_\lambda},
\qquad c_\lambda=\frac{\lambda}{\alpha_\TCL}.
\end{equation*}
For $|c_\lambda|<1$, set
\begin{equation*}
s_\lambda=\sqrt{1-c_\lambda^2},
\qquad
\ket{r_\lambda}=
\frac{(I-\Pi)\widetilde U\ket{v_\lambda}}{s_\lambda}.
\end{equation*}
The projected block gives $\widetilde U\ket{v_\lambda}=c_\lambda\ket{v_\lambda} +s_\lambda\ket{r_\lambda}$. Using $\widetilde U^2=I$ then shows that the span of these two states is invariant and that the walk acts there as
\begin{equation*}
\left.W\right|_{\operatorname{span}\{v_\lambda,r_\lambda\}}
=
\begin{pmatrix}
c_\lambda&s_\lambda\\
-s_\lambda&c_\lambda
\end{pmatrix}.
\end{equation*}
Its eigenvalues are $e^{\pm i\theta_\lambda}$ with $\cos\theta_\lambda=c_\lambda$. Thus the Hamiltonian eigenvalue is encoded in a rotation angle, even though a single walk generally leaves the projected ancilla subspace. At $c_\lambda=1$ or $-1$, $\ket{v_\lambda}$ itself is a walk eigenstate with eigenvalue $1$ or $-1$, so no complementary state is needed.

The $\ket{+}$ projection belongs to the Hermitianizing qubit $h$; the outer selection qubit $s$ is included among the ancillas projected onto zero. The factorization
\begin{equation}
\widetilde U=\left(\proj0_h\otimes U_\TCL
+\proj1_h\otimes U_\TCL^\dagger\right)(X_h\otimes I)
\label{eq:hermitianized-circuit}
\end{equation}
uses one controlled $U_\TCL$, one controlled $U_\TCL^\dagger$, and one Clifford gate. For an externally controlled walk, the external control must enable those operations and the reflection. The outer branch control $s$, Hermitianizing control $h$, and external walk control therefore have distinct roles.

The shared-angle circuit in Appendix \ref{app:angle-explicit} proves that its catalyst returns to its original state and remains unentangled with all other registers. We can therefore exclude it from the reflection on that invariant subspace. The construction below identifies the projected registers and gives a reflection circuit with an explicit scratch allocation. The alias construction in Appendix~\ref{app:alias-rounding} fixes the projected-register width for the retained padded-label layout.

\begin{figure*}
\centering
\input{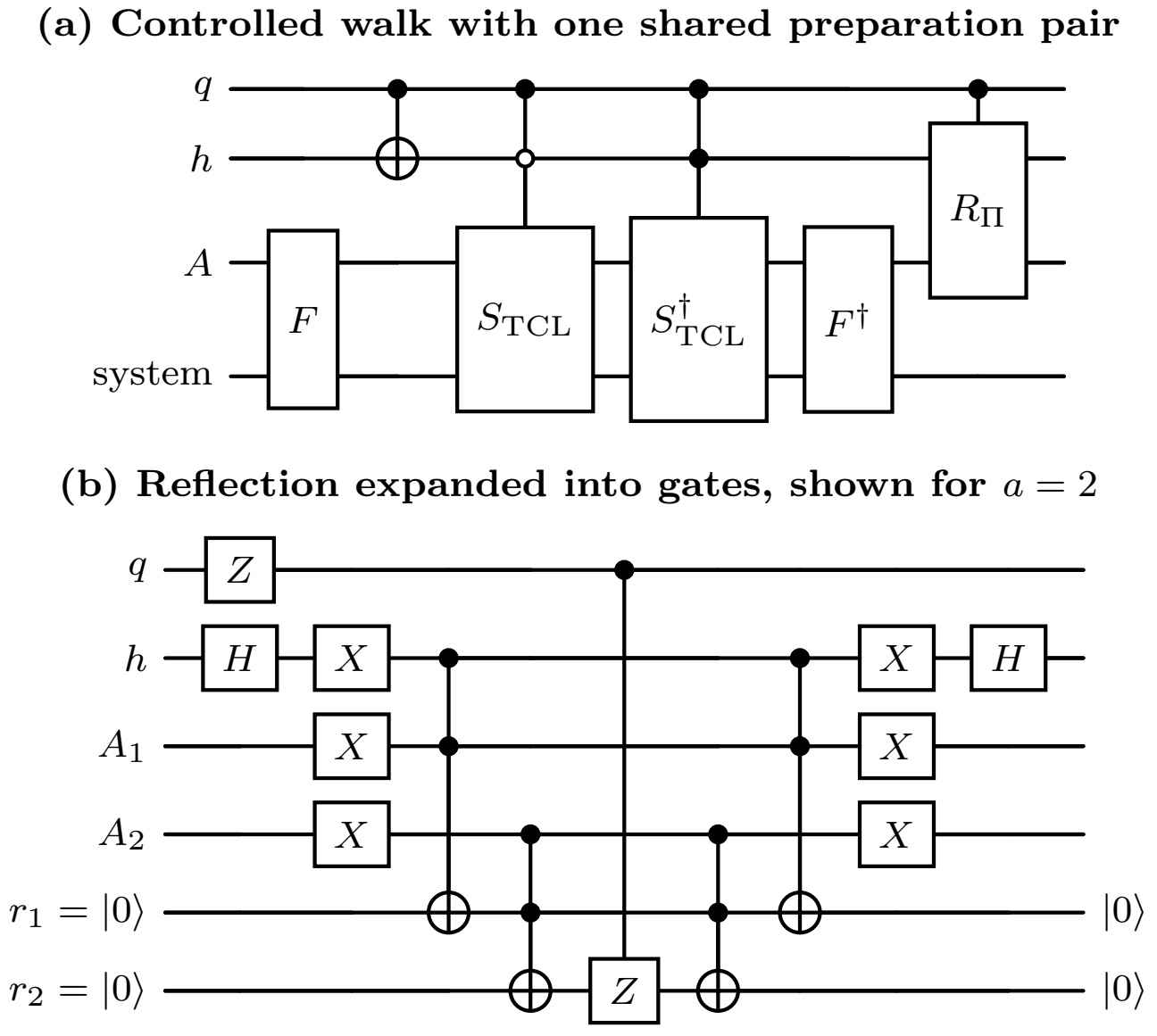}
\caption{ Controlled walk with shared preparation, read from left to right. (a) Apply $F$ before the controlled flip and the two selected branches in \cref{eq:shared-hermitianization}, then apply $F^\dagger$. The two selections depend on $q$ and on the value of $h$ after the flip. When $q=0$, preparation cancels with its inverse and the reflection acts as the identity, including its phase. The phase-gradient catalyst returns to its input state. (b) The reflection computes the joint zero test into clean scratch qubits, applies the phase, and reverses the computation. The $Z_q$ supplies the sign in \cref{eq:walk-reflection-sign}. The system, $h$, and projected register $A$ are not reset between queries.}
\label{fig:walk-explicit}
\end{figure*}

\paragraph{An explicit controlled walk.}
Having established the qubitized walk operator $W$, we now describe its controlled implementation, as required by the GQSP simulation circuit. The construction follows the Hermitianized encoding of \cref{lem:hermitian}, with the additional control qubit $q$ enabling the entire walk. \Cref{fig:walk-explicit}(a) shows the corresponding circuit. Let $S_\TCL$ denote the SELECT unitary of the TCL block encoding. Apply the shared preparation $F$ first, then the controlled flip of $h$. Using the value of $h$ after the flip, apply $S_\TCL$ when $(q,h)=(1,0)$ and $S_\TCL^\dagger$ when $(q,h)=(1,1)$. A single $F^\dagger$ and the controlled reflection $R_\Pi$ complete the walk. This sequence implements
\begin{equation}
C_q(W)=\proj{0}_q\otimes I+\proj{1}_q\otimes W.
\label{eq:controlled-walk-explicit}
\end{equation}
The adjoint is obtained by reversing the logical gate sequence and taking the adjoint of each operation, including the rounded angle lookups and synthesized rotations. It is not compiled independently using a second approximation. For measurement-assisted primitives, the inverse must implement the same logical adjoint, with the corresponding measurements and feed-forward corrections.

The control $s$ inside $U_\TCL$ chooses a Hamiltonian branch; $h$ chooses between an encoding and its adjoint; $q$ enables the whole walk. In particular, the initial $X_h$ must also obey $q$. The registers $h$ and $A$ can leave their projected states during a walk and are not reset between queries.

\paragraph{Controls inside the encoding.}
The local and hopping preparations act on separate label registers. Write
\begin{equation*}
F=\PREP_\hop\PREP_\loc P_\TCL,
\end{equation*}
and let $\SELECT_\hop$ denote the hopping operations between $\PREP_\hop$ and $\PREP_\hop^\dagger$ in \cref{fig:outer}. Then
\begin{equation}
\begin{aligned}
S_\TCL&=\proj0_s\SELECT_\loc+\proj1_s\SELECT_\hop,\\
U_\TCL&=F^\dagger S_\TCL F,\\
C_e(U_\TCL^{\pm1})&=F^\dagger C_e(S_\TCL^{\pm1})F.
\end{aligned}
\label{eq:controlled-encoding-factor}
\end{equation}

The encoding and adjoint branches use the same preparation $F$. We therefore need only one preparation and its inverse in the Hermitianized encoding:
\begin{equation}
\begin{aligned}
\widetilde U_\TCL
={}&(I_h\otimes F^\dagger)
\bigl(\ket0\bra1_h\otimes S_\TCL\\
&\qquad+\ket1\bra0_h\otimes S_\TCL^\dagger\bigr)
(I_h\otimes F).
\end{aligned}
\label{eq:shared-hermitianization}
\end{equation}
This follows by substituting $U_\TCL=F^\dagger S_\TCL F$ in \cref{eq:hermitianized}. To implement the controlled walk, apply $F$ unconditionally, then the controlled flip of $h$ and the two selected branches. Finish with $F^\dagger$ and the controlled reflection. When the walk is disabled, $F^\dagger F=I$. This cancellation remains exact when the preparation uses finite-precision tables, provided that unpreparation reverses the implemented gates.

Here $e$ is the conjunction of the controls that enable an encoding call. In the positive-control walk of \cref{fig:walk-explicit}, the two conditions are
\begin{equation*}
\begin{aligned}
U_\TCL:&\quad q=1,\quad h=0,\\
U_\TCL^\dagger:&\quad q=1,\quad h=1.
\end{aligned}
\end{equation*}
The value of $h$ is taken after the controlled flip. If the complete evolution has an external control $b_{\rm ext}$, add $b_{\rm ext}=1$ to both conditions. Within $S_\TCL$, the local and hopping operations also test $s=0$ and $s=1$, respectively.

The preparations may be unconditioned because they cancel on every inactive branch. The signs, rotations, shifts, and other state-changing selected gates still obey the full branch control. \Cref{fig:dynamics-explicit}(b) shows the forward call; the adjoint uses the exact reversed selected circuits.

\paragraph{The reflection and its scratch space.}
Let $A$ contain the ancillas projected onto zero in the definition of $U_\TCL$. For the separate label registers used in the explicit outer circuit, take
\begin{equation}
A=(s,x,\ell,\eta,\sigma_1,\sigma_2,J_\loc,J_\hop).
\label{eq:walk-projected-registers}
\end{equation}
The registers $J_\loc$ and $J_\hop$ retain the random integer, alternative label, threshold, and comparison flag of Appendix \ref{app:alias-rounding}. With both preparations present and the padded label layout retained, \cref{eq:alias-work-width} gives
\begin{equation}
\begin{aligned}
w_\loc&=p+2b_{\rm alias}+1,\\
w_\hop&=\ell_x+\ell_y+2+2b_{\rm alias}+1.
\end{aligned}
\label{eq:alias-register-widths}
\end{equation}
The number of qubits projected onto zero is therefore
\begin{equation}
\begin{aligned}
a&=p+\ell_x+\ell_y+6+w_\loc+w_\hop\\
&=2(p+\ell_x+\ell_y+2)+4b_{\rm alias}+6.
\end{aligned}
\label{eq:walk-projected-width}
\end{equation}
The Hermitianizing qubit $h$ is additional and is projected onto $\ket{+}$. The separate clean workspace used by address calculations, occupation loads, lookup words, enable masks, carries, and unary flags is excluded because these circuits return it to zero before the complete query ends. In the alias preparation, the lookup and comparison scratch is erased before the conditional swap, while the retained outputs remain in $J_\loc,J_\hop$. These return properties are also preserved by the exact circuit adjoint. \Cref{tab:walk-registers} records the resulting allocation. Removing a zero branch or simplifying a uniform preparation changes this allocation and must be counted separately.

There are $n=a+1$ reflected qubits when $h$ is included. After applying $H$ to $h$, define
\begin{equation}
\begin{aligned}
D_n&=2\proj{0^n}-I,\\
R_\Pi&=(H_h\otimes I_A)D_n(H_h\otimes I_A).
\end{aligned}
\label{eq:walk-zero-reflection}
\end{equation}
The identity on the system is implicit. The controlled zero-state reflection is
\begin{equation}
C_q(D_n)=Z_q\left(I-2\proj{1}_q\otimes\proj{0^n}\right).
\label{eq:walk-reflection-sign}
\end{equation}
The factor $Z_q$ is essential: omitting it changes $C_q(W)$ by a relative sign between the two values of $q$.

For $n\ge2$, the following reversible construction implements \cref{eq:walk-reflection-sign}. Apply $X$ to all $n$ reflected wires. Call their resulting computational bits $z_1,\ldots,z_n$, and initialize $n-1$ scratch qubits $r_j$ to zero. Compute
\begin{equation}
\begin{aligned}
r_1&=z_1z_2,\\
r_j&=r_{j-1}z_{j+1},
\end{aligned}
\label{eq:walk-and-chain}
\end{equation}
using the second recurrence for $2\le j\le n-1$ and one exact Toffoli for each step. Apply $CZ$ between $q$ and $r_{n-1}$, undo the Toffolis in reverse order, and undo the $X$ gates. Apply $Z_q$ and the two Hadamards on $h$ as in \cref{fig:walk-explicit}(b). Each computational component acquires phase $+1$ when $q=0$, and phase $+1$ or $-1$ when $q=1$ according to whether the reflected register passes or fails the projection. Linearity therefore gives the reflection on every superposition, with all scratch qubits returned to zero. The AND computation need not be controlled by $q$, because it is undone around the controlled phase.

This construction uses
\begin{equation*}
\begin{aligned}
\text{Exact Toffoli gates:}&\quad 2a,\\
\text{Clifford gates:}&\quad 1\,CZ,\ 1\,Z,\ 2(a+1)\,X,\ 2\,H,\\
\text{Additional clean scratch qubits:}&\quad a.
\end{aligned}
\end{equation*}
Here $X$ denotes a Pauli-$X$ gate and $H$ a Hadamard. For $a=0$, replace the AND chain by a $CZ$ directly on $q$ and the transformed $h$ wire. The scratch qubits are not part of $A$: their return is explicit. This is one concrete reflection implementation. Its scratch allocation and gate count are explicit, while the complete encoding calls still require their own control and preparation counts.

The phase-gradient register $G$ is also outside $A$. With the prescribed clean arithmetic workspace, \cref{eq:angle-catalyst-return} shows that the signed additions preserve its specified state for every angle, signal state, and control state. Each complete query and its adjoint therefore leave $G$ unentangled with the other registers on the invariant subspace used by the algorithm. If its preparation is approximate, that setup error must be included separately. Neither the catalyst nor the reflection scratch should be counted as a zero-projected signal register merely because it is auxiliary to the system.
\begin{table}[t]
\caption{Registers in the explicit controlled-walk construction. The projected register $A$ contains the registers listed in \cref{eq:walk-projected-registers}; its width is given in \cref{eq:walk-projected-width}. The coefficient-preparation widths $w_\loc,w_\hop$ are given in \cref{eq:alias-register-widths}. The clean arithmetic workspace is shared and excluded from the reflection because its return is proved by the explicit circuits.}
\label{tab:walk-registers}
\squeezetable
\setlength{\tabcolsep}{3.5pt}
\begin{ruledtabular}
\begin{tabular}{p{0.21\columnwidth}p{0.25\columnwidth}p{0.43\columnwidth}}
Register & Qubits & Role \\
\colrule
System & $L\qph+LN$ & Photon and emitter states \\
$q$ & $1$ & GQSP signal and walk control \\
$h$ & $1$ & Projected onto $\ket{+}$ \\
$s$ & $1$ & Local or hopping branch \\
$x$ & $p$ & Local term label \\
$\ell$ & $\ell_x+\ell_y+2$ & Directed hopping label \\
$\eta,\sigma_1,\sigma_2$ & $3$ & Block-encoding signals \\
$J_\loc$ & $w_\loc$ & Local preparation outputs \\
$J_\hop$ & $w_\hop$ & Hopping preparation outputs \\
$G$ & $b+1$ & Reused phase-gradient state \\
$r$ & $a$ & Clean reflection scratch \\
\end{tabular}
\end{ruledtabular}
\end{table}

\paragraph{Combining the controlled-query costs.}
For this count, let $\mathcal C_{\rm Tof}$ denote ordinary Toffoli gates in a fully coherent implementation. Write $cU_\TCL$ for either encoding call in \cref{fig:walk-explicit}(a), enabled jointly by $q$ and the specified value of $h$. Its cost includes those controls. The exact circuit adjoint has the same Toffoli count, and changing a control from zero to one requires only Pauli-$X$ gates. The controlled walk therefore uses
\begin{equation}
\mathcal C_{\rm Tof}(cW)
=2\mathcal C_{\rm Tof}(cU_\TCL)-C_F+2a.
\label{eq:walk-cost}
\end{equation}
The reflection requires $a$ clean scratch qubits in addition to its $a+1$ projected qubits. The initial controlled $X_h$ and the other reflection gates are Clifford operations.

The matched construction in Appendix \ref{app:matched-query} specifies $cU_\TCL$, the common preparation cost $C_F$, and their register allocation. Rotation synthesis is also a separate part of a complete logical resource estimate. In particular, the Toffoli-family ledger of \cref{tab:compare} cannot be inserted directly into this coherent count. A numerical direct/sparse comparison at the walk level requires both circuits to be compiled under the same convention.

\paragraph{From controlled walks to time evolution.}
The relation $W=(2\Pi-I)\widetilde U$ and the Hermiticity of $\widetilde U$ give
\begin{equation}
\frac{W+W^\dagger}{2}
=\Pi\widetilde U\Pi-(I-\Pi)\widetilde U(I-\Pi).
\label{eq:walk-cosine}
\end{equation}
Thus the $h=+$, $A=0$ subspace is invariant under $(W+W^\dagger)/2$, where its action is $H_\TCL^\prime/\alpha_\TCL$. For a walk eigenvalue $z$, the required time-evolution function is therefore $\exp[-i\alpha_\TCL\tau(z+z^{-1})/2]$, which we approximate using a truncated Jacobi--Anger expansion.

Let $z$ be a unit-modulus eigenvalue of $W$. For a Jacobi--Anger cutoff $K$, let $J_k$ be the Bessel function of the first kind and set
\begin{equation}
\begin{aligned}
r_K(z)&=\sum_{k=-K}^{K}(-i)^k J_k(\alpha_\TCL\tau)z^k,\\
\eta_K&=\sum_{|k|>K}\abs{J_k(\alpha_\TCL\tau)},\\
p_K(z)&=\frac{r_K(z)}{1+\eta_K}.
\end{aligned}
\label{eq:gqsp-laurent}
\end{equation}
The tail in $\eta_K$ sums over integer $k$. For $|z|=1$, this scaling gives
\begin{equation*}
\begin{aligned}
|p_K(z)|&\le1,\\
\abs{p_K(z)-\exp[-i\alpha_\TCL\tau(z+z^{-1})/2]}
&\le2\eta_K.
\end{aligned}
\end{equation*}
For the coherent error budget (derived in Appendix \ref{app:error}) above, choose $K$ so that
\begin{equation*}
2\eta_K\le\varepsilon^2/32.
\end{equation*}
Multiplying $p_K(z)$ by $z^K$ removes its negative powers, giving the ordinary polynomial
\begin{equation*}
P_K(z)=z^Kp_K(z)
\end{equation*}
of degree at most $2K$. The GQSP phase sequence for $P_K$ defines the rotations $V_0,\ldots,V_{2K}$ in \cref{fig:dynamics-explicit}(a). To recover $p_K(W)$, the circuit must also supply the compensating factor $W^{-K}$.

To see how this is achieved without additional queries, define $D_0(W)=\proj0_q\otimes W+\proj1_q\otimes I$. A walk-adjoint query controlled on $q=1$ satisfies
\[
C_q(W^\dagger)=(I_q\otimes W^\dagger)D_0(W).
\]
Since the unconditional $W^\dagger$ commutes with the signal rotations, the first $K$ such queries supply $W^{-K}$, while the resulting $2K$ factors of $D_0(W)$ implement $P_K(W)$. Thus, with the optional external control $b_{\rm ext}$ held at $1$ or omitted, the ideal GQSP circuit $U_{\rm poly}$ satisfies
\begin{equation}
\bra0_q U_{\rm poly}\ket0_q
=W^{-K}P_K(W)=p_K(W).
\label{eq:gqsp-projected-block}
\end{equation}
The query schedule is
\begin{equation*}
\begin{aligned}
\text{First }K\text{ queries:}&\quad W^\dagger\text{ controlled on }q=1,\\
\text{Next }K\text{ queries:}&\quad W\text{ controlled on }q=0.
\end{aligned}
\end{equation*}
The former is the exact adjoint of the circuit in \cref{fig:walk-explicit}(a), and the latter is obtained by conjugating its $q$ wire with $X$. Signal rotations follow every query, after an initial $V_0$. There are $2K$ controlled walk queries and $2K+1$ full single-qubit $SU(2)$ rotations.

With $b_{\rm ext}$ held at $1$ or omitted, the coherent Toffoli contribution of the queries is
\begin{equation*}
2K\mathcal C_{\rm Tof}(cW)
\end{equation*}
under \cref{eq:walk-cost}, before rotation synthesis and before adding controls for a varying $b_{\rm ext}$.

The angle values depend on $\tau$, $K$, and the requested error and are obtained by classical phase synthesis. Replacing ideal $V_j$ by synthesized gates gives $U_{\rm sim}$; their error is included in Appendix \ref{app:error}.

The cost in \cref{eq:walk-cost} assumes that the qubitized walk is controlled only by the internal GQSP signal qubit $q$. For a Hadamard test, an additional control qubit $b_{\rm ext}$ is required to enable the complete time-evolution circuit. This includes the controlled walks and the GQSP signal-processing rotations $V_0,\ldots,V_{2K}$ introduced in \cref{fig:dynamics-explicit}(a). The circuit must act as the identity when $b_{\rm ext}=0$, so all controlled operations must preserve the relative phase between the two branches. In particular, the synthesis of controlled $V_j$ rotations and the additional walk controls introduce costs not included in \cref{eq:walk-cost}. If the simulation uses the shifted Hamiltonian $\overline H_\TCL=H_\TCL-C_{\rm id}I$, the phase $e^{-iC_{\rm id}\tau}$ must also be restored for controlled evolution. This is achieved by applying $\operatorname{diag}(1,e^{-iC_{\rm id}\tau})$ to $b_{\rm ext}$.

\section{Proofs for the sampled open-system construction}
\label{app:open-proofs}

We now turn to the open-system simulation, where the Hamiltonian dynamics is supplemented by photon loss, atomic relaxation, and dephasing. As described in \cref{sec:open}, our approach samples individual Lindblad jump channels and implements each through a unitary dilation. Here we establish the corresponding block encoding, including its normalization, and derive the error and resource bounds for the resulting sampled evolution. These results support \cref{lem:open_normalization,thm:open} and follow the energy-shift convention of \cref{sec:open}.

\subsection{Proof of the sampled dilation block-encoding lemma}
\label{app:open-block-encoding}

\begin{proof}[Proof of \cref{lem:open_normalization}]
The two terms of \cref{eq:dilated-split} are block encoded separately and combined by a one-qubit linear combination, exactly as in \cref{sec:full}. The first term uses $U_\TCL$ and has normalization $\sqrt{\Delta t}\alpha_\TCL$. To encode the dilation projector, initialize an additional signal qubit in $\ket0$, apply $U_\TCL$ when the dilation qubit is $\ket0$, and flip the new signal when the dilation qubit is $\ket1$. Include this signal in the common zero-projected ancilla register, whose width is $a_{\rm dil}$. For the second, note that
\begin{equation}
\mathcal K_j=(X\otimes I)
\left(\proj{0}\otimes \widetilde{V}_j +\proj{1}\otimes \widetilde{V}_j ^\dagger\right),
\label{eq:M-factor}
\end{equation}
so given a block encoding $U_{\widetilde{V}_j }$ of $\widetilde{V}_j /\alpha_{\widetilde{V}_j }$ it suffices to apply $U_{\widetilde{V}_j }$ controlled on the dilation qubit being $\ket0$, $U_{\widetilde{V}_j }^\dagger$ controlled on $\ket1$, and then $X$ on the dilation qubit. This is a block encoding of $\mathcal K_j$ with the same normalization $\alpha_{\widetilde{V}_j }$, because $X\otimes I$ is unitary and acts outside the block-encoding ancillas.

The normalizations are tight for each channel of \cref{eq:jumps}. The truncated annihilation operator is block encoded with normalization $\sqrt{\nmax}=\norm{a_i}$ by the ladder construction of \cref{sec:loc}, $\sigma^-$ has Pauli one-norm, $\norm{\sigma^-}=1$, and $\sigma^z$ is unitary. In each case
\begin{equation*}
\alpha_{\widetilde{V}_j }=\norm{\widetilde{V}_j }=\sqrt{\lambda_{\mathcal D}}
\end{equation*}
by \cref{eq:equal-norm}, so each jump branch has the same normalization. Pad the Hamiltonian and jump encodings with identity ancillas to a common projected width. Prepare the outer branch qubit with probabilities
\begin{equation*}
\sqrt{\Delta t}\alpha_\TCL/\alpha_{\rm dil}\qquad\text{and}\qquad\sqrt{\lambda_{\mathcal D}}/\alpha_{\rm dil},
\end{equation*}
respectively, where $\alpha_{\rm dil}$ is their sum as in \cref{eq:alpha-dil}. Select the corresponding normalized block and undo this preparation. The projected block of the resulting encoding is
\[
\frac{\sqrt{\Delta t}\alpha_\TCL}{\alpha_{\rm dil}}
\frac{\proj0\otimes H_\TCL}{\alpha_\TCL}
+
\frac{\sqrt{\lambda_{\mathcal D}}}{\alpha_{\rm dil}}
\frac{\mathcal K_j}{\sqrt{\lambda_{\mathcal D}}}
=\frac{\widetilde H_j}{\alpha_{\rm dil}}.
\]
This proves the normalization and its independence from $j$. Approximation errors in these component encodings are included in the implementation-error allocation of \cref{thm:open}.
\end{proof}

\subsection{Proof of the timestep and arithmetic-cost theorem}
\label{app:open-cost}

\begin{proof}[Proof of \cref{thm:open}]
\textit{Sampling error.} We first quantify the error introduced by replacing the full dissipator with a randomly sampled jump channel. Write
\begin{equation*}
\begin{aligned}
\mathcal D=\sum_j\mathcal D_j,\qquad
\mathcal E_j=p_j^{-1}\mathcal D_j-\mathcal D,\qquad
\mathcal L^{\rm samp}_j&=\mathcal L+\mathcal E_j,\qquad
\sum_jp_j\mathcal E_j=0.
\end{aligned}
\end{equation*}
Applying Duhamel's identity twice gives the exact second-order remainder
\begin{equation}
\begin{aligned}
&\sum_jp_je^{\Delta t\mathcal L^{\rm samp}_j}
-e^{\Delta t\mathcal L}\\
&=\sum_jp_j\int_0^{\Delta t} ds\int_0^{\Delta t-s} dr 
e^{(\Delta t-s-r)\mathcal L^{\rm samp}_j}
\mathcal E_j e^{r\mathcal L}\mathcal E_j e^{s\mathcal L}.
\end{aligned}
\label{eq:samp-expansion}
\end{equation}
For a jump $V$,
\begin{equation*}
\norm{\mathcal D[V]}_\diamond\le2\norm{V}^2,\qquad\text{hence}\qquad\norm{\mathcal E_j}_\diamond\le4\lambda_{\mathcal D}.
\end{equation*}
Every exponential in \cref{eq:samp-expansion} is a completely positive trace-preserving map and has diamond norm one. Therefore,
\begin{equation}
\begin{aligned}
\norm{\sum_jp_je^{\Delta t\mathcal L^{\rm samp}_j}
-e^{\Delta t\mathcal L}}_\diamond
\quad\le\quad\frac{\Delta t^2}{2}\sum_jp_j\norm{\mathcal E_j}_\diamond^2
\le8\lambda_{\mathcal D}^2\Delta t^2.
\end{aligned}
\label{eq:samp-error}
\end{equation}
Independent samples at successive steps make the averaged $N_{\rm step}$-step channel the $N_{\rm step}$th power of the averaged one-step channel. A telescoping sum then bounds the accumulated sampling error by $8T\lambda_{\mathcal D}^2\Delta t$.

\textit{Dilation error.} The implemented one-step dilation must also approximate the sampled exponential. For $a=0,1$, its Kraus operators are
\[
K_{a,j}=(\bra a\otimes I)e^{-i\sqrt{\Delta t}\widetilde H_j}(\ket0\otimes I).
\]
For $\Delta t\Lambda\le1$, expanding the exponential in \cref{eq:dilated} gives
\begin{align*}
K_{0,j}&=I-i\Delta t H_\TCL
-\tfrac12\Delta t\widetilde{V}_j^\dagger\widetilde{V}_j
+O(\Delta t^2\Lambda^2),\\
K_{1,j}&=-i\sqrt{\Delta t}\widetilde{V}_j
+O(\Delta t^{3/2}\Lambda^{3/2}),
\end{align*}
where the remainders are in operator norm and uniform in $j$. The dilation channel $\sum_{a=0}^1K_{a,j}(\cdot)K_{a,j}^\dagger$ thus equals
\begin{equation*}
I+\Delta t\mathcal L^{\rm samp}_j+O_\diamond(\Delta t^2\Lambda^2).
\end{equation*}
Comparing with the exponential and telescoping gives $O(T\Lambda^2\Delta t)$ total dilation error, consistent with the first-order construction of \cite{PRXQuantum.5.020332}.

Since $\lambda_{\mathcal D}\le\Lambda$, the accumulated sampling and dilation errors are bounded by $O(T\Lambda^2\Delta t)$. The timestep choice in \cref{eq:timesteps}, with a sufficiently large constant $c$, assigns a fixed fraction of $\varepsilon$ to these errors and ensures $\Delta t\Lambda\le1$.

\textit{Implementation error.} As in the closed-system allocation, a polynomial projected-block error at most $\varepsilon_{\rm step}^2/32$ and full-circuit synthesis error at most $\varepsilon_{\rm step}/4$ suffice for their respective error budgets.

The accuracy $\varepsilon_{\rm step}$ includes block-encoding, signal-processing, and gate-synthesis errors. If the Hermitian effective dilation Hamiltonian differs from $\widetilde H_j$ by at most $\xi$ in operator norm, Duhamel's inequality bounds the induced unitary error by $\sqrt{\Delta t}\xi$. \begin{samepage} Choose
\begin{equation*}
\sqrt{\Delta t}\xi\le\varepsilon_{\rm step}/2
\end{equation*}
and allocate the remaining budget to polynomial approximation and gate synthesis using the tolerances specified above.
\end{samepage}
With these allocations, the initialized-state isometries of the exact and implemented one-step dilations differ by at most $\varepsilon_{\rm step}$ in operator norm. Denote them by $\mathsf V_j$ and $\widehat{\mathsf V}_j$, placing the exact output in the same space by adjoining clean simulation ancillas. For any input state $\rho$ on the system and a reference, the identity on the reference is implicit, and
\begin{align*}
\widehat{\mathsf V}_j\rho\widehat{\mathsf V}_j^\dagger
-\mathsf V_j\rho\mathsf V_j^\dagger
&=(\widehat{\mathsf V}_j-\mathsf V_j)
\rho\widehat{\mathsf V}_j^\dagger
+\mathsf V_j\rho
(\widehat{\mathsf V}_j^\dagger-\mathsf V_j^\dagger),\\
\bigl\|\widehat{\mathsf V}_j\rho\widehat{\mathsf V}_j^\dagger
-\mathsf V_j\rho\mathsf V_j^\dagger\bigr\|_1
&\le2\|\widehat{\mathsf V}_j-\mathsf V_j\|
\le2\varepsilon_{\rm step}.
\end{align*}
Here both isometries have operator norm one, and tensoring their difference with the reference identity preserves its norm. Partial trace cannot increase the trace norm of this Hermitian difference. Taking the supremum over input states and reference systems therefore gives at most $2\varepsilon_{\rm step}$ diamond-norm error for the implemented dilation channel. Averaging over $j$ preserves this uniform bound, and telescoping over timesteps gives $O(N_{\rm step}\varepsilon_{\rm step})$ accumulated implementation error. Choose the constant in $\varepsilon_{\rm step}=\Theta(\varepsilon/N_{\rm step})$ sufficiently small to fit the remaining channel-error budget.

\textit{Degree and arithmetic cost.} The evolution time for $\widetilde H_j$ is $\sqrt{\Delta t}$, so \cref{eq:alpha-dil} gives the normalized scale in \cref{eq:x-step}. Apply the uniform degree bound of \cref{eq:degree} at polynomial projected-block tolerance $\varepsilon_{\rm step}^2/32$. Squaring the precision target changes its logarithm by a constant factor, so \cref{eq:d-step} follows. Each timestep uses $O(d_{\rm step})$ fully coherent controlled signal queries. Multiplication by $N_{\rm step}$ gives \cref{eq:G-open}; synthesis of the arbitrary signal-processing rotations is additional.

To relate this bound to the total evolution time, note that
\begin{equation*}
N_{\rm step}x_{\rm step}=\alpha_\TCL T+\sqrt{\lambda_{\mathcal D}T N_{\rm step}}.
\end{equation*}
Since $d_0(x,\epsilon)\le1+x+\log(1/\epsilon)$ for the stated domain, substitution of \cref{eq:timesteps} also recovers the expanded bound
\[
G_{\rm open}=\widetilde O\!\left[
\left(1+\alpha_\TCL T
+\sqrt{\lambda_{\mathcal D}}T^{3/2}\Lambda\varepsilon^{-1/2}
+T^2\Lambda^2\varepsilon^{-1}\right)C_{\rm qry}^{\rm samp}
\right],
\]
where $\widetilde O$ suppresses polylogarithmic precision factors. Here the bounded-step contribution $\sqrt{\lambda_{\mathcal D}T}$ is absorbed into $O(1+T^2\Lambda^2/\varepsilon)$, using $\lambda_{\mathcal D}\le\Lambda$. Combining the sampling, dilation, and implementation budgets establishes the stated diamond-norm accuracy for the averaged channel.
\end{proof}

\section{A matched coherent controlled-query construction}
\label{app:matched-query}

This appendix provides an explicit circuit construction for the controlled encoding $cU_\TCL$, whose cost was left symbolic in \cref{eq:walk-cost}. We derive its resource counts
from the circuit components and check the resulting encodings and controlled walks through executable Qualtran decompositions; we do not claim that these constructions are globally optimal. Throughout, we count ordinary CCX gates with coherent uncomputation, treating Clifford gates and arbitrary-rotation synthesis separately. Both Hamiltonian branches, the padded local labels, and the general alias preparations are retained. As in \cref{sec:model}, we assume $L_x,L_y\geq4$ and that $N$ is a power of two. The numerical examples use uniform positive hopping, $g_i=\nu_i=1$, and $\Delta=0$, and simulate $\widehat H_\TCL=H_\hop+H_\intt$. Zero-weight diagonal labels and the original padded widths are retained. To ensure a consistent comparison, the direct-register and sparse-access constructions use identical controls, coefficient tables, precision, and clean-workspace conventions.

We use the notation
\begin{align*}
q&=\qph,&R&=2^q,\\
B&=b_{\rm alias},&\ell&=\log_2 L,\\
E&=LN,&M_\loc&=1+L(q+2N),\\
p&=\lceil\log_2 M_\loc\rceil,&D_\loc&=2^p,\\
D_\hop&=4L,&d_{\rm ex}&=p-\ell-\log_2N.
\end{align*}
We first establish the coherent costs of the lookup, alias preparation, and register-access primitives used in the two constructions. The shift formulas below assume $q\geq2$, with the $q=1$ case treated separately.

\paragraph{QROM and coherent alias preparation.}
Consider a QROM with $D=2^d\geq2$ entries containing classical output words. We implement the lookup using a binary-tree unary traversal, with the root flag fixed to $1$. The remaining $D-2$ internal flags are computed and erased coherently using CCX gates, while the selected classical output bits are written using CNOTs. This gives
\begin{equation}
Q(D)=2(D-2).
\label{eq:v3-qrom}
\end{equation}

The coherent alias preparation of Appendix~\ref{app:alias-rounding} uses Hadamard gates to initialize the column and random-integer registers. The QROM then loads the alternative label and threshold, a reversible comparator computes $v=[r\geq T]$, and $d$ Fredkin gates conditionally swap the two $d$-bit labels. Each Fredkin contributes one CCX and two CNOTs.

For $B$-bit registers $r$ and $T$, we use a reversible lexicographic comparator. Equality-prefix flags identify the first differing bit, at which the condition $[r_i=1,T_i=0]$ contributes to the comparison result. The all-equal flag is also XORed into $v$ before the prefix flags are erased. The resulting CCX count is
\[
\underbrace{1}_{\text{highest bit}}
+\underbrace{3(B-1)}_{\text{lower terms}}
+\underbrace{2(B-1)}_{\substack{\text{equality prefixes}\\
\text{and their erasure}}}
=5B-4.
\]
All comparison scratch is returned to zero, while $v$ is retained. Thus, one alias preparation and its exact adjoint have costs
\begin{equation}
\begin{aligned}
P(D)&=Q(D)+(5B-4)+\log_2D,\\
C_{\rm Tof}(\PREP^\dagger)&=P(D).
\end{aligned}
\label{eq:v3-alias-prep}
\end{equation}
As in \cref{eq:alias-work-width}, the alternative label, threshold, random integer, and comparison flag remain in the projected alias output $J$ until the adjoint preparation. We retain the same endpoint convention for $T=2^B$.

\paragraph{Indexed access and selected photon updates.}
Unlike a classical QROM lookup, reading a selected quantum photon occupation requires an additional control on each data bit. An XOR read of a $q$-bit occupation from one of $L$ cavity registers therefore costs
\begin{equation}
r_{\ph}=Q(L)+Lq,
\label{eq:v3-quantum-read}
\end{equation}
where $Lq$ counts one CCX per leaf and data bit. The inverse read has the same cost.

For emitter access, a Fredkin tree routes the selected bit from the $E=LN$ emitter qubits to a designated location. After copying it with a CNOT, the routing is reversed, giving $2(E-1)$ CCXs per read. An enabled indexed emitter flip likewise costs $2(E-1)$ CCXs.

The coherent selected-site shift constructed in Appendix~\ref{app:shift-explicit} combines unary address selection with a controlled increment or decrement at each leaf. Its cost per enabled update is
\begin{equation}
S_1=2(L-1)+L(2q-3),
\label{eq:v3-one-shift}
\end{equation}
for $q\geq2$. The first term accounts for coherent address traversal, while the second counts the carry- or borrow-based arithmetic at the $L$ leaves. For $q=1$, the arithmetic reduces to a CNOT, so $S_1=2(L-1)$.

\paragraph{Angle rotations and destination calculation.}
The occupation-dependent rotations use the shared-angle construction of Appendix~\ref{app:angle-explicit}. Each enabled rotation consists of a classical angle-table lookup and its inverse, a $b$-bit enable mask and its inverse, and coherent signed addition using the phase-gradient state. The resulting cost is
\begin{equation}
A_b=2Q(R)+2b+2(b-1)=4(R-2)+4b-2.
\label{eq:v3-angle-module}
\end{equation}
Since the angle-table words are classical, their leaf writes require only Clifford gates. The phase-gradient catalyst is prepared once per run and is excluded from these per-query costs.

Finally, the destination-address computation evaluates $i=j+d$ for a selected hopping direction. We first copy $j$ into a clean destination register, then conditionally complement the coordinate fields for negative directions and perform an axis-enabled modular increment on each coordinate. Complementation converts the increment into a decrement without additional Toffoli gates. For $\ell_x,\ell_y\geq2$, the destination computation costs
\begin{equation}
D_{\rm dest}=2(\ell_x+\ell_y)-6=2\ell-6,
\label{eq:v3-destination}
\end{equation}
with the same cost for its inverse.

These primitive costs are combined below to obtain the matched coherent controlled-encoding resource estimates for the direct-register and sparse-access constructions.

\paragraph{Direct selected branches.}
We arrange the local labels with the $E$ exchange terms first, at $0,\ldots,E-1$, followed by the $E$ emitter-$Z$ terms, the $Lq$ photon-$Z$ terms, one identity term, and padding. This ordering preserves the coefficients and padded width $p$, while placing the site and emitter addresses of each exchange term in the lowest $\ell+\log_2 N$ bits.

An enabled unary traversal over the $D_\loc$ labels costs $2(D_\loc-1)$ CCXs, with coefficient signs and diagonal Pauli operators implemented by Clifford $Z$ or $CZ$ gates at the leaves. The exchange-group enable is computed from the remaining $d_{\rm ex}$ high bits and subsequently erased, costing $2d_{\rm ex}$ CCXs.

For an exchange term, we first read the selected emitter bit and compute the enables for the excited-state photon increment and ground-state decrement. Computing and erasing these enables costs four CCXs. Between the two conditional shifts, we load the updated photon occupation, apply the shared angle module, and uncompute the load. We then flip the selected emitter, correct the retained emitter sample by a CNOT controlled on the exchange enable, and erase the sample. This restores all workspace for either emitter state and for inactive exchange labels. The resulting cost is
\begin{equation}
C_\loc=2(D_\loc-1)+2d_{\rm ex}+6(E-1)+4
          +2r_{\ph}+A_b+2S_1.
\label{eq:v3-local-select}
\end{equation}
Here $6(E-1)$ accounts for two emitter reads and one enabled emitter flip.

The hopping branch computes and uncomputes the destination address, loads and erases both selected photon occupations, applies the two factorized angle modules, and performs the source decrement and destination increment. For the uniform negative hopping coefficients used in the numerical examples, the sign is implemented by a Clifford $Z$ gate. Thus,
\begin{equation}
C_\hop=2D_{\rm dest}+4r_{\ph}+2A_b+2S_1.
\label{eq:v3-hop-select}
\end{equation}
For general bond-sign patterns, an enabled unary sign traversal adds at most $2(D_\hop-1)$ CCXs per encoding. The destination address is computed only once and reused throughout the hopping branch.

To implement either controlled encoding call in \cref{fig:walk-explicit}, we introduce the enables
\begin{equation*}
\begin{aligned}
e&=q_{\rm GQSP}\wedge[h=h_0],\\
e_\loc&=e\wedge(1-s),\\
e_\hop&=e\wedge s.
\end{aligned}
\end{equation*}
These require six CCXs for computation and uncomputation. The preparations need not be controlled, since $C_e(U_\TCL)=F^\dagger C_e(S_\TCL)F$ from \cref{eq:controlled-encoding-factor}, and the preparation and its exact adjoint cancel when the selection is inactive. The complete encoding is enabled when $q_{\rm GQSP}=1$ and $h=h_0$, with $h_0=0$ for $U_\TCL$ and $h_0=1$ for $U_\TCL^\dagger$, evaluated after the controlled flip of $h$. Its Toffoli count is
\begin{equation}
\begin{aligned}
C_{\rm dir}&=C_{\rm Tof}(cU_{\rm dir})\\
&=2[P(D_\loc)+P(D_\hop)]
  +C_\loc+C_\hop+6.
\end{aligned}
\label{eq:v3-direct-cu}
\end{equation}
The two outer arbitrary rotations $P_\TCL$ are excluded from this count. The combined Toffoli cost of the shared preparation $F$ and its inverse $F^\dagger$ is $C_F=2[P(D_{\rm loc})+P(D_{\rm hop})]$. Sharing these operations gives Eq.~(E13). The exact gate-reversed adjoint has the same Toffoli cost.

\paragraph{Matched two-register row maps.}
We now construct a matched two-register alternative to the direct-register implementation. Let $x$ denote the $P_\ph=Lq$ photon-system bits and $y$ an additional clean $P_\ph$-bit photon register. Write $S_{\ell,x}$ and $S_{\ell,y}$ for the same selected modulo-$R$ source/destination permutation acting on either register, and let $\mathrm{CX}_{x\to y}$ denote the bitwise CNOT bank. Using the same hopping enable $e$ as above, define the individually controlled unitary extensions
\begin{equation}
\begin{aligned}
O_F^{(e)}&=C_e(S_{\ell,y})C_e(\mathrm{CX}_{x\to y}),\\
O_R^{(e)}&=C_e(\mathrm{CX}_{y\to x})C_e(S_{\ell,x}),\\
T_{\ell}^{\rm row}&=C_e(\mathrm{SWAP}_{x,y})O_R^{(e)}O_F^{(e)}.
\end{aligned}
\label{eq:v3-row-maps}
\end{equation}
With the rightmost factor acting first, the sequence on the initialized subspace is
\begin{equation}
\begin{aligned}
\ket{x,0}&\xrightarrow{O_F^{(e)}}\ket{x,S_\ell x}\\
&\xrightarrow{O_R^{(e)}}\ket{0,S_\ell x}\\
&\xrightarrow{C_e(\mathrm{SWAP})}\ket{S_\ell x,0},
\qquad(e=1).
\end{aligned}
\label{eq:v3-row-action}
\end{equation}
When $e=0$, every factor acts as the identity for arbitrary $x,y$. These maps are reversible on the full binary space, including invalid photon labels, while the $A$ and $B$ coefficient tables assign zero projected amplitude to forbidden physical transitions. They provide unitary extensions of the initialized-subspace forward and reverse maps in Appendix~\ref{app:ledger}, but need not reproduce the standard XOR row-oracle action for arbitrary $y$.

A controlled CNOT bank costs $P_\ph$ CCXs, while a controlled SWAP bank costs $P_\ph$ CCXs together with Clifford CNOTs. Each of $O_F^{(e)}$ and $O_R^{(e)}$ uses one enabled pair of selected shifts, costing $S=2S_1$. Since the remaining selection and preparation components are identical to those of the direct construction, the resulting costs are
\begin{equation}
\begin{aligned}
C_{\rm row}&=C_{\rm dir}+S+3P_\ph,\\
C_{\rm Tof}(cW_{\rm dir})&=2C_{\rm dir}-C_F+2a,\\
C_{\rm Tof}(cW_{\rm row})&=2C_{\rm row}-C_F+2a.
\end{aligned}
\label{eq:v3-matched-walk}
\end{equation}
Both the completed encoding and its exact adjoint preserve the $y=0$ subspace. Hence, the additional register $y$ need not be included in the projected ancillas $A$ or the reflection, although it must be allocated during the two-register update.

For externally controlled walks, we compute
\begin{equation*}
q_{\rm eff}=q_{\rm GQSP}\wedge b_{\rm ext},
\end{equation*}
use this effective control for the complete walk, and subsequently erase it. This requires two additional CCXs and one clean qubit per externally controlled walk. Both the initial flip of $h$ and the reflection respect $q_{\rm eff}$, ensuring that the inactive branch is the identity. For the $W$ queries controlled on $q_{\rm GQSP}=0$ in Appendix~\ref{app:walk}, Clifford $X$ gates conjugate the signal qubit before and after computing the effective control, without changing the Toffoli count.

The row-map construction is independent of the measurement-assisted $O_F/O_R$ costs in \cref{tab:ledger}. It uses coherent uncomputation, shares the destination address, and avoids repeated occupation loads within the individual row maps.

A cheaper paired two-register construction implements the same initialized-subspace transformation while remaining the identity when disabled:
\begin{equation}
\begin{aligned}
T_{\ell}^{\rm pair}
&=\mathrm{CX}_{x\to y}\,
C_e(S_{\ell,x}S_{\ell,y})\,
\mathrm{CX}_{x\to y},\\
C_{\rm pair}&=C_{\rm dir}+S.
\end{aligned}
\label{eq:v3-paired-variant}
\end{equation}
For $e=1$, its action on the initialized registers is
\begin{equation*}
\begin{aligned}
\ket{x,0}
&\xrightarrow{\mathrm{CX}_{x\to y}}\ket{x,x}\\
&\xrightarrow{S_{\ell,x}S_{\ell,y}}
\ket{S_\ell x,S_\ell x}\\
&\xrightarrow{\mathrm{CX}_{x\to y}}\ket{S_\ell x,0}.
\end{aligned}
\end{equation*}
The final CNOT bank clears $y$ because both registers contain the same bit string. When $e=0$, the middle operation is the identity and the two unconditional CNOT banks cancel for every input, including $y\ne0$. These banks require only Clifford gates, leaving the two selected shifts as the additional CCX cost. Unlike \cref{eq:v3-row-maps}, however, this combined update does not provide separately controlled forward and reverse row maps.

A third, stronger convention requires the standard XOR row-oracle action for arbitrary $y$. Conjugating a controlled CNOT bank by an unconditional $S_\ell$ or $S_\ell^{-1}$ in each oracle, followed by a controlled SWAP, gives a selected cost of $4(S-4)+3P_\ph$ for $L\geq2$. This convention is not used in \cref{tab:v3-coherent-compare}. The resource savings therefore depend on the chosen two-register implementation.

\paragraph{A concrete qubit allocation.}
The common projected width is the $a$ of \cref{eq:walk-projected-width}.  Let
\begin{equation*}
c_{\rm dest}=\max(\ell_x-2,\ell_y-2)\qquad\text{and}\qquad c_{\rm shift}=\max(q-2,0).
\end{equation*}
We use a conservative allocation of clean arithmetic workspace, with width
\begin{equation}
w=d_{\rm ex}+\ell+c_{\rm dest}+q+\max(p,q)
   +c_{\rm shift}+3b+2B+4.
\label{eq:v3-work-bank}
\end{equation}
It allocates the enable chain, destination and carries, one reusable occupation word, unary flags, selected-shift carries, the angle word/mask and carries, one emitter sample, and alias-comparator scratch.  This workspace returns to zero before the walk reflection, allowing its qubits to be reused for the $a$ reflection scratch qubits.  One achievable, unoptimized peak allocation is
\begin{equation}
\begin{aligned}
\mathsf Q_{\rm dir}&=L(q+N)+2+a+(b+1)+\max(a,w),\\
\mathsf Q_{\rm row}&=\mathsf Q_{\rm dir}+P_\ph.
\end{aligned}
\label{eq:v3-peak-width}
\end{equation}
where the $+2$ term counts the GQSP signal and Hermitianizing qubit.  The second equality allocates separate physical wires to $y$. Reusing wires after their registers are no longer needed may yield a different peak width. The displayed widths describe achievable allocations and do not establish lower bounds.

\subsection{A logical \texorpdfstring{$T$}{T} count for controlled evolution}
\label{app:matched-t}

An exact coherent CCX has a standard seven-$T$ decomposition, counting $T$ and $T^\dagger$ together~\cite{amy2013meet,selinger2013tdepth}.  We use a second reusable phase-gradient catalyst to implement the classically specified outer and GQSP rotations.  Let $\ket{G_{m+1}}$ be the state in \cref{eq:gradient-state-explicit} with $b$ replaced by $m$.  For a nearest-grid integer $k$ obeying the rounding condition, write the quotient and remainder as
\begin{equation*}
\begin{aligned}
|\theta-2\pi k/2^m|&\leq\pi/2^m,\\
k&=2^mw+r,\qquad 0\leq r<2^m.
\end{aligned}
\end{equation*}
Set clean angle bits to $a_i=e r_i$ by CNOT from the enable $e$, run the signed adder of Appendix~\ref{app:angle-explicit}, and erase these bits by CNOT.  The result is
\begin{equation*}
C_e[R_z(2\pi r/2^m)],
\end{equation*}
exactly the identity for $e=0$, at $2(m-1)$ CCXs; Clifford conjugation gives $R_y$. For odd $w$, apply $Z_e$ because
\begin{equation*}
R_z(\theta+2\pi)=-R_z(\theta).
\end{equation*}
This Clifford correction retains the relative branch phase in a Hadamard test.  The GQSP phase list must describe full $SU(2)$ matrices; any extra $U(2)$ phase must also be controlled. After every use, the catalyst returns to its original state and remains unentangled with the other registers.

For the chosen frame Hamiltonian, let $\alpha=\alpha_{\rm dyn}$ and take $b,B$ from \cref{eq:angle-precision,eq:precision-choice} using the shifted coefficients.  Choose a Jacobi--Anger degree $K$ with $2\eta_K\leq\varepsilon_{\rm sim}^2/32$, and set $n_K=2K+1$. A sufficient grid width for the outer selection and all three Euler factors of each $V_j$ is
\begin{equation}
m\geq\max\left\{1,
\left\lceil\log_2\frac{4\pi|\tau|\alpha}
{\varepsilon_{\rm sim}}\right\rceil,
\left\lceil\log_2\frac{6\pi n_K}
{\varepsilon_{\rm sim}}\right\rceil\right\}.
\label{eq:v3-fixed-angle-grid}
\end{equation}
Each rounded axial rotation has operator-norm error at most $\pi/2^{m+1}$. Write the Euler decomposition as
\begin{equation*}
V_j=R_z(\phi_{j1})R_y(\phi_{j2})R_z(\phi_{j3})\in SU(2).
\end{equation*}
The two outer rotations have propagated error at most the first bound below, while the $3n_K$ controlled axial factors have circuit error at most the second:
\begin{equation*}
\begin{aligned}
\frac{2|\tau|\alpha\pi}{2^{m+1}}
&\le\frac{\varepsilon_{\rm sim}}{4},\\
\frac{3n_K\pi}{2^{m+1}}
&\le\frac{\varepsilon_{\rm sim}}{4}.
\end{aligned}
\end{equation*}
The angle grids, alias thresholds, and polynomial approximation use the other allocations of Appendix~\ref{app:error}. Classical phase-synthesis error must fit within these axial-angle bounds or receive its own budget.

Each controlled walk has an arithmetic cost $C_{W,X}=2C_X-C_F+2a$, excluding the two outer selection rotations. Since GQSP requires $2K$ walk queries, these rotations contribute $8K(m-1)$ Toffoli gates. The $n_K=2K+1$ GQSP signal rotations each decompose into three Euler rotations, contributing a further $6n_K(m-1)$ Toffoli gates. Thus, for $X\in\{\mathrm{dir},\mathrm{row},\mathrm{pair}\}$,
\begin{align}
C_{U,X}(\tau)
&=2KC_{W,X}+(8K+6n_K)(m-1),\\
T_{U,X}(\tau)
&=7C_{U,X}(\tau).
\end{align}

Here, $U$ denotes evolution without an external Hadamard-test control. For externally controlled evolution, the GQSP signal rotations are enabled using the same phase-gradient construction, without increasing their Toffoli counts. Each walk additionally requires computing and uncomputing $q_{\mathrm{eff}}=q_{\mathrm{GQSP}}\land b_{\mathrm{ext}}$, at a cost of two Toffoli gates per query. Consequently,
\begin{equation}
T_{\mathrm{HT},X}(\tau)=T_{U,X}(\tau)+28K.
\end{equation}
The common-evolution interferometer requires no external control on GQSP and therefore uses $T_{U,X}(\tau)$.

Constant-angle arithmetic can simplify this specified construction, so its counts are upper bounds rather than optimal $T$ costs.  A separate $\ket{G_{m+1}}$ occupies $m+1$ persistent qubits; the fixed-angle adder uses at most $2m-1$ clean bits that return to zero and can share the bank in \cref{eq:v3-work-bank} when it is large enough.  One external Hadamard control and its conjunction bit add two more qubits.  Neither catalyst nor this workspace is included in the zero-projected register $A$.

For all three examples below, $2m-1\leq w$, so the fixed-angle adder can reuse the existing workspace. The peak allocations are
\begin{equation*}
\mathsf Q_{U,X}=\mathsf Q_X+m+1,\qquad
\mathsf Q_{{\rm HT},X}=\mathsf Q_X+m+3 .
\end{equation*}
The common-evolution correlator needs one additional Hadamard control and two separate ladder signal qubits. The clean workspace is also large enough for the ladder arithmetic, giving a peak of $\mathsf Q_{U,X}+3$.

\paragraph{Polynomial degree.}
For clarity, a sufficient $K$ can be certified without numerically summing the Bessel tail.  With $x=\alpha|\tau|>0$, the Fourier integral for $J_k(x)$ may be shifted by $iy$; periodicity cancels the vertical edges and gives
\begin{equation*}
|J_k(x)|\leq\exp[-ky+x\sinh y].
\end{equation*}
For $k>x$, optimize at $y=\operatorname{arcosh}(k/x)$ to obtain
\begin{equation}
\begin{aligned}
f_x(k)&=-k\operatorname{arcosh}(k/x)+\sqrt{k^2-x^2},\\
|J_k(x)|&\leq e^{f_x(k)}.
\end{aligned}
\label{eq:v3-bessel-single}
\end{equation}
Since $f_x'(k)=-\operatorname{arcosh}(k/x)$ decreases, $f_x$ is concave. For $K+1>x$, set
\begin{equation*}
y_K=\operatorname{arcosh}[(K+1)/x].
\end{equation*}
The tangent at $K+1$ then gives, for every integer $r\ge0$,
\begin{equation*}
f_x(K+1+r)\le f_x(K+1)-r y_K.
\end{equation*}
Hence the positive-order tail is bounded by a geometric series:
\begin{equation*}
\sum_{k=K+1}^\infty |J_k(x)|
\le e^{f_x(K+1)}\sum_{r=0}^\infty e^{-r y_K}
=\frac{e^{f_x(K+1)}}{1-e^{-y_K}}.
\end{equation*}
Since $|J_{-k}(x)|=|J_k(x)|$, the two-sided tail has twice this bound. The certificate for $2\eta_K$ therefore has a factor of four:
\begin{equation}
2\eta_K=4\sum_{k=K+1}^{\infty}|J_k(x)|
\leq\frac{4e^{f_x(K+1)}}{1-e^{-y_K}}.
\label{eq:v3-bessel-tail}
\end{equation}
The degrees listed in \cref{tab:v3-T-compare} make the bound in \cref{eq:v3-bessel-tail} smaller than $\varepsilon_{\rm sim}^2/32$. These degrees suffice, but we do not claim that they are minimal.

\begin{table*}[t]
\caption{One evolution of $\widehat H_\TCL$ without an external Hadamard control at $|\tau|=1$, $\varepsilon_{\rm sim}=10^{-3}$ and the uniform parameters in \cref{tab:v3-coherent-compare}. The degrees satisfy \cref{eq:v3-bessel-tail}; $m$ satisfies \cref{eq:v3-fixed-angle-grid}. $T_U$ includes shared walks, outer selection rotations, and all GQSP Euler rotations using seven-$T$ CCXs. The totals exclude input-state and catalyst preparation, ladder insertions, and measurement repetitions. The qubit counts include the second catalyst. The correlator circuit adds three qubits. The external-control variant adds $28K$ $T$ gates.}
\label{tab:v3-T-compare}
\squeezetable
\begin{ruledtabular}
\begin{tabular}{c r r r r r r r}
$(L,N,\nmax)$ & $K$ & $m$ & bound on $2\eta_K$ &
$T_{U,\rm dir}$ & $T_{U,\rm pair}$ &
$\mathsf Q_{U,\rm dir}$ & $\mathsf Q_{U,\rm pair}$\\
\colrule
$(16,1,3)$ & 62 & 22 & $2.0870\times10^{-8}$ & $3{,}386{,}082$ & $3{,}545{,}794$ & 320 & 352 \\
$(64,2,7)$ & 485 & 25 & $2.3636\times10^{-8}$ & $99{,}474{,}508$ & $108{,}111{,}388$ & 639 & 831 \\
$(256,4,15)$ & 4863 & 28 & $2.9338\times10^{-8}$ & $5{,}718{,}208{,}314$ & $6{,}205{,}675{,}434$ & 2428 & 3452 \\
\end{tabular}
\end{ruledtabular}
\end{table*}
\Cref{tab:v3-T-compare} summarizes the resulting logical-$T$
resource bounds and peak logical-qubit allocations for the
direct and paired constructions.

\paragraph{Qualtran checks.}
Recursive decomposition of explicit Qualtran circuits
\cite{harrigan2024qualtran} reproduces the $T_U$ and
$\mathsf Q_U$ columns under the stated shared-preparation,
precision, and supplied-catalyst conventions. For the first
instance, numerical GQSP phases are also generated and the
rounded angle-specific circuit is constructed and counted.
The larger instances use the generic phase-independent
rotation-resource bound. Reproduction details are provided
in the accompanying repository.

\paragraph{Catalyst preparation.}
For $d\in\{b,m\}$, the catalyst product state of \cref{eq:gradient-product-preparation} uses Clifford phases $Z,S$, one exact $T$ phase when $d\geq2$, and $d-2$ smaller dyadic phase rotations when $d\geq3$.  Suppose a fixed phase-aware compiler synthesizes a phase rotation of angle $\theta$ with operator-norm error at most $\delta$ using $t_{R_z}(\theta,\delta)$ $T/T^\dagger$ gates.
\begin{samepage}
A sufficient one-time preparation cost with trace-norm error budget $\varepsilon_{G,d}$ is then obtained by assigning each smaller phase rotation an operator-norm error budget
\begin{equation*}
\delta_d=\varepsilon_{G,d}/[2(d-2)]:
\end{equation*}
\begin{equation}
\begin{aligned}
T_G(d,\varepsilon_{G,d})
\quad &\leq \quad\mathbf{1}_{d\geq2}+\sum_{r=4}^{d+1}
t_{R_z}\!\left(\frac{2\pi}{2^r},\delta_d\right),
\quad(d\geq3).
\end{aligned}
\label{eq:v3-catalyst-setup}
\end{equation}
\end{samepage}

For $d=2$, only the exact $T$ phase contributes to the cost; for $d=1$, no $T$ gates are required. The displayed sum is empty for $d\leq2$.  Preparing the two catalysts once per shot adds
\begin{equation*}
T_G(b,\varepsilon_{G,b})+
T_G(m,\varepsilon_{G,m})
\end{equation*}
rather than a cost per query.  Their combined trace-norm preparation error contributes at most $\varepsilon_{G,b}+\varepsilon_{G,m}$ to a shot's final-state error, independent of its query count, by the contractivity argument in \cref{sec:endtoend}.  This setup allowance must be included in the observable error budget.

\paragraph{Ladder insertions.}
For a fixed cavity $i$, an externally controlled block encoding of either $a_i/\sqrt{\nmax}$ or $a_i^\dagger/\sqrt{\nmax}$ can use one enabled angle module directly on its $q$ photon bits and one enabled modular shift on those same bits.  The angle table enforces the physical boundary, so no validity comparator or indexed read is required.  This gives the following upper bound on the coherent Toffoli count
\begin{equation}
C_{\rm lad}=A_b+(2q-3),\qquad T_{\rm lad}=7C_{\rm lad}
\quad(q\geq2),
\label{eq:v3-ladder-cost}
\end{equation}
with a distinct projected signal qubit for each of the two ladder insertions as required by \cref{eq:hadamard-target}.  At finite $b$, this exactly encodes the rounded ladder operator; each normalized ladder block differs from its ideal counterpart by at most $\pi/2^{b+1}$.  If the observable requires ladder accuracy $\varepsilon_a$, choose $b$ to satisfy
\begin{equation*}
b\geq\max\left\{1,
\left\lceil\log_2\frac{\pi}{2\varepsilon_a}\right\rceil\right\}
\end{equation*}
together with the Hamiltonian precision rule, and recompute the counts if this increases $b$.

\paragraph{Shot count.}
\Cref{eq:weighted-spectral-shots} gives the shot count for a spectral statistical error of $e_{\rm stat}$ on $N_\omega$ frequencies. If the complex correlators themselves are requested, a sufficient count for additive error $\varepsilon_C$ at every time is
\begin{equation}
N_{\rm shot}^{\rm point}
=\left\lceil\frac{16\nmax^2}{\varepsilon_C^2}
\ln\frac{8N_t}{\delta}\right\rceil .
\label{eq:v3-shots}
\end{equation}
It allocates statistical error $\varepsilon_C/2$ per complex correlator and follows by applying Hoeffding to the $4N_t$ quadrature means. Requiring this accuracy at every time uses more shots than bounding the weighted spectrum directly.

\paragraph{Total spectral cost.}

Let $T_{\rm prep}$ be the logical-$T$ cost of the input state. With $N_{\rm shot}$ from \cref{eq:weighted-spectral-shots}, the common-evolution workload has
\begin{equation}
\begin{aligned}
T_{\rm spec,X}\leq4N_{\rm shot}\sum_{k=0}^{N_t-1}\bigl[
&T_{\rm prep}
+T_G(b_k,\varepsilon_{G,b,k})\\
&+T_G(m_k,\varepsilon_{G,m,k})\\
&+T_{U,X}(\tau_k)+2T_{{\rm lad},k}\bigr].
\end{aligned}
\label{eq:v3-spectrum-T}
\end{equation}
The factor of four accounts for measuring the real and imaginary parts of both correlators. Each circuit uses one evolution and two controlled ladder insertions, without an eigenenergy estimate. At $\tau=0$, the evolution can be omitted. Precision can be chosen separately at each time or fixed to a value sufficient for the full grid. Each run prepares the catalysts once, and their combined trace-norm error contributes to the correlator error budget. The classically calculated phases must also meet the stated axial-angle accuracy.

\paragraph{Illustrative spectral workload.}
Take $(L,N,\nmax)=(16,1,3)$, $gT_{\max}=50$, $W=5g$, $N_t=90$ equally spaced times including both endpoints, trapezoidal weights, $N_\omega=1000$, and $\delta=0.01$. For the $Q=1$ reference at these uniform parameters, $W$ covers the transition frequencies in $Q=0,2$. A different reference state requires a new bandwidth check. Allocate $e_{\rm stat}=e_{\rm sys}=\varepsilon_A/2$ and $\varepsilon_{\rm sim}=\pi\varepsilon_A/(24\nmax T_{\max})$. Assign at most $\varepsilon_{\rm sim}$ to input-state error, each normalized ladder error, and the sum of catalyst errors, as above. Choosing the degree and precision separately at each time gives
\begin{center}
\begin{tabular}{l r r}
Accuracy convention & circuit runs & evolution + ladder $T$\\
$1\%$ of $T_{\max}/\pi$ & $7.070\times10^7$ & $4.359\times10^{15}$\\
Absolute $\varepsilon_A=0.01$ & $1.791\times10^{10}$ & $1.181\times10^{18}$
\end{tabular}
\end{center}
The first line measures accuracy relative to the height of an isolated unit-weight pole under a hard window. Actual spectral peaks may have different heights. The respective shots per time/correlator/quadrature are $196{,}383$ and $49{,}744{,}145$. Input-state and catalyst preparation must be added using \cref{eq:v3-spectrum-T}. The totals count logical gates and do not include physical error correction. Even with the $328$-fold shot reduction and shared evolution, measurement repetitions account for a large part of this example's cost. These conservative estimates describe the chosen algorithm; they do not establish a lower bound or quantum advantage.

\section{Circuits and resource costs for sampled open-system simulation}
\label{app:open-circuits}

The sampled dilation of \cref{sec:open} uses the lattice Hamiltonian and one local jump operator at each timestep. Here we construct the three jump encodings, combine them with the Hamiltonian encoding, and count the gates and qubits needed for a controlled walk. We follow the coherent arithmetic and register conventions of Appendix~\ref{app:matched-query}. The counts below assume $\alpha_\TCL>0$ and $\lambda_{\mathcal D}>0$; a zero-weight branch can be omitted. {The} Hamiltonian encoding, its normalization, and the dilation error bounds are evaluated using the same energy-frame convention as in \cref{sec:open}.

\subsection{Jump circuits}
\label{app:sampled-jumps}

The sampling probabilities in~\cref{eq:sampling} absorb the individual decay rates into the common normalization $\lambda_D$. The resulting normalized jump operators for the three dissipative processes are,
\begin{equation}
J_j=\frac{\widetilde V_j}{\sqrt{\lambda_{\mathcal D}}}
=\begin{cases}
a_i/\sqrt{\nmax},&\text{photon loss},\\
\sigma^-_{i,r},&\text{emitter relaxation},\\
Z_{i,r},&\text{emitter dephasing}.
\end{cases}
\label{eq:sampled-normalized-jumps}
\end{equation}
The sampled address $(i,r)$ is classical and stays fixed during the timestep. Its gates therefore act directly on the chosen photon word or emitter qubit. No quantum address register or indexed jump-routing circuit is needed.

Write $q=\qph$ and $R=2^q$. For photon loss, apply an occupation-dependent rotation to a signal qubit $\eta$ and then decrement the photon word modulo $R$. The rotation angle is
\begin{equation}
\theta_{\rm loss}(n)=\begin{cases}
2\arccos\sqrt{n/\nmax},&0\le n\le\nmax,\\
\pi,&\nmax<n<R.
\end{cases}
\label{eq:sampled-loss-angle}
\end{equation}

Thus, for $1\le n\le\nmax$, the zero-signal block implements the normalized photon annihilation operator:
\begin{equation*}
\bra{0}_\eta U_{\rm loss}\ket{0}_\eta\ket n
=\sqrt{\frac{n}{\nmax}}\ket{n-1}.
\end{equation*}
The vacuum and unused photon labels have zero amplitude in this block, so the modular decrement introduces no unphysical contribution to the encoded operator.

The occupation-dependent rotation is implemented using the angle-lookup and phase-gradient circuit of Appendix~\ref{app:angle-explicit}, with the angles specified by \cref{eq:sampled-loss-angle}. The angle register is uncomputed before decrementing the photon occupation. For a supplied enable qubit $u$, the Toffoli cost is
\begin{equation}
c_{\rm loss}=A_b+\max(2q-3,0),
\qquad A_b=4(R-2)+4b-2.
\label{eq:sampled-loss-cost}
\end{equation}
The second term accounts for the enabled modular decrement using reversible carry arithmetic. For $q=1$, the decrement requires only a CNOT. Since the selected cavity is known classically, its photon register directly addresses the angle table, without an additional indexed-read circuit.

For emitter relaxation, let $(i,r)$ denote the classically selected emitter, with $\ket0_{i,r}$ and $\ket1_{i,r}$ representing its ground and excited states, respectively. The unitary
\begin{equation}
U_-=X_{i,r}\,
\mathrm{CX}_{(i,r)\to\eta}\,X_\eta
\label{eq:sampled-relaxation-circuit}
\end{equation}
satisfies
\begin{equation*}
\bra0_\eta U_-\ket0_\eta
=\ket0\bra1_{i,r}=\sigma^-_{i,r}.
\end{equation*}
An enabled implementation applies $\mathrm{CX}_{u\to\eta}$, $\mathrm{CCX}_{u,(i,r)\to\eta}$, and $\mathrm{CX}_{u\to(i,r)}$ in that order. It acts as the identity when $u=0$ and requires one Toffoli gate.

\begin{figure*}
\centering
\input{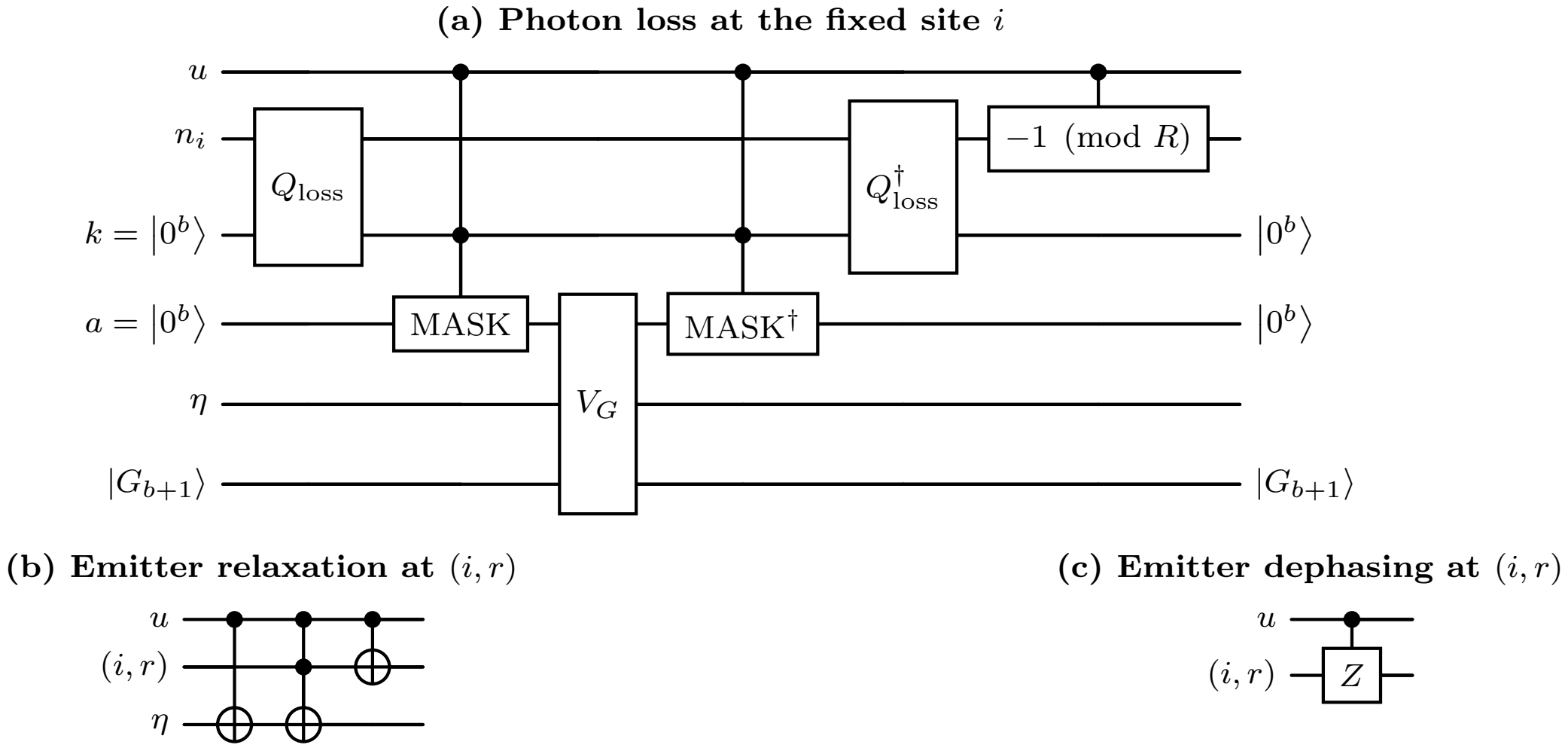}
\caption{Enabled jump encodings, read from left to right. The classically sampled address is fixed for the timestep. (a) The loss-angle lookup $Q_{\rm loss}$ writes the $b$-bit angle from \cref{eq:sampled-loss-angle}. The MASK gates and $V_G$ apply the enabled $R_y$ rotation to $\eta$ as in Appendix~\ref{app:angle-explicit}, and the lookup is erased before the selected photon word is decremented. Both angle words return to zero and the phase-gradient catalyst returns to its input state. (b) Emitter relaxation uses one Toffoli between two CNOTs. (c) Dephasing uses a $CZ$. In all panels $u$ is a supplied enable; every circuit is the identity when $u=0$. The projected jump block uses $\eta=0$, while the full controlled circuits remain unitary for arbitrary signal inputs.}
\label{fig:sampled-jumps}
\end{figure*}

Dephasing is implemented by $U_z=Z_{i,r}$, whose enabled version is a controlled-$Z$ gate. Consequently, the Toffoli costs are $c_{\rm relax}=1$ and $c_{\rm dephase}=0$. \Cref{fig:sampled-jumps} illustrates the three jump circuits. Their adjoints are implemented by reversing the corresponding gate sequences, including the finite-precision angle-lookup and rotation operations.

\subsection{Controlled dilation encoding and walk}
\label{app:sampled-walk}

The dilation qubit $d$ is part of the enlarged system and is initialized in $\ket0$ at the beginning of each timestep. It is not included in the block-encoding ancillas. Let $A_H$ denote the Hamiltonian ancilla register of \cref{eq:walk-projected-registers}, with width $a$. We reuse its signal qubit $\eta$ for the jump encoding and introduce two additional ancillas: $\beta$ selects between the Hamiltonian and jump contributions, while $f$ removes the unwanted Hamiltonian contribution when $d=1$. The complete projected register is
\begin{equation}
A_{\rm dil}=(\beta,f,A_H),
\qquad a_{\rm dil}=a+2.
\label{eq:sampled-projected-registers}
\end{equation}
The Hermitianization qubit $h$ and GQSP signal $q_{\rm GQSP}$ are introduced separately.

Write the Hamiltonian block encoding as
\begin{equation*}
F_H=\PREP_\hop\PREP_\loc P_\TCL,
\qquad
U_\TCL=F_H^\dagger S_HF_H,
\end{equation*}
following \cref{eq:controlled-encoding-factor}. Here, $F_H$ acts only on the Hamiltonian preparation registers and leaves the system and the shared signal qubit $\eta$ unchanged. The jump circuits act on disjoint preparation registers and therefore commute with $F_H$. Define the two selected branches by
\begin{equation}
\begin{aligned}
S_{H,d}&=\proj0_d\otimes S_H+\proj1_d\otimes X_f,\\
S_{J,d}&=X_d\bigl(\proj0_d\otimes U_{J_j}
              +\proj1_d\otimes U_{J_j}^\dagger\bigr).
\end{aligned}
\label{eq:sampled-selected-branches}
\end{equation}
Identities on the remaining registers are implicit. In the Hamiltonian branch, $S_H$ is applied when $d=0$, whereas $X_f$ is applied when $d=1$. Projection onto $\ket0_f$ therefore eliminates the lower-right Hamiltonian block. The jump branch flips $d$ and places $J_j$ and $J_j^\dagger$ in the off-diagonal blocks of the dilation Hamiltonian.

The branch-selection qubit is prepared using
\begin{equation}
P_\beta=R_y\!\left(
2\arcsin\sqrt{\frac{\sqrt{\lambda_{\mathcal D}}}{\alpha_{\rm dil}}}
\right).
\label{eq:sampled-branch-preparation}
\end{equation}
The complete dilation encoding is
\begin{equation}
\begin{aligned}
F_D&=F_HP_\beta,\\
S_D&=\proj0_\beta\otimes S_{H,d}
       +\proj1_\beta\otimes S_{J,d},\\
U_D&=F_D^\dagger S_DF_D.
\end{aligned}
\label{eq:sampled-shared-encoding}
\end{equation}
Because $F_H$ commutes with $S_{J,d}$, its preparation and unpreparation cancel in the jump branch. The projected block is therefore $\widetilde H_j/\alpha_{\rm dil}$, with the normalization in \cref{eq:alpha-dil}. This construction uses the Hamiltonian preparation once, even though both selected branches share $A_H$.

Apply the Hermitianization of \cref{lem:hermitian} to $U_D$ and form $W_j=R_\Pi\widetilde U_D$, where
\begin{equation}
\Pi=\proj+_h\otimes\proj{0^{a+2}}_{A_{\rm dil}}
\otimes I_{d,\rm sys},\qquad R_\Pi=2\Pi-I.
\label{eq:sampled-walk-projector}
\end{equation}
The physical dilation qubit is outside the projector. Reflecting it about $\ket0$ would instead remove the off-diagonal jump blocks from the simulated system.

\begin{figure*}
\centering
\input{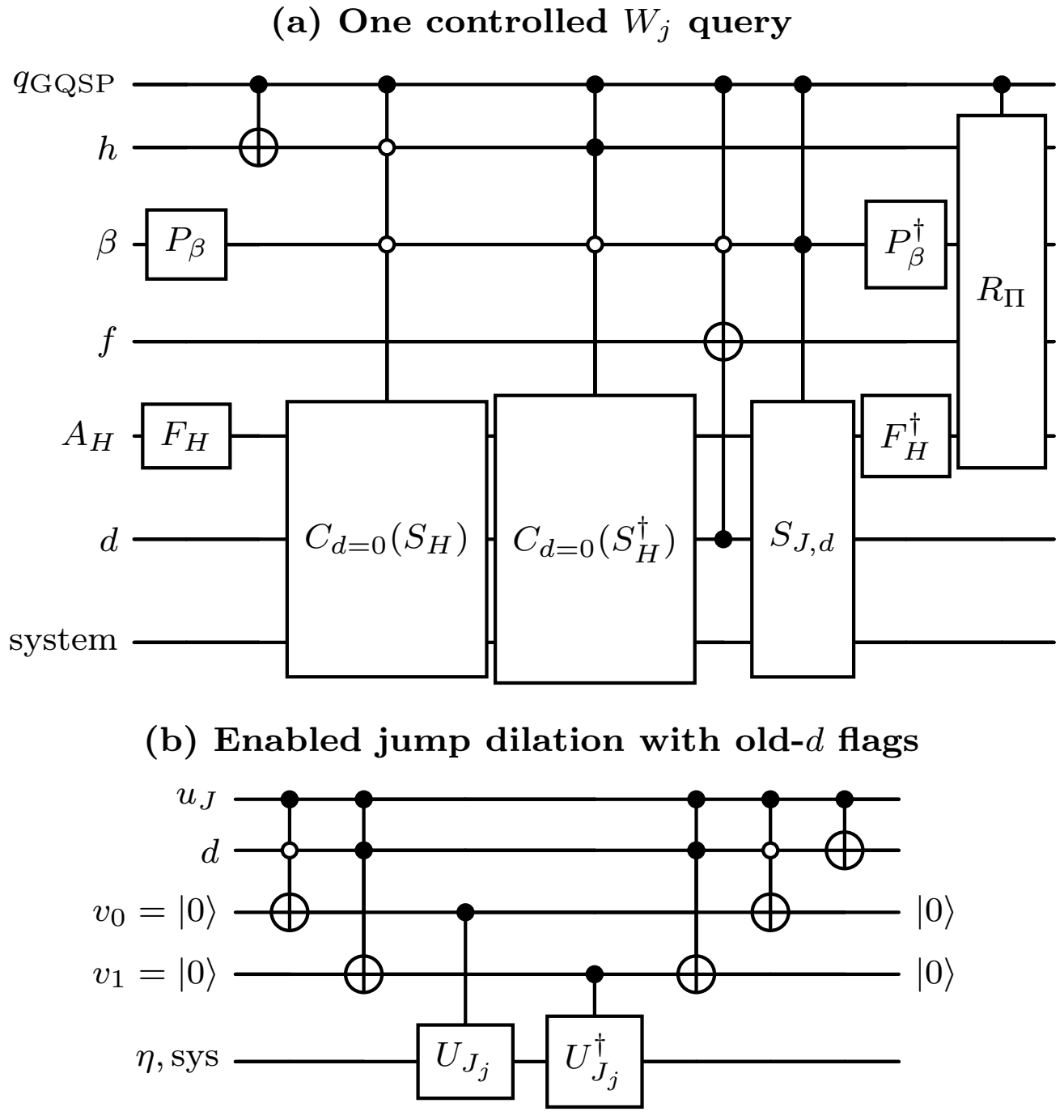}
\caption{Controlled sampled-dilation walk with shared preparation, read from left to right. (a) The Hamiltonian and adjoint selections use the value of $h$ after its controlled flip. The $C_{d=0}$ gates act only for $d=0$. The filter flip and the self-adjoint jump selection are independent of $h$ and appear once. The final reflection acts on $h,\beta,f,A_H$, leaving the physical dilation qubit $d$ and the lattice system outside the projector. (b) Expansion of the enabled $S_{J,d}$ for loss or relaxation. Two flags are computed from the old $d$, used to enable the jump and its exact adjoint, and erased before $d$ is flipped. The supplied $u_J$ is computed and erased in the outer circuit. For dephasing this panel reduces to $\mathrm{CX}_{u_J\to d}$ and {$\mathrm{CZ}_{u_J,(i,r)}$}. Clean scratch and the returned catalysts are omitted from panel (a).}
\label{fig:sampled-walk}
\end{figure*}

There is a further simplification. The jump selection satisfies $S_{J,d}^\dagger=S_{J,d}$ exactly, since it uses the adjoint of the implemented $U_{J_j}$. The filter flip is also self-adjoint. These operations need not be repeated in the two $h$ branches. \Cref{fig:sampled-walk}(a) gives the resulting controlled walk. Apply $F_D$, then flip $h$ controlled on $q_{\rm GQSP}$. Using $h$ after this flip, apply $S_H$ for $(q_{\rm GQSP},\beta,d,h)=(1,0,0,0)$ and $S_H^\dagger$ for $(1,0,0,1)$. Apply $X_f$ for $(q_{\rm GQSP},\beta,d)=(1,0,1)$, and apply $S_{J,d}$ once for $(q_{\rm GQSP},\beta)=(1,1)$. Finish with $F_D^\dagger$ and the controlled reflection. When $q_{\rm GQSP}=0$, all selections and the reflection are inactive, and $F_D^\dagger F_D=I$ on the full register space.

To implement the Hamiltonian controls, compute $u_H=q_{\rm GQSP}\wedge[\beta=0]$ and $v_H=u_H\wedge[d=0]$. Compute and erase $e_H=v_H\wedge[h=h_0]$ around each of the two Hamiltonian selections. Their internal local and hopping controls then use $e_H$ as the supplied enable. The filter flip is $\mathrm{CCX}_{u_H,d\to f}$. Clear $v_H$ and $u_H$ before the jump branch changes $d$.

For the jump branch, compute $u_J=q_{\rm GQSP}\wedge[\beta=1]$. \Cref{fig:sampled-walk}(b) computes two flags from the old value of $d$, applies $U_{J_j}$ or its adjoint, erases the flags, and then flips $d$ controlled on $u_J$. The flags must be erased before the flip because their definitions use the old $d$. Finally erase $u_J$. For dephasing, $S_{J,d}=X_d{Z_{i,r}}$, so a CNOT and a $CZ$ implement the entire enabled jump selection without these two flags. The controlled reflection uses the circuit in \cref{fig:walk-explicit}(b), now with $a+2$ zero-projected qubits. It includes the control-dependent sign of \cref{eq:walk-reflection-sign}. Adjoint walk queries reverse this full circuit. Changing the GQSP control to zero uses Clifford $X$ conjugation of its signal wire.

\subsection{Gate counts and precision}
\label{app:sampled-gates}

Let $C_{\rm dir}$ and $C_F$ be the controlled Hamiltonian encoding and preparation-pair counts in \cref{eq:v3-direct-cu}. The cost of $S_H$ with a supplied enable is
\begin{equation}
C_S=C_{\rm dir}-C_F-2.
\label{eq:sampled-H-select-cost}
\end{equation}
The subtraction removes the preparation pair and the two Toffolis that compute and erase the original joint $(q_{\rm GQSP},h)$ enable. The internal local and hopping enables remain in $C_S$. With a supplied jump enable, the dilation selection costs
\begin{equation}
{C_{J,j}}=\begin{cases}
2c_{\rm loss}+4,&\text{photon loss},\\
6,&\text{emitter relaxation},\\
0,&\text{emitter dephasing}.
\end{cases}
\label{eq:sampled-jump-dilation-cost}
\end{equation}
For loss and relaxation, the four additional Toffolis compute and erase the two old-$d$ flags. Dephasing uses the direct $X_d{Z_{i,r}}$ circuit.

The Hamiltonian branch uses eight Toffolis to compute and erase $u_H$, $v_H$, and the two successive $e_H$ flags, and one for the filter flip. Computing and erasing $u_J$ costs two more. The reflection costs $2(a+2)$ Toffolis. Hence one controlled walk has arithmetic cost
\begin{equation}
\begin{aligned}
C_{W,j}&=C_F+2C_S+11+{C_{J,j}}+2(a+2)\\
       &=2C_{\rm dir}-C_F+2a+11+{C_{J,j}}.
\end{aligned}
\label{eq:sampled-walk-cost}
\end{equation}
This count uses ordinary coherent Toffolis and excludes rotation synthesis and catalyst preparation. Relative to the shared closed-system walk for the same Hamiltonian encoding, the additional arithmetic cost is $11+{C_{J,j}}$. The worst-case query cost in \cref{thm:open} can therefore be taken as $C_{\rm qry}^{\rm samp}=\max_j C_{W,j}$ over the channels with nonzero sampling probability.

{For the retained-label $(16,1,3)$ instance of \cref{tab:v3-coherent-compare}, with $(b,B)=(20,19)$, the construction in Appendix~\ref{app:matched-query} gives $C_{\rm dir}=2308$, $C_F=1142$, and $a=108$. Thus $a_{\rm dil}=110$ and $c_{\rm loss}=87$. The resulting counts are
\begin{center}
\begin{tabular}{l r r r}
Sampled channel & $c_j$ & $C_{J,j}$ & $C_{W,j}$\\
Photon loss & 87 & 178 & 3879\\
Emitter relaxation & 1 & 6 & 3707\\
Emitter dephasing & 0 & 0 & 3701
\end{tabular}
\end{center}
The corresponding shared closed-system walk is the 3690-Toffoli walk of \cref{tab:v3-coherent-compare}, so at the same precision the sampled walk costs 0.3--5.1\% more. The coefficients and padded widths are those of Appendix~\ref{app:matched-query}; the diagonal labels are retained with zero weight. These widths match the closed-system comparison; a requested open-system accuracy requires the precision choices below.}

Let $\epsilon_s=\varepsilon_{\rm step}$ and $x=\alpha_{\rm dil}\sqrt{\Delta t}$. The Hamiltonian angle errors propagate for time $\Delta t$, while the loss-angle error is multiplied by $\sqrt{\lambda_{\mathcal D}\Delta t}$. A common bound is
\begin{equation}
\Phi_b=\Delta t(2\alpha_\hop+\alpha_\intt)
          +\sqrt{\lambda_{\mathcal D}\Delta t}.
\label{eq:sampled-angle-scale}
\end{equation}
The last term can be dropped when photon loss is absent. Using the angle and alias bounds of Appendix~\ref{app:error}, sufficient grid widths are
\begin{equation}
\begin{aligned}
b&\ge\max\left\{1,\left\lceil
\log_2\frac{4\pi\Phi_b}{\epsilon_s}\right\rceil\right\},\\
B&\ge\max\left\{1,\left\lceil
\log_2\frac{8\alpha_\TCL\Delta t}{\epsilon_s}
\right\rceil\right\}.
\end{aligned}
\label{eq:sampled-data-precision}
\end{equation}
These choices assign at most $\epsilon_s/8$ to all occupation-angle errors and $\epsilon_s/8$ to the alias preparations after propagation through one timestep. An absent table or preparation has no corresponding constraint.

Use the fixed-angle circuit of Appendix~\ref{app:matched-t} for $P_\TCL$, $P_\beta$, and the GQSP rotations, with a second phase-gradient catalyst of width $m+1$. Choose a Jacobi--Anger cutoff $K$ such that $2\eta_K(x)\le\epsilon_s^2/32$, and set $n_K=2K+1$. One sufficient grid width is
\begin{equation}
m\ge\max\left\{1,
\left\lceil\log_2\frac{8\pi x}{\epsilon_s}\right\rceil,
\left\lceil\log_2\frac{6\pi n_K}{\epsilon_s}\right\rceil
\right\}.
\label{eq:sampled-fixed-precision}
\end{equation}
Each axial rotation has operator-norm error at most $\pi/2^{m+1}$. The first logarithm bounds the propagated error of each preparation pair by $\epsilon_s/8$; using $x$ for the inner Hamiltonian pair is conservative. Together with \cref{eq:sampled-data-precision}, these four contributions use at most half of the timestep budget. The polynomial contributes at most $\epsilon_s/4$ to initialized-state error, including leakage, by Appendix~\ref{app:error}. The second logarithm assigns the remaining $\epsilon_s/4$ to the $3n_K$ Euler factors of the signal rotations. These bounds assume that classical angle-generation errors fit the stated angle tolerances; otherwise their contribution must also be included.

The shared preparation pair contains four axial rotations: two $P_\TCL$ rotations and two $P_\beta$ rotations. Each uses $2(m-1)$ Toffolis in the fixed-angle construction. The GQSP schedule of Appendix~\ref{app:walk} uses $2K$ controlled walks and $n_K$ signal rotations. Converting each coherent Toffoli to seven logical $T$ gates gives the timestep bound
\begin{equation}
T_{{\rm step},j}\le
7\bigl[2K C_{W,j}+(16K+6n_K)(m-1)\bigr].
\label{eq:sampled-step-T}
\end{equation}
Constant-angle simplifications can lower this specified construction's cost. With common degree and precision for all sampled channels, the worst-case evolution cost is $N_{\rm step}\max_jT_{{\rm step},j}$, while its mean is $N_{\rm step}\sum_jp_jT_{{\rm step},j}$. Initial-state and catalyst preparation are added once per run. The measurement repetitions in \cref{sec:g2-resources} multiply the cost of each complete run.

The transient-statistics example of \cref{sec:g2-resources} has $N_{\rm step}=251{,}658{,}240$ steps of size $\Delta t=3.97\times10^{-8}$ and $\epsilon_s=e/(6N_{\rm step}) =1.03\times10^{-12}$, so $x=5.06\times10^{-4}$. \Cref{eq:sampled-data-precision,eq:sampled-fixed-precision} then give $(b,B)=(33,24)$, $K=6$, and $m=48$, with $m$ set by the Euler-factor term. At these widths, $C_{\rm dir}=2564$, $C_F=1242$, $a=128$, and $c_{\rm loss}=139$, and the shared closed-system walk costs $4142$ Toffolis. \Cref{eq:sampled-walk-cost,eq:sampled-step-T} give
\begin{center}
\begin{tabular}{l r r r}
Sampled channel & $C_{J,j}$ & $C_{W,j}$ & $T_{{\rm step},j}$\\
Photon loss & 282 & 4435 & 429{,}786\\
Emitter relaxation & 6 & 4159 & 406{,}602\\
Emitter dephasing & 0 & 4153 & 406{,}098
\end{tabular}
\end{center}
Dephasing has zero rate in this example, so loss and relaxation are sampled with probabilities $3/4$ and $1/4$. The mean is $4.24\times10^{5}$ $T$ gates per step. The arithmetic part of the $12$ controlled walk queries accounts for $86$--$87\%$ of the timestep's logical $T$-gate count; the axial rotation circuits account for the remainder.

\subsection{Qubit allocation and timestep resets}
\label{app:sampled-qubits}

Three additional clean enable qubits suffice for the control sequence. During the Hamiltonian selection they hold $u_H$, $v_H$, and $e_H$. After these are cleared, the same wires hold $u_J$ and the two old-$d$ flags. Let $w$ be the arithmetic-workspace width in \cref{eq:v3-work-bank}. The encoding workspace and these three flags are clean before the reflection, so their qubits can also supply its $a+2$ scratch bits. The fixed-angle adder needs at most $2m-1$ clean bits when the query flags are idle. Keeping both phase-gradient catalysts on separate persistent wires gives the conservative allocation
\begin{equation}
\begin{aligned}
{\mathsf Q_{\rm samp}}\le{}&L(q+N)+3+(a+2)\\
&+(b+1)+(m+1)\\
&+\max\{a+2,w+3,2m-1\}.
\end{aligned}
\label{eq:sampled-qubit-allocation}
\end{equation}
The $+3$ counts $q_{\rm GQSP}$, $h$, and the physical dilation qubit $d$. The catalysts and clean scratch are outside the zero projector. This is an achievable allocation for the stated circuits; other schedules may use fewer qubits. For the example of Appendix~\ref{app:sampled-gates}, \cref{eq:v3-work-bank} gives $w=167$, so the maximum in \cref{eq:sampled-qubit-allocation} is $w+3=170$ and $\mathsf Q_{\rm samp}\le434$, including $L(q+N)=48$ system qubits. At the same widths, the closed-system allocation $\mathsf Q_{U,\rm dir}=\mathsf Q_{\rm dir}+m+1$ of Appendix~\ref{app:matched-t} is $428$. The six additional qubits are $d$, $\beta$, $f$, and the three enable flags.

\begin{figure*}
\centering
\input{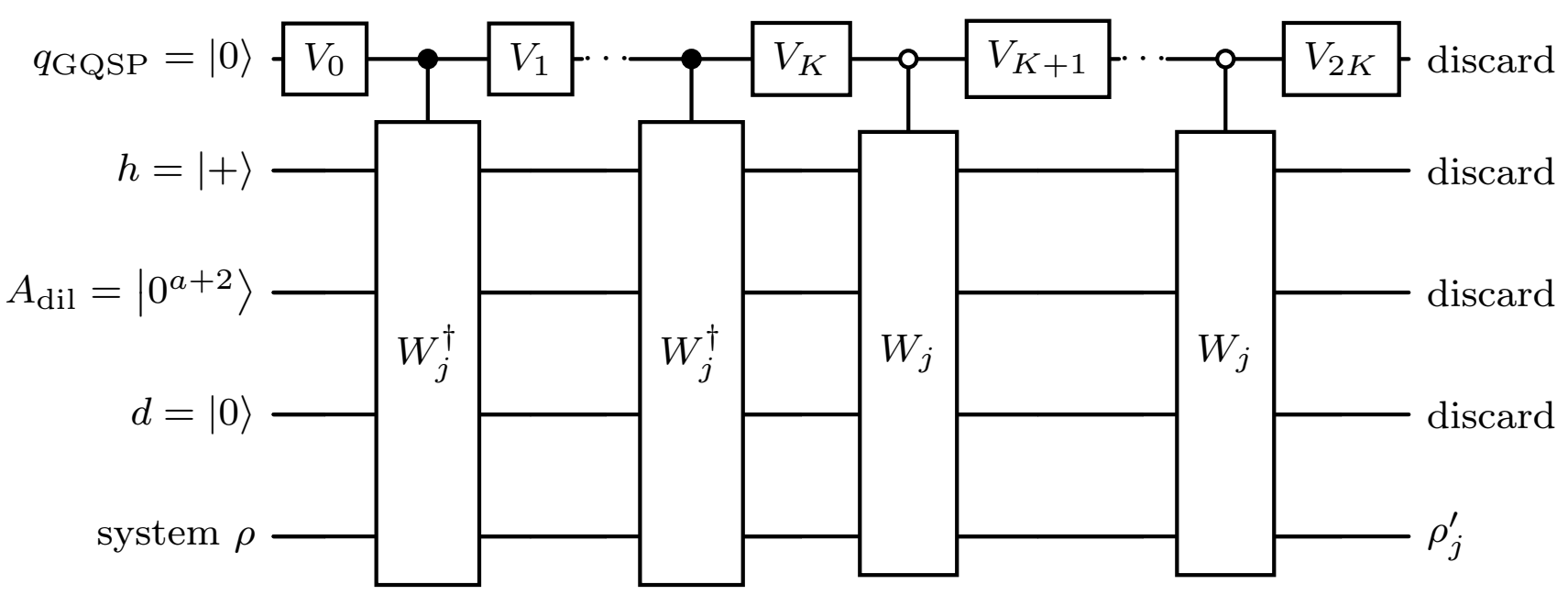}
\caption{One physical timestep after drawing the jump index $j$ classically. The first $K$ walk queries are $W_j^\dagger$ controlled on $q_{\rm GQSP}=1$; the next $K$ are $W_j$ controlled on $q_{\rm GQSP}=0$. The $2K+1$ signal rotations are chosen for $x=\alpha_{\rm dil}\sqrt{\Delta t}$ and supply the polynomial sequence of Appendix~\ref{app:walk}. The displayed initial states are prepared once for the whole timestep. After the final rotation, $q_{\rm GQSP},h,A_{\rm dil},d$ are discarded, giving the reduced system state $\rho'_j$; they are reinitialized before the next physical timestep. No signal or dilation reset occurs between walk queries. The clean workspace and both persistent phase-gradient catalysts are omitted.}
\label{fig:sampled-timestep}
\end{figure*}

At the start of each physical timestep, initialize $q_{\rm GQSP}$ in $\ket0$, $h$ in $\ket+$, $A_{\rm dil}$ in $\ket{0^{a+2}}$, and $d$ in $\ket0$. \Cref{fig:sampled-timestep} shows the complete GQSP sequence. The simulation signals and $d$ are retained throughout its walk queries. They can leave their initial states during a query, so resetting them between queries would change the evolution. After the complete sequence, trace out $d$ and the simulation signals, then initialize them for the next timestep. With finite precision, the signals can retain small components outside their initial states. The initialized-state bound above includes this leakage and gives at most $2\epsilon_s$ diamond-norm error for the reduced timestep channel, as derived in Appendix~\ref{app:open-cost}.

The clean arithmetic registers return to zero after each completed query. Exact phase-gradient catalysts return to their input states after every use and remain available for the next timestep. If catalyst preparation is approximate, its joint trace-norm error is counted once per run: contractivity bounds its effect through all subsequent unitaries, discarded ancillas, and classical jump samples.

We checked the jump blocks, shared dilation encoding, and controlled walk by matrix calculations using one two-bit photon word, one emitter, and a coefficient label. The check includes an unused photon label and a Hamiltonian selection whose full unitary is not self-inverse. For both values of the GQSP control, the displayed sequence agrees with the reference walk and returns every enable flag to zero. These small-instance calculations check operator actions andcontrol logic; the resource counts are checked separately by the executable circuit decompositions described in Appendix~\ref{app:sampled-gates}.

Recursive decomposition of executable Qualtran circuits reproduces the per-channel timestep counts above and the logical-qubit allocation of \cref{eq:sampled-qubit-allocation}, using the stated fixed-angle rotation circuits and seven-$T$ coherent Toffolis. This verifies the specified circuit accounting; it does not generate the numerical GQSP phase sequence or resolve the outstanding dilation-error allowance.

\FloatBarrier

\makeatletter
\if@filesw
\immediate\write\@auxout{\string\citation{TCLupdatedv1Control}}\fi
\makeatother

\bibliographystyle{apsrev4-2}
\bibliography{references}

\end{document}